# EPIDEMIOLOGY OF LATENCY AND RELAPSE IN

# *PLASMODIUM VIVAX* MALARIA

## Andrew A. Lover

(BA, MSc, MPH)

A THESIS SUBMITTED IN PARTIAL FULFILLMENT OF

THE REQUIREMENTS FOR THE DEGREE OF

DOCTOR OF PHILOSOPHY

Public Health (Epidemiology)

SAW SWEE HOCK SCHOOL OF PUBLIC HEALTH

NATIONAL UNIVERSITY OF SINGAPORE

2015

## Declaration

I hereby declare that this thesis is my original work and it has been written by me in its entirety. I have duly acknowledged all the sources of information that have been used in this thesis. This thesis has also not been submitted for any degree in any university previously.

\_\_\_\_\_\_\_\_\_\_\_\_\_\_\_\_\_\_\_\_\_\_\_\_\_\_\_\_\_\_\_\_\_\_\_\_\_

Andrew A. Lover

29 April 2015

When different fields of inquiry have been separately cultivated for a while, the borderland between them often provides fertile ground for new investigations.

- Allyn A. Young, 1924; quoted in (Granados 2003).

As scientists and public health workers most of us suffer from a touch of schizophrenia. Though we may rejoice that there are still a few malaria parasites available for basic research we must not forget that we are dedicated to the campaign against a disease which, until recently, kept half the world in servitude and today still divides the rich world from the poor. Malaria eradication in spite of its technical setbacks must succeed, and this alone merits all our efforts.

- Leonard J. Bruce-Chwatt (Bruce-Chwatt 1965).

# Acknowledgements

Science never occurs in a vacuum, and this work is obviously no different. I am extremely grateful for the entire community at SPH/NUS and beyond that has made these efforts possible.

First and foremost, I am truly indebted for my doctoral committee for all of their suggestions, prodding and penetrating queries along this rather meandering research path. Richard Coker has been a fantastically supportive mentor, and was always willing and able to find time to discuss research progress and pitfalls. Moreover, when things veered off into overly-esoteric parasitology, he made sure to pull it back into direct public health relevance- 'Great, but what are the policy implications?' which has been a critically important lesson.

Kee Seng Chia was instrumental in creating an environment where the first amorphous ideas could take shape and become a thesis, and moreover fostered travel to endemic areas and conferences to connect this work to the larger malaria community. David Heymann was a critical sounding board for this and other studies, and his wealth of experience and advice about 'where the rubber hits the road' helped to root this work in the practical realities of infectious disease control. Finally, Li Yang Hsu was always interested and supportive in allowing this work to run in parallel with my SPH official duties.

Many thanks to Alex Cook for providing a sounding board and sage advice for many statistical nuances, plus hard-core editorial assistance for all of these studies.

Finally, and most importantly, I am grateful for the support, patience, and endless understanding from my wife Leontine during this entire process.

# Table of Contents

# Summary

Malaria is a major contributor to morbidity and mortality throughout the regions where it is endemic; there are six species that commonly infect humans: *Plasmodium falciparum, P. vivax, P. ovale* (two sympatric species), *P. malariae*, and *P. knowlesi*. Historically, it was believed that there was limited morbidity and essentially no mortality associated with *P. vivax*, and so this parasite was not a major contributor to disease burden on a global scale. This paradigm is being rigorously re-evaluated, and evidence from diverse settings now suggests that infections with *P. vivax* can be both severe and fatal.

This increasing awareness has highlighted a critical gap: the vast majority of research has been directed towards *P. falciparum*, and so there exists a decades-long neglect of epidemiological and clinical studies of *P. vivax*. As efforts towards global malaria elimination have progressed, two facets have become clear: programs directed toward decreasing *P. falciparum* transmission may have very limited impact on *P. vivax*, and the biology of this parasite (especially that of hypnozoites, the dormant liver stages) will be a major barrier to elimination.

There exists a large body of historical data on human experimental infections with *P. vivax* from two major sources: pre antibiotic-era treatment for neurosyphilis ('malariotherapy'), and antimalarial drug trials in prison volunteers. These studies in controlled settings provided a wealth of wide-ranging statements based on expert opinion, which form the basis for much of what is currently known about *P. vivax*. In this thesis, portions of this evidence base have been re-examined using modern epidemiological analyses with two primary aims: to critically examine this accumulated knowledge base, and to inform current research agendas towards global malaria elimination for all species of *Plasmodium*.

Specifically, Chapter 1 provides an overview of malaria, including the parasitology and epidemiology of *P. vivax*, and discussion about malariotherapy and related studies. Chapter 2 examines geographic variation in the epidemiology of *P. vivax*, especially the timing of incubation periods and of relapses, by broad geographic regions determined by origin of the parasites. Chapter 3 reassesses the impact of sporozoite dosage upon incubation and pre-patent periods (a critical consideration in modern vaccine trials); Chapter 4 provides well-defined mathematical distributions for incubation and relapses periods in experimental infections, and explores the epidemiological impacts of these distributions using simple mathematical models of transmission. Chapter 5 examines the epidemiology of mixed-strain *P. vivax* infections and compares these results with studies in diverse murine malaria models and general ecological theory; and Chapter 6 clarifies the origin of the Madagascar strain of *P. vivax*, to potentially provide data to explore the emerging awareness of *P. vivax* transmission in sub-Saharan Africa. Finally, Chapter 7 concludes the thesis with suggestions for future research.

# List of Tables

# List of Figures

**Abbreviations**

ACT: artemisinin-combination therapy.

EIR: entomological inoculation rate.

HR: hazard ratio (effect size estimate from survival models).

IRR: incidence rate ratio (effect size estimate from Poisson models).

IRS: indoor residual spraying.

LLITN: long lasting insecticide-treated bed net.

OR: odds ratio (effect size estimate from logistic models).

RDT: rapid diagnostic test.

RMST: restricted mean survival time.

RR: risk ratio (effect size estimate from binomial generalized linear models).

SIR: Susceptible-infected-recovered disease model.

# Chapter 1: Introduction

## 1.1 Malaria within a global context

Despite decades of control measures and intensive interventions, malaria continues to cause extensive morbidity and mortality throughout the widespread regions where it is endemic. There are six species that commonly cause malaria infections in humans: *Plasmodium falciparum*, *P. vivax*, *P. ovale* (two sympatric species), *P. malariae*, and *P. knowlesi*; and recently *P. cynomolgi* has been implicated as a zoönosis (Ta *et al*. 2014). The vast majority of research has been directed towards *P. falciparum*, which is the primary contributor to disease burden throughout sub-Saharan Africa.

The most recent reporting from the Global Burden of Disease study estimated that in 2013 there were ~166 million (95% Bayesian credible interval: 95 to 284 million) incident cases, and ~855,000 (95% CI: 703,000 to 1.03 million) deaths globally (Murray, C.J.L. and Global Burden of Disease Group 2014). The most recent World Malaria report estimated that in 2012 there were 207 million (95% CI: 135 to 287 million) cases, and 627,000 deaths (95% CI: 473,000 to 789,000) (World Health Organization 2013a). The extensive uncertainty in these estimates highlights major limitations in data sources in many high burden countries, as well as differing model assumptions, especially concerning the use of verbal autopsy methods.

From 1955-1972, the World Health Organization spearheaded a Global Malaria Eradication campaign. These tightly coordinated regional- and national-level programs succeeded in greatly reducing the malaria prevalence in many epidemiological settings, but numbers quickly rebounded upon tapering of control activities in many areas. Consequently, research into malaria control and elimination languished for decades. However, this situation improved in 1999 with the establishment of the Roll Back Malaria programme, and then changed again in 2007 with the dramatic announcement of a renewed push for the goal of malaria *eradication* by the Bill and Melinda Gates Foundation. This clarion call has greatly revitalized malaria research; however, large gaps remain in knowledge of malaria transmission and epidemiology (Baird 2007, Cotter *et al*. 2013).

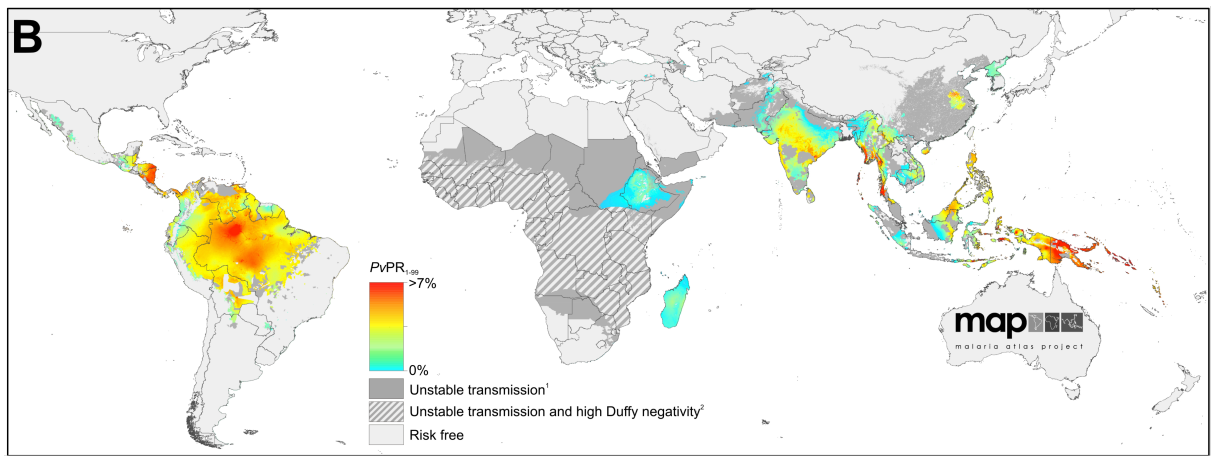

**Figure 1. Modelled geographic range of *Plasmodium vivax*, 2010.**
Source: (Gething *et al*. 2012) ('CC BY' license).

## 1.2 Malaria caused by *Plasmodium vivax*

*Plasmodium vivax* is the major parasite outside of Sub-Saharan Africa, with extensive burden in South and Southeast Asia, and in the Pacific regions. There is very large uncertainty in the global burden of *P*. *vivax* due to reporting biases; the most recent estimates are 19 million (95% CI: 16 to 22 million) cases per year (World Health Organization 2013a). More importantly, this parasite has a much wider geographic range than *P*. *falciparum*, and it is estimated that a total of 2.85 billion people (40% of the world population) are at risk for infection, the majority being in South and Central Asia (Guerra *et al*. 2010). Detailed models have also been produced that predict the geographic range on a global scale (see Figure 1) (Gething *et al*. 2012).

Historically, it was believed that there was limited morbidity, and essentially no mortality associated with *P. vivax*, and hence it was assumed that this parasite was not a major contributor to overall disease burden on a global scale. However, this paradigm of 'benign tertian' malaria is now being rigorously re-evaluated, and evidence from diverse settings now suggests that infections with *P. vivax* can be both severe and fatal. Moreover, several authors suggest that malariologists in the early 20[th] century were in fact well aware of potentially fatal outcomes from *P. vivax* (Baird 2013).

The general clinical course of infections with *P. vivax* are similar to those from *P. falciparum*- fever, headache, nausea, chills, and rigors (Warrell and Gilles 2002), and severe disease (as assessed by WHO standard definitions for severe malaria) has been documented in infections with *P. vivax* in a range of transmission settings. Case series in Papua New Guinea, Indonesia, Thailand, and India have found that 20-27% of patients with severe malaria had PCR-confirmed *P. vivax* mono-infection (Price *et al*. 2009), and village cohorts in Papua New Guinea found no difference in the odds of severe disease amongst patients with *P. falciparum and P. vivax* mono-infections, respectively (OR 0.99; 95% CI: 0.78 to 1.24) (Baird 2013).

More critically, there are several aspects that differ in important ways from severe disease in *P. falciparum* infections- hyperparasitemia is only very rarely observed in *P. vivax* infections (where it rarely exceeds 5%); and there is no substantial evidence for cyto-adherence or rosetting of red blood cells in *P. vivax* infections (Anstey *et al*. 2012). These findings suggest that the pathogenesis of severe disease may have fundamentally different mechanism in the two species. An immune response-dependant pathway has been suggested, based on the observation that the pyrogenic threshold (density of parasites at first sign of fever) may be significantly lower in *P. vivax* than in *P. falciparum* infection (Anstey *et al*. 2012).

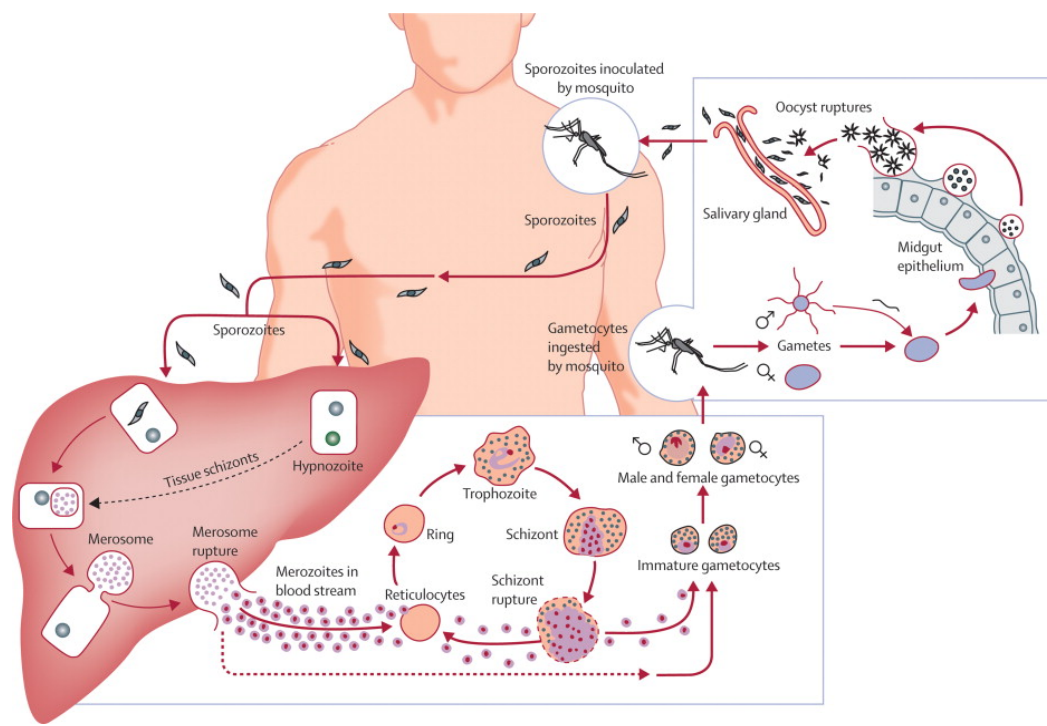

**Figure 2. Schematic lifecycle of *Plasmodium* in human and anopheline hosts.**
Source: (Mueller *et al*. 2009) (see Appendix for copyright approval).

## 1.3 Challenges towards elimination of *P. vivax* malaria

The major knowledge gaps in the biology, clinical presentation, ecology, and epidemiology of *P. vivax* have been highlighted (Galinski and Barnwell 2008, Mueller *et al*. 2009, WHO Malaria Policy Advisory Committee and Secretariat 2013, World Health Organization 2013b), and a recent publication outlines some of the main gaps in current control strategies for vivax malaria, but provides limited suggestions for further research, especially regarding epidemiological studies (Shanks 2012).

Several aspects of the biology and parasitology of *P. vivax* have major implications specifically for malaria control and elimination programs. First and foremost, the existence of a dormant liver stage, the hypnozoite (Greek for 'sleeping animal'), is a critical concern for surveillance towards elimination (Markus 2012); see Figure 2 for a detailed view of the parasite lifecycle. This quiescent phase can reactivate and allow onward transmission to new vectors in the absence of importation events (White and Imwong 2012).

Secondly, and in contrast to *P. falciparum*, gametocytemia may occur very early in the clinical course of infections with *P. vivax*. These parasite stages are infective to mosquitoes, and so early production leads to short serial intervals with consequent rapid evolution of epidemics (Sivagnanasundram 1973, Bousema and Drakeley 2011).

Finally, *P. vivax* also exhibits long-latency, where the incubation period (time from mosquito exposure to first clinical symptoms) may stretch to 6-9 months; importantly this phenotype has been recently identified in tropical, as well as temperate climates (Warrell and Gilles 2002, Brasil *et al*. 2011, Kim *et al*. 2013). This facet of parasite epidemiology is likely to have large impacts on the long-term feasibility of the intensive surveillance programs required for malaria elimination.

## 1.4 Current strategies for control of *P. vivax*

There are currently no vivax-specific control measures, aside from a single drug regimen- primaquine. This 8-aminoquinoline is currently the only approved drug that targets hypnozites (another, tafenoquine, is currently enrolling for Phase III trials).

While effective, both drugs can cause severe haemolysis in patients who are deficient in glucose-6-phosphate dehydrogenase (G6PD); moreover, there are diverse polymorphisms with a continuum of enzyme activity throughout *P. vivax* endemic areas (Howes, Dewi, *et al*. 2013). There are currently also concerns about cytochrome P-450 effects in some populations (Marcsisin *et al*. 2014).

However, large efforts are underway to field-test and deploy rapid tests for G6PD deficiencies in *P. vivax* endemic areas moving towards elimination. Beyond primaquine usage, the general 'pillars' of modern malaria control are applicable to *P. vivax* control and elimination. Primarily, this consists of correct and consistent usage of insecticide-treated bed nets (LLITNs), prompt parasitological diagnosis using rapid diagnostic tests (RDTs), treatment with quality-assured artemisinin combination therapy (ACTs), and integrated vector control measures (World Health Organization 2013a).

It has only recently become apparent that as countries with co-transmission of both major species move towards control and elimination, *P. vivax* control lags, and this parasite generally then predominates (World Health Organization 2013a). Indeed, even the historical Malaria Eradication Program had very limited consideration of the potential role of *P. vivax* in continued transmission, with an implicit assumption that control would be equally effective for the two species (Nájera 1989, 1999). It has been suggested that current interventions will be insufficient to eliminate vivax malaria, and that a deeper understanding of both the parasite and the disease, combined with *P. vivax*-specific interventions, will be required (Mueller *et al*. 2009). To address this policy gap, the World Health Organization is currently developing a global plan for malaria control and elimination that explicitly considers the unique aspects of *P. vivax*; this plan is expected to be released in 2015 (WHO 2014).

## 1.5 Malariotherapy and related studies

Much of what is currently known about the epidemiology and detailed clinical picture of *P. vivax* comes from historical human challenge studies conduced from ca. 1920-1970. The earliest studies focused on patients undergoing treatment using malariotherapy: intentional infection with malaria parasites for the treatment of terminal neurosyphilis (Chernin 1984). Later studies that were directed towards the development of novel antimalarial therapies began during WWII, and utilized prison volunteers at several research centres in the USA; the inherent ethical issues of studies within these populations have been examined by multiple authors (Weijer 1999, Harcourt 2011, Snounou and Pérignon 2013).

From a study design viewpoint, these experimental studies utilized institutionalized patients, with complete observation and follow-up; however, standards of data reporting and analysis within the original publications are inherently limited by the era in which they were published. The conclusions in studies published from ~1930-1950 utilized very rudimentary statistical analyses, or relied on qualitative comparisons between groups.

Beyond the published analyses, these studies do have several important limitations. The exclusion/inclusion criteria employed by the original investigators are unknown for these studies, and treatments were not randomly allocated. In fact, malariotherapy patient selection was based on overall clinical appraisal, and was generally contra-indicated for older patients or those with major comorbidities beyond syphilis (Winckel 1941, Snounou and Pérignon 2013); the prison volunteers were healthy adult males (Coatney *et al*. 1948). As such, the reported data represent generally healthy, non-immune patients, and therefore may not represent clinical outcomes in endemic areas or populations.

## 1.6 Specific aims of this thesis

The central questions to be addressed within this thesis are: can data from historical human infections provide evidence to inform research & policy agendas for *Plasmodium vivax* towards global malaria elimination? Moreover, these studies were aimed to generate new insights and knowledge about the neglected epidemiology of *P. vivax*.

This thesis aims to address some of the limitations within original published analyses, and to systematically examine the evidence base for multiple aspects of *P. vivax* epidemiology that have been assumed, or that have become accepted as 'clinical wisdom,' with limited consideration of the underlying data. Specifically, latency and relapse are fundamental and neglected aspects of *P. vivax* transmission, but the biological basis and epidemiology of both processes are poorly understood. Decades of expert opinion and accumulated clinical knowledge have suggested a wide range of potentially important factors including stress, diet, parasite strains, sporozoite dosage, seasonality of infection, drug prophylaxis, and host genetics. However, rigorous evidence to prioritize any of these within research or control programs is currently lacking.

# Chapter 2: Quantifying effects of geographic location on the epidemiology of *Plasmodium vivax* malaria

## 2.1 Abstract

The recent autochthonous transmission of the malaria parasite *Plasmodium vivax* in previously malaria-free temperate zones, including Greece, Corsica, the Korean Peninsula, Central China, and Australia, has catalysed renewed interest in *P. vivax* epidemiology. To inform surveillance and patient follow-up policies requires accurate estimates of incubation period and time-to-relapses, but these are currently lacking. Utilizing historic data from experimental human infections with diverse strains of *Plasmodium vivax*, survival analysis models have been used to provide the first quantitative estimates for the incubation period, and for distribution of time-to-first-relapse for *P. vivax*, by broad geographic regions. Quantitative evidence is presented that shows clinically significant divergent responses in non-immune patients, by both latitude and hemispheres. Specifically, Eurasian temperate strains show longer incubation periods, and Eastern hemisphere parasites (both tropical and temperate) show significantly longer times-to-relapse relative to Western hemisphere strains. All parasite populations show much longer median time-to-relapse than conventional wisdom ascribes. Finally, the distribution of the number of relapses is significantly different between the parasite populations in the Western and Eastern hemispheres. These estimates of key epidemiological parameters for *P. vivax* strongly suggest that the primary evidence basis for surveillance and control of this parasite needs to be rigorously reassessed. Specifically, active surveillance and patient follow-up need to be tailored to local strains to be effective, and that the origin of infection must be accurately ascertained for cases of imported *P. vivax* malaria. These results provide an evidence-based framework for effective disease control with inherently limited health resources in support of global malaria elimination. These results also provide epidemiological evidence to support earlier molecular and entomology-based proposals for the designation of two sub-species, *P. vivax vivax* (found in the Eastern hemisphere) and *P. vivax collinsi* (Western hemisphere).

## 2.2 Introduction

After decades of limited research attention, the malaria parasite *Plasmodium vivax* has moved onto the global health agenda for two primary reasons- it will be more difficult to eliminate than *P. falciparum* due to dormant liver stages and broader geographic range, and the parasite has now re-emerged in previously malaria-free temperate zones, including Greece, Corsica, the Korean peninsula, central China, and Australia (Brachman 1998, Hanna *et al*. 2004, Armengaud *et al*. 2006, Danis *et al*. 2011, Lu *et al*. 2011). Moreover, there is increasing evidence that *P. vivax* infections are not benign, but can be both severe and fatal (Price *et al*. 2007, Mueller *et al*. 2009).

A large body of epidemiological and clinical data exists to support the existence of discrete strains of Plasmodium within each species. Many studies have provided descriptive evidence to suggest that latitude of isolation has large impacts on the total number and spacing of relapses: tropical strains generally have a larger number of closely spaced relapses, while temperate strains generally have evolved to have a long incubation period, allowing the survival of the parasite as dormant hypnozoites during colder months (Myatt and Coatney 1954, White 2011). Additionally, malariologists have long recognized that that the incubation period also showed variation with strain and latitude (Wernsdorfer and Gregor 1988).

Modern research of temperate strains from the Republic of Korea supports these observations, and also suggests that chemoprophylaxis may contribute to extended latency period (Nishiura *et al*. 2007, Moon *et al*. 2009). However, these observational studies have been unable to estimate the time from infection to relapse or relapse periodicities due to uncertain exact infection times, and no direct comparisons have been reported with other strains. A second related study examined geographic differences in relapse rate during primaquine therapy for three strains utilizing logistic regression, but did not examine relapse time explicitly (Goller *et al*. 2007).

Numerous *P. vivax* classification schemes have been suggested based on observed clinical and epidemiological characteristics, including temperate/tropical and temperate/sub-tropical/temperate, Northern/Southern/Chesson-type, among others, but quantitative data to support these distinctions are sparse (Winckel 1955, World Health Organization 1969, Krotoski 1989). Additionally, recent molecular and entomological data suggest that *P. vivax* may consist of two separate subspecies, one in the Old World/Eastern hemisphere (*P. vivax vivax)* and the other in the Americas (*P. vivax collinsi*) (Li *et al*. 2001); however, research with other isolates has not confirmed these results (Prajapati *et al*. 2011). These two strains/subspecies show remarkable differences in their infectivity to different species of *Anopheles* vectors; however, impacts on human epidemiology have not been demonstrated. At least two other subspecies have also been reported- *P. vivax hiberans* and *P. vivax multinuclatum*- both notable for extremely long incubation periods (> 200 days). Although not universally accepted, these sub-species may still exist in northern

Eurasia, and their clinical behaviour has been reported to be similar to currently circulating strains from the Korean peninsula (Shute *et al*. 1976, Song *et al*. 2007). From these observations, it has been suggested that *P. vivax* should correctly be considered a species complex (Mouchet *et al*. 2008).

A large body of historical data exist from deliberate laboratory infections in two populations: institutionalized patients from pre-antibiotic era neurosyphilis treatments, and healthy prison volunteers from malaria drug trials. Using these experimental infection data in controlled settings, the relationship between parasite origin and epidemiological variables for parasites by location of isolation is explored in this study.

We quantified the impact of geography on *P. vivax* infections: identifying statistically different epidemiological parasite sub-populations, which reinforces the hypothesis that distinct subspecies exist. We provide evidence that parasite sub-populations show clinically divergent epidemiology.

## 2.3 Materials and methods

*Data sources / selection criteria*

Search strategy: a comprehensive literature search using PubMed and Google Scholar was performed in English, searching for ('vivax' OR 'benign tertian') AND ('induced' OR 'human' OR 'experimental'); more limited searches were performed in Dutch, German and French. The citations in these initial papers were examined, and from these a large number of non-indexed papers were identified.

Data inclusion criteria: malaria-naïve human cases, with defined inoculation dates (mosquito infection only; sporozoite injection and blood transfers were excluded), explicit symptomatic-only drug treatments, protection from re-infection, and named, traceable stains with defined place of origin. For the incubation period study, only this metric was utilized, and not pre-patent periods; for the relapse study, only papers with explicit follow-up periods were included. Incubation periods of > 50 days have not been included in the exact incubation period analysis due to very small numbers in well-defined studies (7 individual cases excluded).

| Characteristic | Number (%) |
|---|---|
| Neurosyphilis / neurotreatment patient | 433 (64.0) |
| Exact incubation period recorded | 460 (68.0) |
| Time-to-relapse recorded | 320 (47.3) |
| Exact incubation and time-to-relapse recorded | 104 (15.4) |
| Infected with tropical strains | 275 (40.6) |
| Infected with temperate strains | 402 (59.4) |
| Infected with New World strains | 181 (26.7) |
| Infected with Old World strains | 496 (73.3) |
| **Total** | 677 |
| Length of follow-up in weeks: mean, (SD) [Range] | 82.5, (31.7) [2.0 to 173.0] |

**Table 1. Study population, historical *P. vivax* studies.**
Note: Studies conducted ca. 1920-1980.

Available covariates were extracted from these studies, and the individual records were digitized with PlotDigitizer as needed, to create individual case-patient records *(Huwaldt 2012)*. Data for the incubation period study consists of 453 patients, infected with 11 strains, from 19 studies; data for first relapse include 320 patients, 18 strains, from 15 studies (see Table 1). Details of the parasite strains included in this analysis are shown in Table 2. Age and gender were not recorded for the majority of the neurosyphilis patients; all prison volunteers were Caucasian men. Several sources had interval censoring, that is, infections were reported as occurring within "month # 1" or "week # 16" from infection. To facilitate comparisons between studies, reports with more specific relapse times were then converted up to the next integer week from infection. Latitude of isolation was determined from the site of isolation in the original reports, and coded as a binary variable (dividing at +/- 23.5 degrees S/N) for determining tropical and temperate; "New World" includes the Americas, and "Old World" consists of Eurasia, Africa and the Pacific, as previously suggested (Li *et al*. 2001).

*Statistical analysis*

Statistical analysis was performed using Stata 12.1 (College Station, Texas); all tests were two-tailed. Kaplan-Meier analysis was used to examine the unadjusted relationship between parasite origin and event times; differences were assessed using a log rank test. Multivariate models were used to overcome the limitations of Kaplan-Meier analyses and to provide adjustment for the impact of neurological treatment and other covariates, and to produce hazard ratios to gauge to strength of association. The standard Cox proportional hazards model is unable to provide confidence intervals for predicted survival times; therefore, more complex models (flexible parametric Royston-Parmar models) were used to provide both covariate-adjusted hazard ratios (HR) and covariate-adjusted median survival times for sub-populations (Royston and Parmar 2002). These models extend Cox methods by adding parameters that model the underlying hazard of a disease event, which allows more comprehensive predictions to be made. The predicted survival times from this analysis have all been made for a neurological treatment-free population, to provide estimates that are relevant to natural human infections. Detailed methods can be found in Appendix A.

| Strain | Place & date of origin | Malariotherapy (% of total, if Y) | Zone | n |
|---|---|---|---|---|
| Chesson | Papua New Guinea, c. 1944 | N | Trop | 145 |
| Hlebnikovo | Moscow Oblast, 1948 | Y (100%) | Temp | 19 |
| Holland | Netherlands, c. 1928 | Y (100%) | Temp | 52 |
| Korea | North Korea, 1953 | Y (100%) | Temp | 21 |
| Leninabad | Tajikistan, 1950 | Y (100%) | Temp | 33 |
| Madagascar | Madagascar, 1925 | Y (100%) | Trop | 83 |
| McCoy | Florida, USA, 1931 | Y (100%) | Temp | 70 |
| Moscow | Moscow, 1950 | Y (100%) | Temp | 55 |
| NICA | Nicaragua, c. 1970 | N | Trop | 6 |
| Nahicevan | Azerbaijan, 1937 | Y (100%) | Temp | 5 |
| Naro-Fominsk | Moscow Oblast, 1946 | Y (100%) | Temp | 21 |
| Panama | Panama, c. 1970 | N | Trop | 10 |
| Rjazan | Ryazan, Russia c. 1945 | Y (100%) | Temp | 21 |
| St. Elizabeth | Southern USA, c. 1925 | Y (34.2%) | Temp | 73 |
| Salvador I | El Salvador, c. 1970 | N | Trop | 11 |
| Salvador II | El Salvador, c. 1970 | N | Trop | 11 |
| South Vietnam | Southern Vietnam, c. 1972 | N | Trop | 5 |
| Vietnam (North) | Northern Vietnam, 1954 | Y (100%) | Trop | 4 |
| Volgograd | Volgograd, Russia, 1945 | Y (100%) | Temp | 24 |
| West Pakistan | Pakistan, 1968 | N | Temp | 5 |
| *P. vivax multinulatum* | Central China, 1965 | N | Temp | 3 |
| **Total** | | **64.0%** | | **677** |

**Table 2. *P. vivax* strains included in analysis of geographic variation.**
Note: Temp = temperate; Trop = tropical (see text for definitions).

## 2.4 Results

*Incubation period*

The Kaplan-Meier plot of the 461 included case-patients shows a wide separation of groups, by tropical/temperate (Figure 3-A) and latitude/hemisphere (Figure 3-C), both of which are statistically discernable (tropical/temperate log-rank test for equality, $\chi^2 = 131.7$, 1 df, p < 0.0001; by region, $\chi^2 = 206.4$, 3 df, p < 0.0001)

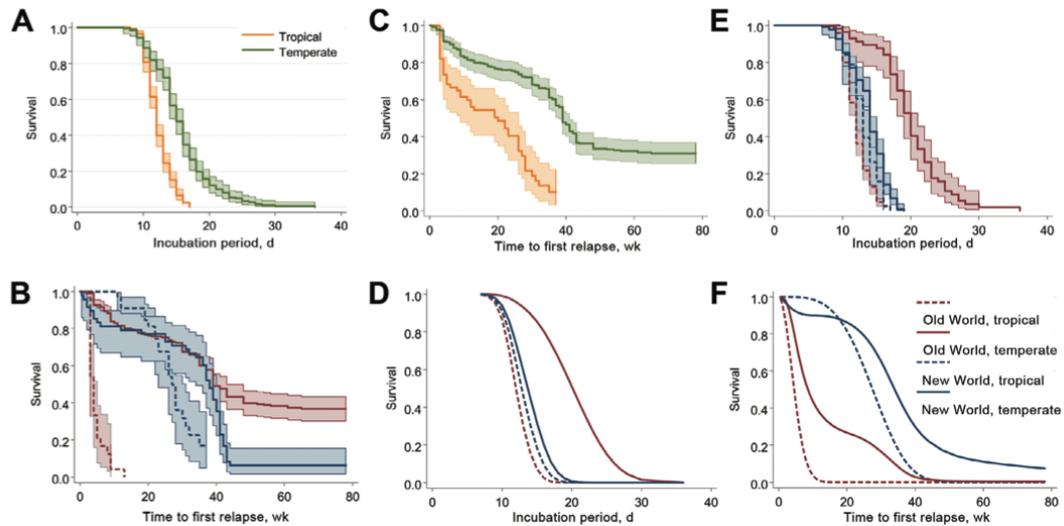

**Figure 3. Kaplan-Meier and flexible parametric models, incubation period and time-to-first relapse, historical *P. vivax* experimental infections.**

Notes: Panel A. Kaplan-Meier estimates for incubation period, temperate/tropical strains. Panel B. Kaplan-Meier estimates for time-to-first relapse, temperate/tropical strains. Panel C. Kaplan-Meier estimates for incubation period, by region. Panel D. Kaplan-Meier estimates for time-to-first relapse, by region. Panel E. Flexible parametric survival model, incubation period projected for neurotreatment free populations, by region. Panel F. Flexible parametric survival model, time-to-first relapse projected for neurotreatment free populations, by region.

The combined tropical strains have an unadjusted median incubation period of 12 days (95% CI: 12 to 12) vs. 15 days (95% CI: 14 to 16) for the temperate. When stratified by Old World/New World, the tropical strains remain essentially unchanged, but a large separation occurs in the temperate strains, with a median survival times of New World, Temperate 14 (95% CI: 14 to 15), and Old World, Temperate 20 (95% CI: 19 to 21) days.

In the full multivariate model, after adjustment for strain effects and neurological treatment status, the parametric survival estimates for the entire population are a median of 13.6 days (95% CI: 12.5 to 14.7) (Figure 3-E). The 95[th] percentile of the incubation period is 17.9 days (95% CI: 16.6 to 19.1). Differences are observed for the times between all regions, except for the New World tropical/temperate categories, which did not achieve significance (p = 0.30) (see Appendix A). The predicted median and 95th percentile survival times are shown in Figure 4. There are no significant differences within the confidence intervals for predicted median survival times with the exception of Old World Temperate (20.1 (95% CI: 17.8 to 22.5)); the 95% percentile for temperate strains are significantly longer than tropical strains. The hazard ratios for the regions: Old World, Tropical 16.8 (95% CI: 7.6 to 36.9); New World, Tropical 10.8 (95% CI: 4.6 to 25.2); New World Temperate 7.3 (95% CI: 3.8 to 14.0); and Old World, Temperate (reference). The case-patients infected with Old World tropical strains (the shortest incubation periods) show 16.8 times (7.6 to 36.9) higher hazard relative to Old World temperate strains (the longest time to event).

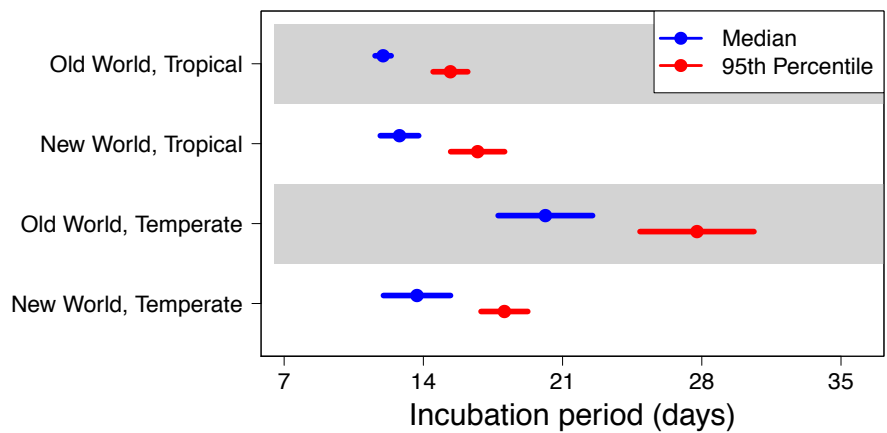

**Figure 4. Median and 95th centile survival times, incubation period in days.**
Note: Flexible parametric survival models, adjusted for neurotreatment status.

*Time-to-first relapse*

The time-to-first relapse for all studies was measured from the reported primary infection. The Kaplan-Meir survival curves for the 312 included case-patients show large, statistically significant differences by both tropical/temperate (Figure 3-B) and latitude/hemisphere (Figure 3-D) (tropical/temperate, log rank test for equality, $\chi^2 = 61.5$, 1 df, p < 0.0001; by regions, $\chi^2 = 145.2$, 3 df, p < 0.0001). Comparing the curves in Figure 3-D shows that the two New World categories fall between the wider-ranging Old World tropical and temperate categories. In the full multivariate model adjusted for strains and neurotreatment status, a distinct separation between the hemispheres is observed (Figure 3-F). The estimated total population values from this model show a median time-to-relapse of 13.5 weeks (95% CI: 8.9 to 18.1), and a 95[th] percentile of 48.8 weeks (18.2 to 79.3).

However, these aggregate values hide substantial heterogeneity in time-to-relapse. Both of the Old World categories show shorter time-to-relapse relative to New World parasites, with the tropical strains for both hemispheres exhibiting short times relative to the corresponding temperate strains, although this difference is not significant in the New World (Figure 5).  The median times-to-relapse, in weeks, are: Old World Tropical 4.5 (95% CI: 3.2 to 5.7); New World Tropical 26.8 (95% CI: 18.4 to 35.2); Old World Temperate 13.7 (95% CI 8.3 to 19.1); and New World Temperate 29.7 (95% CI 28.3 to 31.1). The corresponding 95[th] percentile times are: Old World Tropical 10.0 weeks (95% CI: 7.2 to 12.9); New World Tropical 60.3 (95% CI: 39.1 to 81.5); Old World Temperate 30.9 (95% CI: 19.9 to 41.9); and New World Temperate 66.8 (95% CI: 59.5 to 74.2). The hazard ratios from the survival models, after adjusting for neutrotreatment, are: Old World, Tropical 30.7 (95% CI: 14.1 to 66.6) (p < 0.001); New World, Tropical 1.2 (95% CI: 0.67 to 2.12) (p = 0.53); Old World, Temperate 4.0 (95% CI: 2.1 to 7.6) (p < 0.001); and New, Temperate (reference).

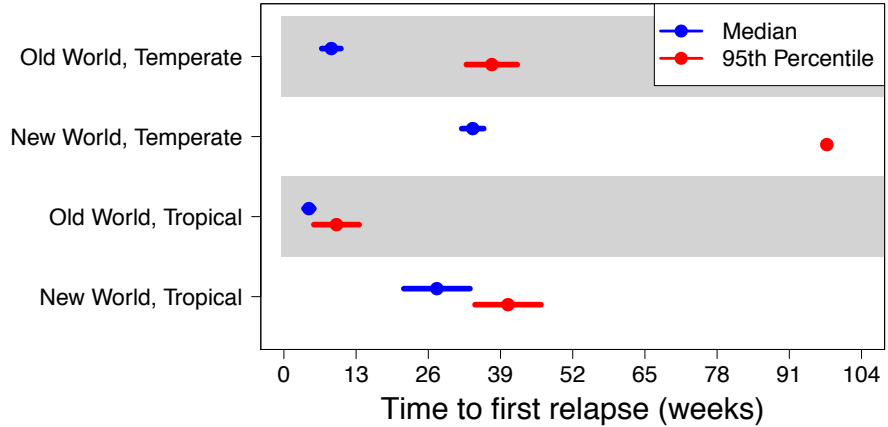

**Figure 5. Median and 95<sup>th</sup> centile survival times, primary attack to first relapse, in days.**
Note: Flexible parametric survival models, adjusted for neurotreatment status.

*Distribution of total relapses*

The total number of relapses was compared by latitude and hemisphere (see Appendix A). A time interval of 48 weeks was chosen to ensure equivalent follow-up periods between the regions [Oneway ANOVA, follow-up time by region, $F$ (3, 43) = 1.07 ($p = 0.37$)]. The Kolmogorov-Smirnov test for equality of distributions shows statistically significant differences between both tropical and temperate strains (p < 0.00001); and Old World and New World Strains (p = 0.00009). These differences remain highly significant when stratified by zones of latitude: Temperate strains, Old World vs. New World, (p = 0.00006); and Tropical strains, Old World vs. New World: (p < 0.00001).

## 2.5 Discussion and conclusions

*Comparison with prior estimates*

Malariologists have made a large number of observations concerning the geographic epidemiology of *P. vivax* (Young *et al*. 1947, Coatney *et al*. 1971). Malariotherapy treatments initially used a range of local strains, from the UK, the Netherlands, and US, but these were quickly replaced with the Madagascar strain among others, which exhibited shorter incubation periods and more reliably produced infections (Chernin 1984). A range of estimates have been reported in the literature for the incubation period of *P. vivax*: a mean of 13.88 days (S.D. 3.7); 14 ± 3 days after the mosquito bite; 12 to 17 days (mean 15); and a minimum of 8 days, extending up to 17 days (Boyd and Kitchen 1937, Baird *et al*. 2007, Warrell and Gilles 2002, Russell 1963). A modern quantitative analysis of malariotherapy data with a single strain (Madagascar) provided estimates for the pre-patent period of ~ 10.3 to 16.9 days (reported to be generally 3 days *longer* than the incubation period within this study) (Glynn and Bradley 1995). After adjustment for strain and neurological status, the parametric survival estimates for the population in this study are a median of 13.6 days (95% CI: 12.5 to 14.7) and the 95th percentile of the incubation period is 17.9 days (95% CI: 16.6 to 19.1). The minimal differences between the Kaplan-Meier estimates and those from the multivariate models for incubation period strongly supports earlier opinions that data from malariotherapy are indeed applicable to natural infections (Winckel 1941).

Similarly, a range of observations has been reported concerning relapse, with strain-specific relapse behaviour (Warrell and Gilles 2002). Relative to temperate strains, tropical strains are reported to relapse more, have a shorter relapse period of 17-45 days, with a higher proportion having more than 2 relapses (Baird *et al*. 2007). It has been estimated that tropical strains relapse every 3-4 weeks, whereas temperate strains show longer periods between relapses with greater variability (Douglas *et al*. 2012).

The estimates from this study suggest that prior estimates for time-to-relapse have been based primarily on data for Old World tropical strains (4.5 weeks (95% CI: 3.2 to 5.7)). The results from the other regions are all significantly longer than earlier estimates, with the exception of a single study from El Salvador, which reported a median relapse interval of 28 weeks (Mason 1975). The arithmetic median interval for all tropical strains (including only exact, non-interval censored times) is 19.5 weeks (95% CI: 15.5 to 23.4). The unadjusted Kaplan-Meier estimates, and parametric models adjusted for strain and neurological treatment status in the full dataset also indicate much longer mean and median times-to-relapse: the parametric survival estimates for tropical strains are a median of 17.7 weeks (95% CI: 12.2 to 23.2) and mean of 17.2 weeks (95% CI: 10.0 to 24.5). The differences in relapses are striking: patients infected with Old World temperate strains have 30.7 times (95% CI: 14.1 to 66.6) higher hazard to relapse relative to those infected with New World temperate strains. The substantial differences between Figures 3-D and 3-F suggest that neurological treatment has significant impacts on the course of relapse; therefore unadjusted relapses times from malariotherapy should be interpreted with caution.

The range of relapses recorded within 48 weeks (for equivalent follow-up) is also consistent with prior estimates: tropical strains show a median of 2.6 relapses (95% CI 1.9 to 3.3) range (0-9); with temperate having a median of 0.68 (95% CI 0.55 to 0.80), range (0-6). The total percentage of case-patients with relapses in these data, 68.5% (95% CI: 64.9 to 71.8) is broadly consistent with previous work that suggests about 60% of untreated cases relapse (Warrell and Gilles 2002).

The general agreement of prior knowledge of incubation period and number of relapses with the results from this study support the conclusion that these patients do not represent a substantially different population from natural *P. vivax* infections. However, these aggregate cohort values obscure large regional differences in epidemiology.

*Geographic differences*

The qualitative impact of geography of the parasites has been long recognized, with conspicuous differences in incubation period and latency that have been broadly correlated with climatic zones (Warrell and Gilles 2002). However, there have also been persistent difficulties in classifying these patterns; some studies suggested two different types of temperate strains, North American (St. Elizabeth) and European (Netherlands), plus tropical groups (Winckel 1955). Inconsistencies have also been noted, including tropical Central American *P. vivax* strains which showed anomalous 'temperate zone' epidemiology, leading to a suggestion that temperature alone might be an insufficient predictor (Contacos *et al*. 1972). These conflicting observations are fully consistent with the results from this study (Figure 3-F), where the most noticeable feature is that both New World populations displaying significantly longer times-to-relapse than Old World parasites.

Our results suggest that the epidemiology of *P. vivax* infection has been occluded by inherent differences between sub-populations, and that both hemisphere and latitude are strong drivers of clinical presentation and epidemiology. This strongly suggests that current paradigms for *P. vivax* clinical follow-up and surveillance may be based on erroneous assumptions. Malaria should not be discounted as a diagnosis even in the presence of long incubation periods, and the geographic origin of the parasite has critical impacts on clinical presentation and should not be ignored in case histories.

Our results show that the mean incubation period of Eurasian temperate strains is statistically, and clinically significantly longer than generally considered. This, plus the inherently longer extrinsic incubation period for Old World temperate strains suggests that an active surveillance period of 31 days after potential exposure is the minimum necessary to capture the 95[th] percentile of new cases. However, this potential surveillance burden is balanced by a shorter median time-to-relapse for these parasites relative to New World strains.

These three sets of independent measures (incubation period, time-to-first-relapse, and distribution of total relapses) across the entire course of illness all suggest that *P. vivax* should not be considered a single parasite, but is in fact several discrete, and clinically distinct populations with unique and measurable characteristics. These data, plus the prior entomological and molecular evidence, support the delineation of sub-species within the range of the parasite: *P. vivax vivax* within the Eastern hemisphere, and *P. vivax collinsi* within the Western. This conclusion is supported by a recently published phylogenetic analysis of global *P. vivax* strains, which shows both high diversity and clustering of isolates by hemispheres (Neafsey *et al*. 2012).

*Public health impacts*

Parasite origin has been shown to have large impacts on both prophylaxis and treatment: while Korean-strain infections respond to standard dosing of primaquine, even higher doses did not fully suppress strains from New Guinea (Chesson), and the Chesson strain required twice the dosing of quinine base relative to a North American strain (McCoy) *(World Health Organization 1969)*. A related study on the impact of primaquine also found large regional effects, and additionally, that Thai strains were more likely to relapse, and required higher doses to suppress relapses relative to parasites of Indian or Brazilian origin (Goller *et al*. 2007). The impact of parasite population differences should be considered in the planning and analysis of interventional trials, and in potential vaccine trials.

A recent analysis of malaria imported into the US and Israel found that a large proportion of cases exhibited long-latency: of 721 *P. vivax* cases with insufficient/non-existent prophylaxis, 46.5% (95% CI: 42.8 to 50.2%) had a delayed incubation period of > 2 months, whereas this was observed in 80.0% (95% CI: 77.0 to 82.8%) of travellers with sufficient prophylaxis (authors' calculations, from (Schwartz *et al*. 2003)). However, no information is provided about the geographic source of these parasites, and the date of exposure is assumed to be the end of travel period, making exact calculation of incubation period impossible. The results from our study are broadly consistent with these values: 31.3% (95% CI: 26.2 to 36.6%) of cases-patients in the time-to-relapse data have an incubation period of greater than 8 weeks. The comparability of these drug-free populations suggests long-term stability in the global parasite populations, and also supports the relevance of these historical challenge studies for modern surveillance programs. Finally, the potential for prolonged incubation due to prophylaxis suggests that the estimates from our analysis should be considered as minimum values for travellers with adequate antimalarial drug compliance.

*Conclusions*

Control and elimination programs for *P. vivax* should be reconsidered in light of these findings. Two major stumbling blocks of the First Global Malaria Eradication Campaign (1955-1972) were an assumption that control methods could be universally applied, and combined personnel and funding fatigue for continued surveillance at increasingly lower levels of infection (Nájera 1999). Addressing these issues in the current elimination campaign will require detailed elucidation of differences in pharmacodynamics among parasite sub-populations, and locale-specific epidemiology, including estimation of incubation periods and time-to-relapse to maximize surveillance efficiency.

Our results suggest that considerable complexity among *P. vivax* populations has been obscured by data aggregation; however, these divisions appear along defined geographic gradients. The long time interval of these studies (1920's to 1980's) implies relatively stable parasite populations; however the impact of greatly increased airplane travel and migration should be explored. The diversity of study sites and investigators included in this study strongly suggests the observed epidemiological variations are not due to differences in study protocols or vector species.

The existence of sub-populations along Eastern/Western hemispheres allows conflicting historical and epidemiological data to be formulated into a consistent and coherent picture, especially with the incorporation of phylogenetic approaches. Additionally, the parasite origin should be considered in drug prophylaxis and treatment, and the pharmacokinetic differences in the parasites should be more fully elucidated.

'Local malaria problems must be solved largely on the basis of local data. It is rarely safe to assume that the variables in one area will behave in the same way as they do in another area, however closely the two regions may seem to resemble each other in topography and climate. Large sums of money have been wasted in attempted malaria control when malariologists have forgotten this fundamental fact.'
  - Paul F. Wallace, 1946 (Russell *et al*. 1946).

# Chapter 3: Re-assessing the relationship between sporozoite dose and incubation period on *Plasmodium vivax* malaria: a systematic re-analysis

## 3.1 Abstract

Infections with the malaria parasite *Plasmodium vivax* are noteworthy for potentially very long incubation periods (6–9 months), which present a major barrier to disease elimination. Increased sporozoite challenge has been reported to be associated with both shorter incubation and pre-patent periods in a range of human challenge studies. However, this evidence base has scant empirical foundation, as these historical analyses were limited by available analytic methods, and provides no quantitative estimates of effect size. Following a comprehensive literature search, we re-analysed all identified studies using survival and/or logistic models plus contingency tables.

We have found very weak evidence for dose-dependence at entomologically plausible inocula levels. These results strongly suggest that sporozoite dosage is not an important driver of long-latency. Evidence presented suggests that parasite strain and vector species have quantitatively greater impacts, and the potential existence of a dose threshold for human dose-response to sporozoites. Greater consideration of the complex interplay between these aspects of vectors and parasites are important for human challenge experiments, vaccine trials, and epidemiology towards global malaria elimination.

## 3.2 Introduction

Malaria caused by *Plasmodium vivax*, after decades of research neglect, is being re-assessed as a major contributor to morbidity and mortality in the widespread regions where it is endemic (Price *et al*. 2007, Galinski and Barnwell 2008). However, large gaps still exist in knowledge of the epidemiology, entomology, and ecology of this parasite (Mueller *et al*. 2009, Gething *et al*. 2012). One of these gaps is the phenomenon of extended incubation periods (> 28 weeks) (Warrell and Gilles 2002). These phenotypes have been observed in modern parasite strains from diverse global settings including Brazil and the Korean peninsula (Brasil *et al*. 2011, Kim *et al*. 2013) and antimalarial drug prophylaxis has also been implicated in prolonged latency (Schwartz *et al*. 2003). The biological basis of delayed onset infections after sporozoite inoculation remains unclear; whether the persisting parasites are hypnozoites, quiescent merozoites or both is unknown (Markus 2012). Persistent and infective dermal sporozoites have also been suggested (Guilbride *et al*. 2012, Ménard *et al*. 2013).

The incubation period is a key parameter in epidemiological and clinical studies; in malaria, it is defined as the time from exposure to infected anopheline vectors to febrile illness. The pre-patent period is similarly calculated to the time when parasites are first visible in the peripheral blood. Gaps in these areas have been highlighted as key research needs for *P*. *vivax* human vaccine development (Targett *et al*. 2013).

Historical human experimental challenges during malariotherapy for terminal neurosyphilis (1920s–1950s) and prison volunteer experiments for antimalarial prophylaxis suggested an inverse relationship between the size of sporozoite inocula and time-to-infection; modern reviews have supported this view (Krotoski *et al*. 1986, Glynn 1994, White 2011, Vanderberg 2014).

Specifically, it has been reported that small sporozoite doses (10–100 sporozoites) of *P. vivax* strains isolated from temperate regions resulted in the primary attack generally being delayed for 9–10 months or longer, whereas illness occurred after a 'normal' incubation period of about two weeks (White 2011) when larger inocula (≥ 1000 sporozoites) were used. Conversely, for tropical parasite strains in these historical studies, no relationship between sporozoite doses and the latent period was observed.

However, there is also considerable ambiguity in relation to dose-dependence in the literature; some authors suggest no differences by latitude of parasite origin, with dose dependence of time-to-infection being a general feature of *P. vivax* (Vanderberg 2014). Pampana reported that long-latency was due to small inocula, but also quoted Russian researchers who described four strains that invariably showed long-latency (Pampana 1969). Other researchers reported that the length of incubation of a North Korean strain was dependant on the number of mosquito bites (Tiburskaja and Vrublevskaja 1977). Finally, recent research has suggested that parasite strain itself may independently influence incubation period (Herrera *et al.* 2009).

These historical studies utilized analytical methods that were very limited and inappropriate for time-to-event data, including simple linear regressions, differences in means between groups, or qualitative reporting of trends between groups. For example, in foundational work James reported that in a large series of 700+ cases analysed by linear regression, the incubation period was inversely correlated with the number of mosquitoes biting (James 1931); other researchers suggested an inverse relationship using mosquito-transmitted St. Elizabeth strain infections based on a visual examination of the trend writing, "Correlation is apparent and all of the 3 greatly delayed primary attacks occurred after relatively small inocula" (Coatney,

Cooper, Ruhe, *et al*. 1950). These studies, while limited by experimental design and reporting, form the core evidence for the alleged dose-dependent basis of long-latency in both experimental and natural *P. vivax* infections, but their validity has heretofore not been formally investigated.

Inherent issues with all methods of quantifying malaria parasite doses have been previously reviewed (Glynn 1994). Briefly, the use of bite-based metrics assumes that all mosquitoes inject the same dose. The use of qualitative gland infection metrics is even more problematic, as a count of 6 'pluses' could be a single heavily infective, 2 moderately infective, or 3 sparsely infective bites; with larger values of 20–30 'pluses' this becomes even more uncertain. Lastly, even 'quantitative' doses are estimates; not all sporozoites may be viable, and dosing may be compromised by adhesion to glass syringes, among other experimental issues (Glynn 1994).

In light of the severe limitations in these analyses, we sought to both critically review and to re-analyse these historical studies using appropriate statistical methods, with the intent to provide a solid evidence base for associations between sporozoite dosage and incubation/pre-patent period in *P. vivax* infections.

### 3.3 Methods

*Data inclusion*

A comprehensive search was performed in Medline and Google Scholar, using ["vivax" AND ("sporozoite" OR "inoculation") AND ("latent" OR "incubation" OR "pre-patent")] to identify all publications and grey literature reports that examined the relationship between sporozoite exposure and either incubation or pre-patent period. The references in these papers were then consulted, and un-indexed papers were identified. Inclusion criteria were malaria-naïve subjects (except for a set of studies in non-human primates); explicit follow-up without chance of re-infection; and inclusion of only primary infections ('latencies' due to presumed hypnozoite activation were not included). Exclusion criteria were poorly documented studies with insufficient experimental detail. Tropical and temperate strains were separated at 27.5° N/S, based on the reported origin of the parasites. All available covariates were extracted for multivariate analysis.

*Analysis*

A full meta-analysis was not possible due to wide variations in reported dose measurements, times-to-events, and extensive data aggregation in many reports. Three complementary sets of analyses are presented: those from studies that reported semi-quantitative biting-based exposure metrics; quantitative inoculations in humans; and finally quantitative inoculations in non-human primate models. Contingency tables ($\chi^2$ tests using Fisher's exact test) were utilized to assess if any relationship exists between dose and incubation period or pre-patent period, but this analysis was unable to assess the direction or magnitude of response. Because of these limitations, after examination of Kaplan-Meier plots we used Cox proportional hazard models to address both of these issues.

Each of the identified studies was analysed with the optimal methods considering the available data; for aggregated data, only contingency tables could be analysed. These tables were calculated for all studies to allow comparison between them. For more detailed reports, both the log-rank test for trend in dose categories, and hazard ratios (HR) from Cox survival models are presented with survival times, with sub-group analyses where possible. In this study, survival model analyses are focused on consideration of the time span from inoculation to all observed (clinical or parasitological) infections. Each of the exposure categories has a rate of progression to infection within the time interval of clinical follow-up (the 'hazard rate'); the hazard ratio follows as the ratio of any of these rates. A hazard ratio of 1.50 can be interpreted as having 1.5 times greater likelihood of presenting with infection at all time points throughout the study, relative to the reference (low-dose) group. These models allow for adjustment of covariates, and provide measures of effect size and associated confidence intervals. Lastly, for one study logistic regression was also utilized to examine a binary outcome of clinical infection.

We used the exposure categories presented in the original reports where possible; doses reported as continuous values were divided into tertiles for consistent analysis and event times were also divided into tertiles for contingency table analysis. In studies with non-explicit follow-up, unsuccessful infections were censored one day after the final reported infection (Collins *et al*. 1994). Cox survival models were tested for proportional hazard violations using Schoenfeld residuals; logistic regression model fit was assessed using the le Cessie-van Houwelingen-Copas-Hosmer unweighted sum of squares test (Hosmer *et al*. 1997). Statistical analysis was performed using Stata 12.1 (College Station, Texas, USA) and all tests were two-tailed with $\alpha = 0.05$.

### 3.4 Results

The majority of studies utilizing bite-based metrics showed no association between the number of infected bites and the incubation/pre-patent period in either contingency table or survival model analysis (Tables 3 and 4). We will therefore focus on those that showed evidence of statistically significant effects of dose on time-to-infection.

Contingency table analysis of studies with the tropical Chesson strain (Table 3, study III) used a dose metric of 'pluses' to grade the salivary gland estimated sporozoite density in the infecting mosquitoes on a qualitative scale from 1 to 6. We found a significant association ($p < 0.001$) between the tertiles of prepatent period and tertiles of the reported 'mean pluses'; this remained unchanged when analysed for all explicit dose categories (mean pluses of 22, 25, 27, 29, 30, 32, and 36; range of 12–36). This range of 'pluses' would indicate 2 to 36 bites (Whorton *et al*. 1947).

A study using the temperate St. Elizabeth strain (Table 3, study IV) showed significant dose-dependence in both contingency table and survival analyses. In Cox models, relative to 11-20 pluses, a dose of 21–30 pluses had a hazard ratio of 5.7 (95% CI 1.6 to 19.8; $p = 0.007$), and a dose of 31–40 pluses ($\sim$ 3–10 infected bites) led to an hazard ratio of 8.4 (95% CI 2.5 to 28.7; $p = 0.001$) (Coatney, Cooper, Ruhe, *et al*. 1950). That is, at all times throughout the study, the higher dose groups had a 5.7 and 8.4 times greater likelihood respectively of presenting with infection relative to the lowest-dose group. The relationship between dose classes and tertiles of the pre-patent period was also significant in contingency table analysis ($p < 0.001$). Analysis by contingency tables of long-term malariotherapy studies that used a North Korean strain (Table 3, study VI) showed evidence of dose-dependence for doses of up to 23 bites ($p = 0.012$); however, when limited to dose categories < 11 bites, there was no evidence for an association between bites and the length of incubation periods ($p = 0.256$).

| Study | N | Strain | Dose metric | Hazard ratio | 95% CI for hazard ratio | p for hazard ratio | p value for $\chi^2$ test, dose metric vs. time-to-event (tertiles) | Reference |
|---|---|---|---|---|---|---|---|---|
| I. | 13 | Southern US | 1 bite<br>2–5 bites<br>6–10 bites | ref.<br>0.30<br>0.41 | -<br>0.066 to 1.41<br>0.093 to 1.79 | -<br>0.128<br>0.235 | 0.259 † | (Mayne 1933) |
| II. | 78 | Dutch | 3–5 bites<br>6–10 bites<br>11–20 bites<br>21–30 bites<br>> 30 bites | - | - | - | Reported as Long/short, 0.250 † | (Swellengrebel and De Buck 1938, Verhave 2013) |
| III. | 87 | Chesson | 22–27 mean pluses<br>29–30 mean pluses<br>32–36 mean pluses | - | - | - | **< 0.001 #** | (Whorton *et al*. 1947) |
| IV. | 53 | St. Elizabeth | 11–20 pluses<br>**21–30 pluses**<br>**31–40 pluses** | ref.<br>**5.65**<br>**8.41** | -<br>**1.61 to 19.80**<br>**2.47 to 28.67** | -<br>**0.007**<br>**0.001** | **< 0.001 #** | (Coatney, Cooper, Ruhe, *et al*. 1950) |
| V. | 15 | Chesson | 2 pluses<br>3 pluses<br>4 pluses | ref.<br>0.79<br>2.93 | -<br>0.16 to 3.95<br>0.80 to 10.73 | -<br>0.78<br>0.10 | 0.189 # | (Coatney, Cooper, and Young 1950) |
| VI. | 77<br><br>subset<br>(n = 67) | N. Korean | 1–2 bites<br>3–5 bites<br>6–23 bites<br><br>1–2 bites<br>3–4 bites<br>5–10 bites | | - | - | **Reported as Long/short, 0.012 †**<br><br>Long/short 0.256 † | (Tiburskaja and Vrublevskaja 1977) |
| VII. | 24 | Yunnan | 1–2 bites<br>3–5 bites<br>**7–10 bites** | ref.<br>2.27<br>**4.00** | -<br>0.74 to 6.95<br>**1.23 to 12.96** | -<br>0.152<br>**0.021** | **0.038 †** | (Yang 1996) |
| VIII. | 17 | Venezuela | 2–4 bites<br>6–7 bites<br>8–10 bites | ref.<br>1.67<br>1.69 | -<br>0.47 to 5.98<br>0.45 to 6.30 | -<br>0.43<br>0.43 | 0.678 # | (Herrera *et al*. 2009) |
| IX. | 16 | Venezuela | 2 bites<br>3 bites<br>4 bites | ref.<br>2.73<br>1.74 | -<br>0.76 to 9.81<br>0.20 to 14.94 | -<br>0.12<br>0.61 | 0.249 # | (Herrera *et al*. 2011) |

**Table 3. Analysis of sporozoite effects in historical human *P. vivax* malaria challenge studies; vector bite-based exposures.**
(Studies I – IX), assess the relationship between sporozoite dose and pre-patent or incubation period (bold-faced entries are significant with p < 0.05). <u>Notes</u>: 'Pluses' refers to the grand total of the estimated sporozoite density (qualitative grading from 1 to either 4 or 6, depending on the study, see text) for the infected mosquito or mosquitoes used in challenge, per volunteer. For $\chi^2$ tests not using tertiles, 'short' refers to incubation periods of less than 8 weeks, and 'long' as anything longer (generally, 6–9 months).
Study end point: † = incubation period; # = prepatent period.

More recent experimental infections with a *P. vivax* strain from Yunnan (Table 3, study VII) showed no significant difference between 1–2 bites (reference) and 3–5 bites (HR = 2.3, 95% CI 0.74 to 7.0; p = 0.152), but there was evidence for dose-dependence at higher doses of 7–10 bites (HR = 4.0, 95% CI 1.2 to 13.0; p = 0.021). Finally, the overall $\chi^2$ test and the log-rank test for trend for an association between bites and incubation period were both marginally significant (p = 0.038 and p = 0.032).

Analysis of data from a large series of studies reporting discrete (integer) bite exposures is presented in Table 4 (study Xa). Log-rank test for trend in survival curves from Kaplan-Meier analysis showed no difference between category doses of 1–5 bites inclusive (p = 0.486) in the incubation period; however, inclusion of the full data (reported as 1 to 10 individually and 11-20; 21-30; 31-40 bites) suggested a significant difference in incubation period (p < 0.0001). In a multivariate Cox model adjusted for mosquito batch oöcyst grading (as reported in the original publication, as either: all in the infecting lot having < 50 sporozoites; or at least one in infecting lot having > 100 sporozoites) while there were no significant differences in the range of doses from a single bite to five inclusive, statistically significant differences in the hazard ratios were apparent from six infected bites onwards. The highest exposure category of 31-40 infected bites had an adjusted hazard ratio of 23.9 (95% CI: 3.1 to 186.8; p = 0.002). These results were consistent with contingency table analysis using all the reported dose groups (p < 0.001), while analysis of five bites and below did not show a significant association with incubation period (p = 0.098).

| Risk factor | Value | Hazard ratio | 95% CI for hazard ratio | p for hazard ratio | p for χ² test, tertiles of incubation period |
|---|---|---|---|---|---|
| Infected mosquito biting dose | 1 | ref. | - | - | **1 to 31–40 bites**<br>**< 0.001**<br>1 to 5 bites<br>0.098 |
| | 2 | 0.99 | 0.54 to 1.79 | 0.964 | |
| | 3 | 0.75 | 0.46 to 1.21 | 0.236 | |
| | 4 | 1.13 | 0.66 to 1.93 | 0.669 | |
| | 5 | 1.29 | 0.75 to 2.22 | 0.366 | |
| | **6** | **2.26** | **1.33 to 3.84** | **0.002** | |
| | 7 | 1.17 | 0.67 to 2.07 | 0.576 | |
| | **8** | **2.23** | **1.34 to 3.70** | **0.002** | |
| | **9** | **2.01** | **1.09 to 3.68** | **0.025** | |
| | 10 | 1.76 | 0.91 to 3.39 | 0.093 | |
| | **11–20** | **2.02** | **1.12 to 3.63** | **0.019** | |
| | 21–30 | 1.38 | 0.42 to 4.57 | 0.590 | |
| | **31–40** | **23.92** | **3.07 to 186.78** | **0.002** | |
| Mosquito batch oöcyst grading | All in lot < 50 sporozoites | ref. | - | - | |
| | **At least one in lot > 100 sporozoites** | **1.41** | **1.06 to 1.89** | **0.019** | |

**Table 4. Analysis of sporozoite effects in historical human *P. vivax* malaria challenge studies; strains from the Southern US.**
(Study Xa), to assess the relationship between sporozoite dose and incubation period (bold-faced entries are significant with p < 0.05) (N = 261). Source: (Boyd 1940).

Analysis of the most rigorous experiments identified, which used quantitative intravenous/intradermal sporozoite dosing, is shown in Table 5 (studies XI and XII).

The dose of 10,000 sporozoites failed due to experimental issues in the North Korean strain study and was not reported by the original authors, and the dose of 100,000 was not included for the Chesson strain. The two highest doses of 10,000 and 100,000 have been combined for analysis after testing for heterogeneity of effect.

| Risk factor | Value | Hazard ratio | 95% CI for hazard ratio | p for hazard ratio | χ² test, tertiles of prepatent period, overall | χ² test, tertiles of prepatent period, Korean only | χ² test, tertiles of prepatent period, Chesson only |
|---|---|---|---|---|---|---|---|
| Estimated Sporozoite Dose | 10 | ref. | - | - | | | |
| | 100 | 1.10 | 0.36 to 3.33 | 0.863 | | | |
| | **1,000** | **3.59** | **1.19 to 10.79** | **0.023** | 10 to 1,000 p = 0.846 | 10 to 1,000 p = 1.00 | 10 to 1,000 p = 0.444 |
| | **10,000 / 100,000** | **68.54** | **7.20 to 651.79** | **< 0.001** | 10 to 100,000 p = 0.081 | **10 to 100,000 p = 0.014** | 10 to 10,000 p = 0.491 |
| Parasite strain | N. Korean | ref. | - | - | | | |
| | **Chesson** | **49.62** | **6.01 to 409.16** | **< 0.001** | | | |

**Table 5. Cox model analysis of sporozoite effects in historical human *P. vivax* malaria challenge studies; quantitative sporozoite dosing.**
Notes: (Studies XI and XII) Contingency table and survival model analysis of historical human to assess the relationship between sporozoite dose and pre-patent period (bold-face entries are significant with p < 0.05) (N = 36). Source: (Shute *et al*. 1976, Ungureanu *et al*. 1976).

| Risk factor | Value | Incidence rate ratio (IRR) | 95% CI for IRR | p for IRR |
|---|---|---|---|---|
| Estimated Sporozoite Dose | 10 | ref. | - | - |
| | 100 | 0.96 | 0.32 to 2.86 | 0.941 |
| | 1,000 | 2.62 | 0.92 to 7.50 | 0.071 |
| | **10,000 / 100,000** | **13.87** | **4.84 to 39.72** | **< 0.001** |
| Parasite strain | N. Korean | ref. | - | - |
| | **Chesson** | **39.37** | **10.04 to 85.96** | **< 0.001** |

**Table 6. Poisson model analysis of sporozoite effects in historical human *P. vivax* malaria challenge studies, quantitative sporozoite dosing.**
Notes: as in Table 5.

Kaplan-Meier curves for these experiments are shown in Figure 6. The two highest doses are evident at the far left, with lower doses tapering out to longer time intervals. The log-rank test for trend among doses was marginally significant in the full dataset (p = 0.03); inclusion of the lowest three doses (10, 100 and 1000) was significant for the Korean strain (p = 0.008), but not for the Chesson strain (p = 0.120).

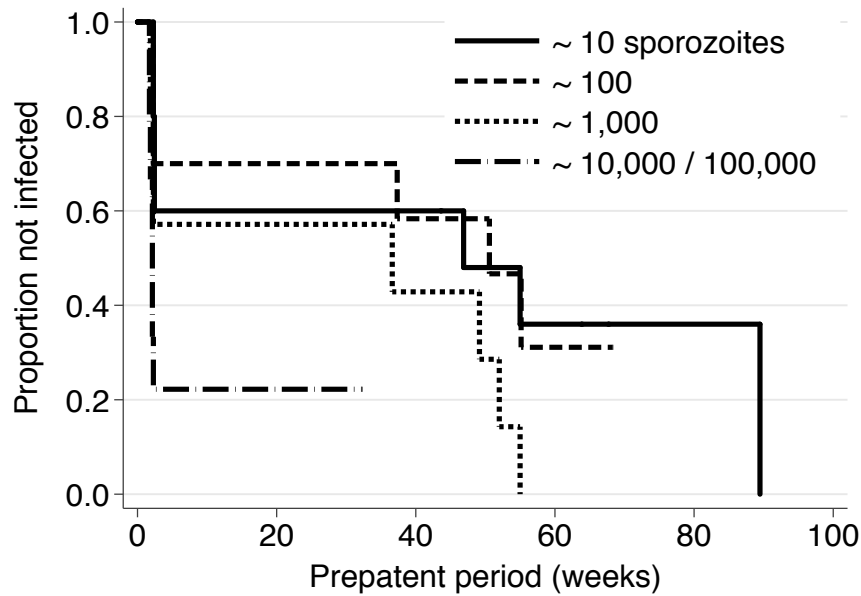

**Figure 6. Kaplan-Meier plot for the relationship between prepatent period and quantitative sporozoite doses in human *P. vivax* infections.**
Note: (N = 36) (Shute *et al*. 1976, Ungureanu *et al*. 1976).

In multivariate Cox survival models adjusted for parasite strain with 10 sporozoites as the references, doses of 100 sporozoites showed no significant impact on the pre-patent period, with an hazard ratio = 1.1 (95% CI: 0.36 to 3.3; p = 0.863); dosing with 1,000 sporozoites showed an increased HR of 3.6 (95% CI: 1.2 to 10.8; p = 0.023); and a dose of 10,000 / 100,000 sporozoites showed a significant effect, with an HR = 68.5 (95% CI: 7.2 to 651.8; p < 0.001). Parasite strain itself was significant in this analysis: with the North Korean strain as the reference, the tropical Chesson strain showed an HR of 49.6 (95% CI: 6.0 to 409.2; p < 0.001). That is, at all time points, case-patients infected with the Chesson strain had 49.6 times greater risk of presenting with infection relative to those infected with the North Korean strain.

This Cox model analysis was complemented with a parallel Poisson regression analysis, split at each failure time (Table 6), which showed consistent estimates, except that the 1,000 sporozoite dosage became not significant (IRR = 2.63; 95% CI 0.92-7.50, p = 0.071). When analysed using contingency tables, there was no evidence for a relationship between dose and time-to-event, including the highest dosages; the results were unchanged with inclusion of only the Chesson strain. Analysis of the North Korean strain alone in contingency tables showed no effect with analysis of the 10 to 1,000 doses (p = 1.00), but a significant effect was evident with inclusion of the full range of up to 100,000 sporozoites (p = 0.014).

Analysis of data from a series of experiments carried out over several years using non-human primates with the US CDC 'Salvador I' *P. vivax* strain is shown in Table 7 (studies XIII–XV). The log-rank test for trend in unadjusted Kaplan-Meier analysis with the full dosing arms (which ranges from 1,000 to 2.4 million sporozoites) was significant (p = 0.011), but not for doses ≤ 10,000 (p = 0.083) or for doses ≤ 100,0000 sporozoites (p = 0.500). In the full Cox multivariate model after adjustment for primate host species, previous malaria infections in individual non-human primates, and sporozoite vector source, with 1,000 sporozoites as the references, doses of > 1,000 to 75,000 showed an adjusted HR of 0.60 (95% CI: 0.25 to 1.4; p = 0.101), and doses from > 75,000 to 2.4 million showed an HR of 2.2 (95% CI: 1.1 to 4.3; p = 0.029).

Examination of the anopheline source of the sporozoites shows that relative to those from *An. stephensi*, those harvested from *An. freeborni* had an adjusted HR of 12.5 (95% CI: 2.6 to 60.4, p = 0.002), while all other vector sources were not significant. In contingency table analysis, no association was found between tertiles of dose and prepatent period (p = 0.065), and there was also no evidence of association between the 1,000 and > 1,000 to 10,000 sporozoites dose categories and tertiles of event times (p = 0.29). All of these results were unchanged if only the larger sub-population of *Saimiri* monkeys (n = 65) was included; evidence has suggested that these primate species react differently to infection (Collins 2002).

| Risk factor | Value | Hazard ratio | 95% CI for hazard ratio | p value for hazard ratio | $\chi^2$ test, tertiles of prepatent period |
|---|---|---|---|---|---|
| Estimated Sporozoite Dose | 1,000 | ref. | - | - | Tertiles of dose, p = 0.065 |
| | > 1,000 to 75,000 | 0.60 | 0.25 to 1.43 | 0.101 | |
| | **> 75,000 to 2.4 million** | **2.16** | **1.08 to 4.30** | **0.029** | 1000 dose versus > 1000 p = 0.290 |
| *Anopheles* species | *An. stephensi* | ref. | - | - | |
| | *An. gambiae* | 1.26 | 0.13 to 12.46 | 0.841 | |
| | *An. dirus* | 2.12 | 0.86 to 5.55 | 0.101 | |
| | ***An. freeborni*** | **12.45** | **2.57 to 60.38** | **0.002** | |
| | *An. stephensi / dirus* | 1.60 | 0.58 to 4.45 | 0.364 | |

**Table 7. Analysis of sporozoite effects in *P. vivax* malaria challenge studies in splenectomised *Saimiri* and *Aotus* non-human primate models.**
Notes: (Studies XIII – XV), assessing the relationship between sporozoite dose and incubation period (bold-face entries are significant with p < 0.05) (N = 105).
US-CDC Salvador I strain used; models adjusted for primate host species and previous experimental malaria infections in individual non-human primates.
Source (Collins *et al*. 1988, 1994, 1996).

Finally, these survival analyses were complemented with a different series of studies from the same publication as study Xa (data not shown; study Xb), examining the number of 'takes;' that is, bites from an infected vector which produced an infection (time span not specified) among 394 inoculations (Boyd 1940). Logistic regression was utilized to examine the relationship between the number of infected mosquito bites and proportion of exposures producing an infection. Mosquito dose was not significantly associated with progression to infection when measured as either discrete exposure categories (1 to 10 individually and 11–20; 21–30; 31–40; all p > 0.06) or for biting dose as a continuous variable (using the mid-point of the higher categories; p = 0.88).

### 3.5 Discussion

*Biologically plausible exposure to sporozoites*

A crucial aspect of these historical studies that has not been previously considered is the plausible range of exposure to sporozoites under field conditions. Beyond informing experimental studies, these historical studies also serve as the basis for understanding natural infections (White 2011). A key measure of malaria exposure in field settings is the entomological inoculation rate (EIR, number of *Plasmodium*-infective mosquito bites per person per time period). The highest monthly *P. vivax*-specific EIRs during high transmission seasons we are aware of are 14.5 from Ethiopia (Animut *et al*. 2013); 44.6 in Sri Lanka; and 46.8 from Thailand [authors' calculations from hourly rates reported in (Ramasamy *et al*. 1992, Rattanarithikul *et al*. 1996)].

Studies using *P. falciparum* have estimated the inocula of sporozoites per mosquito bite, with a median of 15 sporozoites (range 0–978) in three different studies reviewed (Rosenberg 2008); remarkably similar estimates have been obtained with multiple rodent malarias (Frischknecht *et al*. 2004, Medica and Sinnis 2005, Jin *et al*. 2007).

Taken together, these suggest that while the absolute maximum exposures under natural conditions (assuming maxima of 2 bites per night and inocula of 978 sporozoites) are ~ 2,000 sporozoites, likely exposures are less than 100 sporozoites. Historical work was predicated on vastly higher inocula; for example, 500 bites was assumed to inoculate ~ six million sporozoites or 12,000 per bite (Shute *et al*. 1976).

*Impact of sporozoite dose on incubation/prepatent periods*

A summary of results from our analysis is presented in Table 8, with low dose defined as exposures with entomological plausibility and includes doses of ≤ 5 infective bites, or ~ 1,000 sporozoites; while these values are not strictly comparable, they likely represent similar levels of exposure.

| Study | N | Strain type | Dose metric | Evidence for dose-dependency, High dose | Evidence for dose-dependency, Low dose | Reference |
|---|---|---|---|---|---|---|
| I | 13 | Temperate | bites | No | No | (Mayne 1933) |
| II | 78 | Temperate | bites | No | No | (Swellengrebel and De Buck 1938, Verhave 2013) |
| III | 87 | Tropical | mean 'pluses' | YES | - | (Whorton *et al.* 1947) |
| IV | 53 | Temperate | 'pluses' | YES | - | (Coatney, Cooper, Ruhe, *et al.* 1950) |
| V | 15 | Tropical | 'pluses' | No | No | (Coatney, Cooper, and Young 1950) |
| VI | 84 | Temperate | bites | YES | No | (Tiburskaja and Vrublevskaja 1977) |
| VII | 24 | Tropical or Subtropical | bites | YES | No | (Yang 1996) |
| VIII | 17 | Tropical | bites | No | No | (Herrera *et al.* 2009) |
| IX | 16 | Tropical | bites | - | No | (Herrera *et al.* 2011) |
| Xa | 261 | Tropical | bites | YES | No | (Boyd 1940) |
| Xb | 394 | Tropical | bites | No | No | (Boyd 1940) |
| XI–XII | 36 | Temperate and Tropical | quantitative sporozoites | YES (Temperate only) | YES | (Shute *et al.* 1976, Ungureanu *et al.* 1976). |
| XIII–XV | 103 | Tropical | quantitative sporozoites | YES | - | (Collins *et al.* 1988, 1994, 1996). |

**Table 8. Summary of evidence for association between sporozoite dose and incubation or prepatent period in *P. vivax* challenge studies.**
Notes: 'Low' dose refers to ≤ 5 infected bites or ~ 1,000 sporozoites; 'High' dose is any greater exposure. 'Pluses' refers to the grand total of the estimated sporozoite density (qualitative grading from 1 to either 4 or 6, depending on the study) for the infected mosquito or mosquitoes used in challenge, per volunteer.

The studies that show strong evidence for an inverse dose-dependence utilize uncertain 'plus' metrics that correspond to very high sporozoite exposure: the highest exposures in studies III and IV corresponded to 10–40 and 32–36 bites respectively (Whorton *et al.* 1947, Coatney, Cooper, Ruhe, *et al.* 1950). A similar trend is observed in studies XIII-XV, with only doses of > 75,000–2.4 million sporozoites having a significant inverse relationship with incubation period.

The quantitative dosing experiments in studies XI/XII suggest dose-response: the effects of dosing at 1,000 sporozoites were significant relative to 10 or 100, and the combined 10,000/100,000 dose was highly significant. However, when examined by strains individually, the Chesson strain showed no evidence of dose-dependence in any of our analyses. The Korean strain also showed no dose-dependent effect at 100 sporozoites relative to 10; doses of 1,000 and higher were associated with shorter pre-patent periods. However, as likely natural inocula lie in the range from zero to ~ 1,000 sporozoites, and there were no experimental challenge doses between 100 and 1000, it is unclear what the relevance of these data are to natural *P. vivax* infections.

The second major limitation in these data is that challenges in studies XI/XII utilized both intradermal and intravenous dosing; however, only aggregated data was reported. These routes have been shown to have divergent infectivity in experimental human *P. falciparum* infections (Sheehy *et al.* 2013). Additionally, the non-randomized nature of treatment arms, and uncertain inclusion/exclusion criteria suggest that we cannot discount the possibility that these findings were due to chance.

The studies included in this analysis were powered from 0.5–0.96 to detect a minimum hazard ratio of 2.0 for dose tertiles. The smallest study reported to have acceptable power of 0.83 in survival models was study VII (N = 24); the combined quantitative dosing experiment (XI /XII; N = 36) was adequately powered at 0.96. The very large confidence intervals for the hazard ratios in several of the analyses (e.g. studies X, and XI/XII; Tables 3, 5 and 6) reflect the limited number of subjects per strata; however, the lower bounds of these CIs represent a lower limit of effect size for significant factors.

Our results suggest that the general theory of dose-dependence has exceedingly limited statistical support that does not meet modern standards of evidence. While broad dose-dependence was found in a subset of the studies, experimental details and the biologically implausible doses utilized make even these results highly suspect.

*Threshold effects*

Our analysis of data from studies VI and X (Tables 3 and 4) suggests that mosquito challenge of ~ five bites may represent a fundamentally different biological response from greater sporozoite exposure (Boyd 1940, Tiburskaja and Vrublevskaja 1977). At doses lower than this threshold, there was no evidence of any dose-response, but a clear inflection point occurred at higher doses, where a dose-dependent response is apparent. Additional evidence comes from early malariotherapy studies where 6 bites led to incubation periods that were slightly longer than average, while those from 30+ bites were much shorter (James 1931).

The notable consistency of these results suggests a fundamental biological limit, and that dosing beyond this threshold potentially involves divergent biological pathways. If this inflection point occurs at 5–6 bites, and assuming a median of 15 and a maximum of 978 sporozoites (Rosenberg 2008), then bites in this range would lead to doses of approximately (zero) to 90–5,868 sporozoites. The data from the quantitative analysis in this work also suggests that doses above this range of ~ 1,000 sporozoites give rise to clinically divergent outcomes (studies XI/XII). Importantly, saturation of a biological pathway could also obviate the need for the postulated existence of several different types of sporozoites that have been suggested by multiple researchers (Shute *et al*. 1976, Collins *et al*. 1988).

**3.6 Conclusions**

Understanding the basis of long-latency in *P. vivax* malaria infections has important implications for the planning of control programs, and both the design and long-term sustainability of surveillance programs in the context of global malaria elimination (Mueller *et al*. 2009, Shanks 2012). Moreover, an improved understanding of dose-dependence is crucial for planning human vaccine studies-investigators in a recent study were surprised at finding no impact of increasing bite challenge on the pre-patent period (Herrera *et al*. 2009). Our study strongly suggests that sporozoite dose is not a main driver of incubation period in naturally or experimentally acquired infections, and that the true causes remain to be discerned.

The extremely large effect size between the two strains (Chesson and North Korean) in this study suggests that parasite strain (and hence genetics) are far more important factors in long-latency than sporozoite dose itself. Additionally, the large differences by mosquito species in studies XIII–XV suggest important parasite-vector interactions; these results bolster the suggestion that *Anopheles* are far more than just simple vectors (Paul *et al*. 2004). Our results also provide effect estimates that reinforce earlier entomological and epidemiological studies suggesting sub-speciation within *P. vivax* by hemispheres (Li *et al*. 2001, Lover and Coker 2013).

Our findings highlight research gaps that should be addressed to ensure that costly and technically demanding challenge experiments truly mirror natural infection pathways, thereby leading to accurate conclusions about the effectiveness of prophylaxis and vaccine candidates.

# Chapter 4: The distribution of incubation and relapse times in experimental infections with the malaria parasite *Plasmodium vivax*

## 4.1 Abstract

The distributions of incubation and relapse periods are key components of infectious disease models for the malaria parasite *Plasmodium vivax*; however, detailed distributions based upon experimental data are lacking. Using a range of historical, experimental mosquito-transmitted human infections, Bayesian estimation with non-informative priors was used to determine parametric distributions that can be readily implemented for the incubation period and time-to-first relapse in *P. vivax* infections, including global subregions by parasite source. These analyses were complemented with a pooled analysis of observational human infection data with infections that included malaria chemoprophylaxis and long-latencies. The epidemiological impact of these distributional assumptions was explored using stochastic epidemic simulations at a fixed reproductive number while varying the underlying distribution of incubation periods.

Using the Deviance Information Criteria to compare parameterizations, experimental incubation periods are most closely modelled with a shifted log-logistic distribution; a log-logistic mixture is the best fit for incubations in observational studies. The mixture Gompertz distribution was the best fit for experimental times-to-relapse among the tested parameterizations, with some variation by geographic subregions. Simulations suggest underlying distributional assumptions have critically important impacts on both the time-scale and total case counts within epidemics.

These results suggest that the exponential and gamma distributions commonly used for modelling incubation periods and relapse times inadequately capture the complexity in the distributions of event times in *P. vivax* malaria infections. In future models, log-logistic and Gompertz distributions should be utilized for general incubation periods and relapse times respectively, and region-specific distributions should be considered to accurately model and predict the epidemiology of this important human pathogen.

## 4.2 Introduction

Malaria caused by *Plasmodium vivax* has recently entered the global health agenda in the context of global malaria elimination. This has followed a re-evaluation of the long-held opinion that this parasite causes limited morbidity and essentially no mortality; a range of recent studies suggest that it is a major contributor to both in the wide-spread regions where it is endemic (Gething *et al*. 2012, Baird 2013). Furthermore, the presence of dormant liver forms (hypnozoites) which can re-activate infection is an important barrier in disease control towards global malaria elimination (White 2011, Battle *et al*. 2012).

Mathematical and statistical models are an important area of research in malaria given the complex dynamics of the parasite-host-vector system (McKenzie 2000, Mandal *et al*. 2011). The majority of malaria models have focused on the species common in sub-Saharan Africa, *P. falciparum*; only recently have efforts been directed towards *P. vivax* (Ishikawa *et al*. 2003, Pongsumpun and Tang 2007, Chamchod and Beier 2013, Roy *et al*. 2013). The distributions of event times including incubation period have an important role in modelling infectious disease (Sartwell 1950), and realistic assumptions about the distributions are crucial for accurate models. Published *P. vivax* models have made a range of implicit and explicit assumptions about the functional form of incubation periods and relapse intervals with limited empirical justification. Earlier work has focused on statistically and clinically significant differences in the epidemiology of sub-populations of this parasite (Lover and Coker 2013, Battle *et al*. 2014), but these epidemiological models do not provide well-defined parametric distributions for application within mathematical or statistical models. The purpose of this study is to use data synthesis to provide accurate, realistic and readily implementable parameters for modelling *P. vivax* infection event times using several historical human infection datasets.

### 4.3 Methods

We have utilized data from our earlier study of historical human challenge studies in two populations: patients receiving pre-antibiotic era neurosyphilis treatments, and prison volunteers in experiments for malaria prophylaxis. These two groups of institutionalized patients had mosquito-transmitted infections with defined exposure dates and complete follow-up (Lover and Coker 2013). The composition of these populations can be found in Tables 9 and 10; CONSORT diagrams can be found in Figure 7.

Individuals without a recorded incubation or relapse (censored observations) have not been included in this analysis. 'Failed' primary infections were generally not reported within the original studies and may represent experimental difficulties. In analysis of relapses, our primary consideration was to determine the underlying distribution of events for modelling; non-parametric Kaplan-Meier plots of both the full and the censored cohorts (N=320 and N=222), including the proportion with relapses, can be found in Appendix B.

| Characteristic | | Number ( %) |
|---|---|---|
| parasite origin | New World, Temperate | 139 (30.6) |
| | New World, Tropical | 38 (8.4) |
| | Old World, Temperate | 57 (12.6) |
| | Old World, Tropical | 220 (48.5) |
| neurological treatment patient | No | 224 (49.3) |
| | Yes | 250 (50.7) |
| **N** | | 454 |

**Table 9. Study population for analysis of time-to-event distributions, incubation period (experimental studies).**

| Characteristic | | Number (%) |
|---|---|---|
| parasite origin | New World, Temperate | 45 (20.3) |
| | New World, Tropical | 21 (9.5) |
| | Old World, Temperate | 130 (58.6) |
| | Old World, Tropical | 26 (11.7) |
| neurological treatment patient | No | 87 (39.2) |
| | Yes | 135 (60.8) |
| **N** | | 222 |

**Table 10. Study population for analysis of time-to-event distributions, relapse period (experimental studies).**

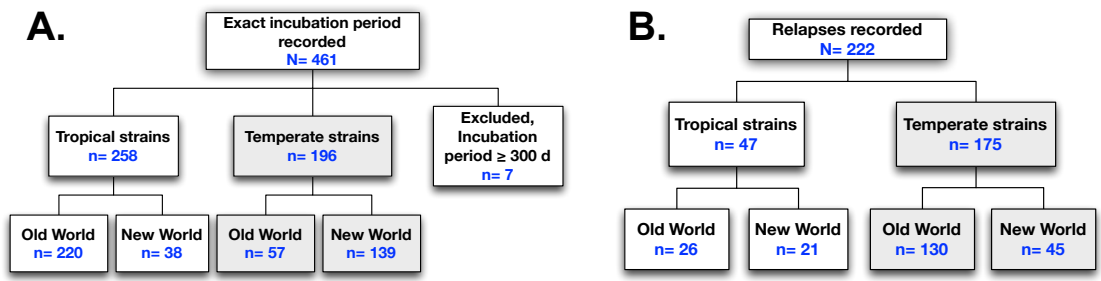

**Figure 7. CONSORT diagram, study populations for event time distribution analysis, *P. vivax* malaria.**
Panel A: experimental incubation period study; Panel B: experimental time-to-first relapse study.

Data for the infections with long-latency (extended incubation periods) are taken from three published studies that involved two unrelated temperate strains; one involved drug prophylaxis (Coatney, Cooper, Ruhe, *et al*. 1950), two were observational studies with inferred exposure dates (Brunetti *et al*. 1954, Kim *et al*. 2013), and all include extensive interval censoring in reported event times (Table 11).

| Characteristic | | Number (%) |
|---|---|---|
| **malaria chemoprophylaxis** | No | 109 (20.6) |
| | Yes | 191 (36.1) |
| | Unknown | 229 (43.3) |
| **parasite strain** | Korean | 262 (49.5) |
| | St. Elizabeth | 267 (50.5) |
| **N** | | 529 |

**Table 11. Study population for analysis of time-to-event distributions in *P. vivax* incubation periods (observational studies).**

Case-patients were exposed to parasites from a range of geographic locations, which were characterized by hemisphere and latitude. As in prior studies and historical precedence, the sub-populations from the Western hemisphere are referred to as the New World, and Old World region consists of the Eastern hemisphere and Pacific regions (Li *et al*. 2001); and temperate and tropical regions have been split at ± 27.5° N/S. Many of these data include interval censoring; that is, the event was reported as occurring within a specified time interval, but the exact time in unknown (Lindsey and Ryan 1998).

This study analyses de-identified, secondary data published in the open literature (in the public domain); no ethics review was required. Analysis of data from patients at the same neurosyphilis treatment centres has been published with an extensive discussion of the ethical issues (Weijer 1999); the issues inherent to the prison volunteers in these studies have also received extensive attention (Harkness JM 1996, Harcourt 2011, Howes, Battle, *et al*. 2013).

The incubation period refers to the time from parasite exposure to onset of clinical symptoms; prepatent periods, which refer to the identification of blood-stage parasites, were not included in this analysis. For the experimental studies, all patients received only symptomatic treatments; all cases with malaria prophylaxis or radical cure were excluded. Relapses were measured from the primary infection as reported by the original study authors, and correspond to the onset of new clinical symptoms after parasites are no longer visible in the peripheral blood following the primary infection (Bruce-Chwatt 1984). These data were examined using survival models, to specifically address the non-normal distribution of event times.

In this analysis, we examined a range of distributions including exponential, gamma, Gompertz, log-logistic, log-normal, Weibull; time-shifted distributions from these respective families; and mixture distributions from each distribution family. The general forms of these distributions are shown in Figure 8.

Model fitting and parameter estimation utilized the Markov-Chain Monte-Carlo algorithm (Gelman *et al*. 2003), and interval censored data were addressed using data augmentation methods (Tanner and Wong 1987). Specifically, the parameter space was extended to include both unknown parameters and unknown event times, with estimation of the joint distribution of both of them. The marginal distribution for the parameters of interest was then obtained by integration over the unknown times. Two complementary sets of analyses were performed for each of the experimental incubation and relapse datasets. In the first set, the best-fit distribution was found using the aggregate data, and then parameters for this optimal distribution were determined for each of the subregions of interest. In the second analysis, the best-fit distribution was found for each of the subregions independently.

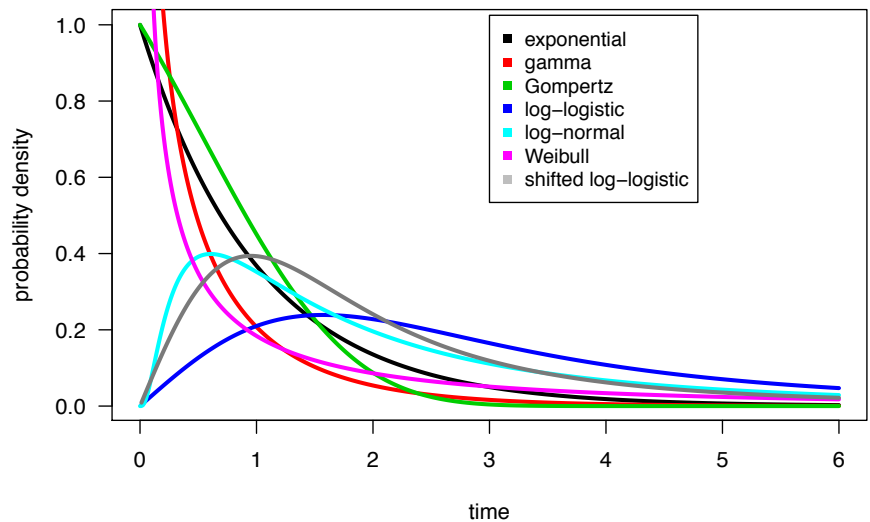

**Figure 8. Comparison of general probability distributions included within time-to-event analysis.**
Note: [shape=0.5, rate=1; shift=0.5 (shifted log-logistic only)].

Deviance Information Criterion (DIC) was used for model comparison (Spiegelhalter *et al.* 2002), with standard thresholds to determine strength of evidence. That is, an absolute difference between models of < 2 DIC units was taken as indicating little difference; from 2-7 units indicating large differences; and > 7 DIC units indicating clear evidence of superiority.

To examine the sensitivity of the model selection procedure, we multiplied all time points by log-normal noise, with mean of 0 on the log scale, plus 0.01 standard deviation, i.e. randomly scaled up or down by ± 2%. Model sensitivity was then assessed by comparing the DIC from best-fit model with the DIC value from fitting the same distributional model to the generated pseudodata.

To assess the epidemiological and practical impacts of identified distributions, we performed a series of stochastic compartmental (SIR) models at fixed $R_0$ values while varying the underlying distributions. The distributions were implemented using the best-fit parameters from our data augmentation process. For these epidemic simulations, we utilized the R0 package in R (Obadia *et al*. 2012) for discrete-time models, running 10,000 stochastic simulations, reporting the mean values for each of the resulting sets of epidemics. These simulations were run for 200 days, with a maximum count of 250 cases per day, with Poisson-distributed new cases. We have not incorporated uncertainty in the extrinsic incubation period due to lack of reliable data, and we have made the assumption that the incubation period distribution is proportional to the generation interval, as *P. vivax* infections produce infective gametocytes rapidly upon onset of clinical symptoms (Bousema and Drakeley 2011).

We used non-informative priors for all parameter estimations. Proposal distributions were adjusted using estimated means and covariances from pilot runs in an iterative process to accelerate convergence; assessment of convergence was performed using Geweke's diagnostic (Geweke 1992). All statistical analyses were performed in R (version 2.15.2) (R Core Team 2013), using the packages fitdistrplus, flexsurv, grid, MASS, seqinr, FAdist, stats4, R0 and custom-built code for the MCMC algorithm.

### 4.4 Results

*Incubation periods*

The study of experimental incubation period included 454 case-patients and overlaid distributions can be found in Figure 9; DIC comparisons for these distributional families for both the aggregate and for subregion-specific incubation period data are shown in Table 12. It should be noted for all the results that 'flattening' of the fitted curves relative to the data-based histograms arises from the data augmentation processes. The Deviance Information Criterion (DIC) indicates that the shifted log-logistic distribution has a substantially better fit than the second best shifted log-normal distribution, as evidenced by a DIC difference of 0.4. Mixture distributions of two gammas had limited levels of support (Δ DIC <3) while all other distributions were not supported by DIC. The increased complexity of mixture distributions does not explain any greater variation in these data, and are also not supported by DIC.

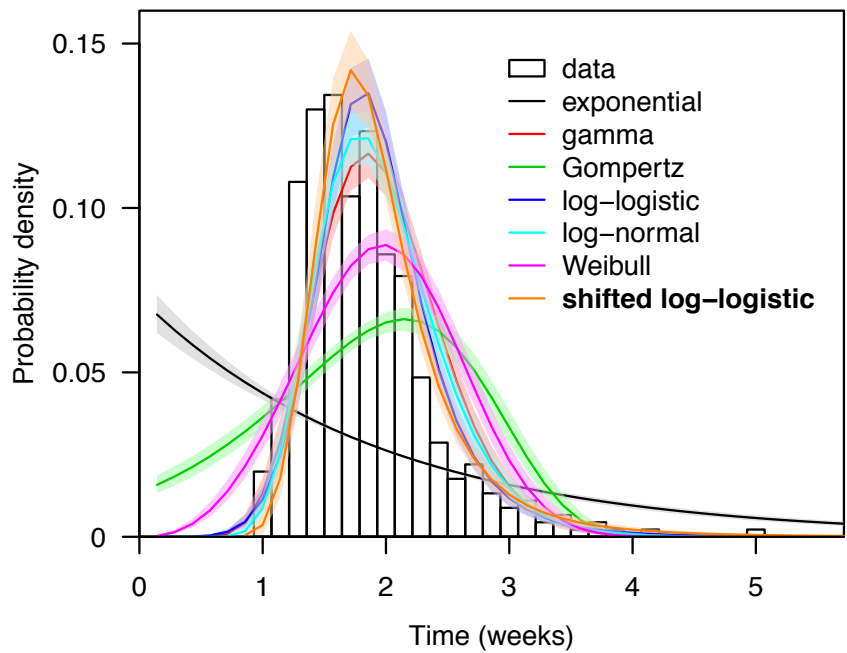

**Figure 9. Comparison of crude (non-data augmented) data and estimated parametric models of experimental incubation times, *P. vivax* malaria.**
Note: (N= 454); Experimental data are in black outlines, and parametric model fits are shown with 95% confidence intervals, along with the overall best parametric fit.

The subregion-specific distributions show some differences from the best-fit distribution (shifted log-logistic) from the aggregate data. Among the New World, tropical parasites there is support for a shifted Gompertz distribution (Δ DIC = 2.5), and in the New World, temperate strains there is very strong evidence for a shifted Weibull distribution ($\Delta$ DIC = 18.3). Quadrant-specific plots of Kaplan-Meier curves with best-fit distributions can be found in Appendix B.

| Distribution | Old World, Tropical | Old World, Temperate | New World, Tropical | New World, Temperate | Global fit, All regions |
|---|---|---|---|---|---|
| | $\Delta$ DIC | $\Delta$ DIC | $\Delta$ DIC | $\Delta$ DIC | $\Delta$ DIC |
| exponential | 583.1 | 111.8 | 109.8 | 332.9 | 917.2 |
| shifted exponential | 114.3 | 45.5 | 1.8 | 141.5 | 275.0 |
| mixture exponential | - | - | - | - | 919.1 |
| Weibull | 76.2 | 3.7 | 1.6 | 0.1 | 166.4 |
| shifted Weibull | 12.3 | 2.3 | 0.2 | **0.0** | 33.7 |
| mixture Weibull | - | - | - | - | 40.1 |
| log-normal | 4.7 | 4.1 | 1 | 15.4 | 16.3 |
| shifted log-normal | 2.5 | 5.5 | 1.1 | 17.6 | 0.4 |
| mixture log-normal | - | - | - | - | 18.3 |
| log-logistic | 3.1 | **0.0** | 3.4 | 16.6 | 14.2 |
| **shifted log-logistic** | 0.0 | 1.5 | 2.5 | 18.3 | **0.0** |
| mixture log-logistic | - | - | - | - | 16.1 |
| gamma | 11 | 1.3 | 0.6 | 9.2 | 40.3 |
| shifted gamma | 6.5 | 1.6 | 1.4 | 11.1 | 12.5 |
| mixture gamma | - | - | - | - | 2.7 |
| Gompertz | 159.4 | 17.3 | 0.8 | 10.6 | 366.0 |
| shifted Gompertz | 42.6 | 15.5 | **0.0** | 4.9 | 142.1 |
| mixture Gompertz | - | - | - | - | 100.7 |
| DIC of best-fit model | 958.8 | 345.5 | 163.6 | 680.1 | 2374.6 |

**Table 12. Best-fit distributions for experimental incubation times, *P. vivax* malaria.** Note: mixture models showed non-normal posteriors and are not reported). Best fitting distributions are shown in bold.

The distribution of the 529 cases in confounded and observational studies with longer-term incubations is shown in Figure 10 and Table 13; a bimodal peak is clearly evident. The studies with the St. Elizabeth strain involved a range of chemoprophylaxis regimens, and the Korean strain infections were all observational studies that largely included chemoprophylaxis. These results also overwhelmingly support a log-logistic distribution; in this case a mixture of two log-logistic distributions accurately capture the bimodal distribution commonly observed in temperate zone epidemiology. Shifted distributions showed extremely poor fit and are not reported. A Kaplan-Meier curve for these data can be found in the Appendix B.

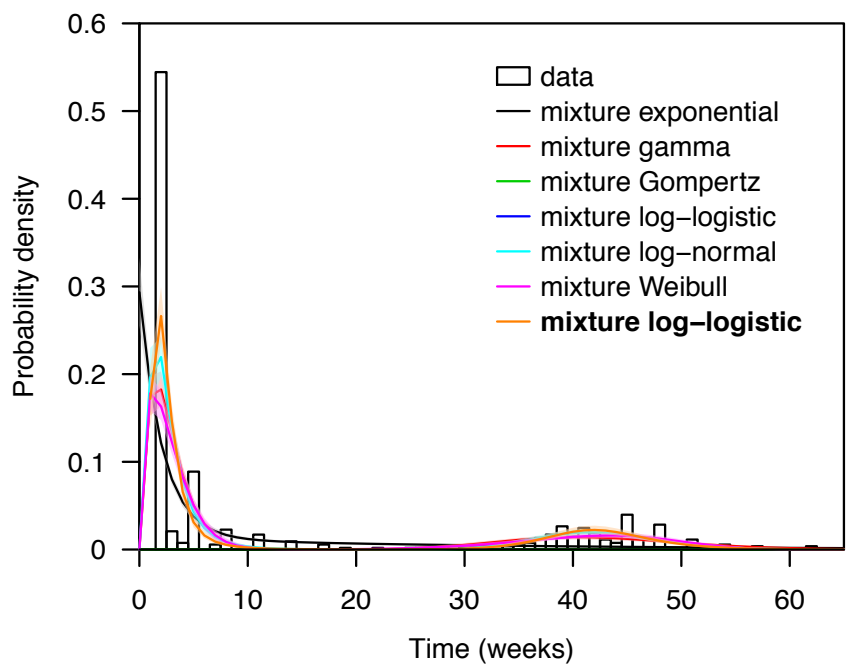

**Figure 10. Comparison of crude (non-data augmented) data and estimated parametric model of observational incubation times, *P. vivax* malaria.**
Notes: (N=529), Observational data are in black outlines, and parametric model fits are shown with 95% confidence intervals, along with the overall best parametric fit.

*Times to first relapse*

The results of the time-to-relapse analysis (primary infection to the first relapse) are shown in Table 14. We find that mixture distributions provide better fit for the total dataset than standard families; specifically, we find the best fit with a Gompertz mixture, followed by the log-logistic mixture ($\Delta$ DIC = 7.1), log-normal mixture ($\Delta$ DIC = 6.9) and Weibull mixture ($\Delta$ DIC = 7.3). While the differences among these three distributions are very minor, all capture the event times poorly relative to the Gompertz. The gamma and exponential mixtures both fit poorly ($\Delta$ DIC >7). Figure 11 shows these distributions compared with the experimental data; the district bimodal peak is captured by the Gompertz mixture. There is some limited support for a shifted Gompertz in the New World Tropical region ($\Delta$ DIC = 2.9), but the remaining regions, and the global fit to aggregate data, all show strong statistical support for a mixture Gompertz distribution.

| Standard distributions | | Mixture distributions | |
|---|---|---|---|
| Base model | $\Delta$ DIC | Base model | $\Delta$ DIC |
| exponential | 952.4 | exponential | 551.5 |
| gamma | 570.6 | gamma | 171.8 |
| Gompertz | 874.2 | Gompertz | 297.7 |
| log-logistic | 719.3 | **log-logistic** | 0.0 |
| log-normal | 688.4 | log-normal | 41.5 |
| Weilbull | 611.9 | Weilbull | 193.4 |
| DIC of best-fit model | | | **2876.7** |

**Table 13. Fitted distributions for observational incubation time studies, *P. vivax* malaria.**
Notes: shifted distributions showed extremely poor fit and are not reported; best fitting distribution is shown in bold.

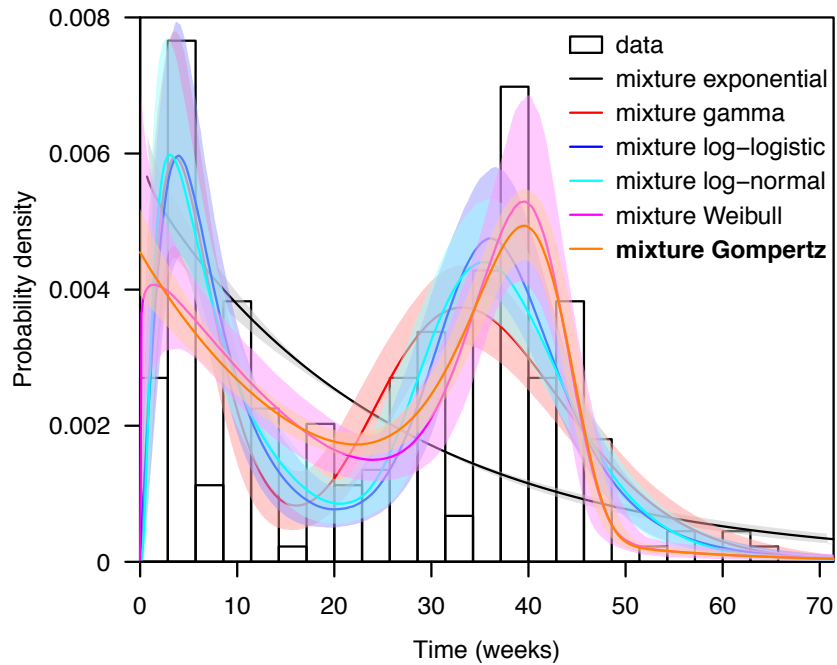

**Figure 11. Comparison of crude (non-data augmented) data and estimated parametric model of first relapse times, *P. vivax* malaria.**

Notes: (N= 222). Experimental data are in black outlines, and parametric model fits are shown with 95% confidence intervals, along with the best overall parametric fit.

*Sensitivity analyses*

A sensitivity analysis was performed for all three datasets and each of the subregions individually, and show strong evidence that the models provided good fits for the pseudodata by comparisons of the DIC values. Detailed results, estimated posterior distributions for model parameters overall and by quadrant, are presented in the Appendix B.

| Distribution | Old World, Tropical | Old World, Temperate | New World, Tropical | New World, Temperate | Global fit, All regions |
|---|---|---|---|---|---|
| | ΔDIC | ΔDIC | ΔDIC | ΔDIC | ΔDIC |
| exponential | 112.3 | 121.8 | 39.0 | 106.2 | 147.4 |
| shifted exponential | 86.2 | 80.1 | 13.3 | 103.9 | 134.1 |
| mixture exponential | 109.1 | 121.1 | 40.2 | 106.4 | 144.9 |
| Weibull | 138.1 | 221.2 | 111.8 | 212.4 | 190.7 |
| shifted Weibull | 55.6 | 79.2 | 13.2 | 104.5 | 120.0 |
| mixture Weibull | 80.5 | 3.8 | 4.2 | 14.3 | 5.4 |
| log-normal | 98.5 | 111.1 | 9.0 | 126.4 | 178.9 |
| shifted log-normal | 60.4 | 151.0 | 34.3 | 146.6 | 219.8 |
| mixture log-normal | 75.4 | 13.6 | 11.1 | 17.5 | 4.7 |
| log-logistic | 93.7 | 125.2 | 8.2 | 121.4 | 189.3 |
| shifted log-logistic | 60.5 | 131.9 | 25.2 | 129.8 | 190.1 |
| mixture log-logistic | 94.5 | 126.8 | 9.5 | 122.3 | 4.1 |
| gamma | 112.0 | 94.3 | 9.7 | 104.7 | 130.5 |
| shifted gamma | 54.3 | 80.8 | 15.0 | 105.5 | 131.0 |
| mixture gamma | 86.4 | 25.7 | 9.3 | 1.1 | 7.5 |
| Gompertz | 114.0 | 107.8 | 2.7 | 64.1 | 74.9 |
| shifted Gompertz | 89.4 | 79.1 | **0.0** | 63.9 | 73.4 |
| **mixture Gompertz** | **0.0** | **0.0** | 2.9 | **0.0** | **0.0** |
| **DIC of best-fit model** | 144.5 | 1402.0 | 219.6 | 467.17 | 2596.2 |

**Table 14. Fitted distributions for experimental relapse times, *P. vivax* malaria.**
Notes: Best fitting distributions are shown in bold.

*Epidemic simulations*

The results of stochastic epidemic simulations can be found in Figure 12. These results suggest that at a reproductive number ($R_0$) of 5, the time scale of a modelled epidemic varies dramatically based upon the distribution of the incubation period. Use of an exponential distribution, as is extremely common in SIR compartmental simulations, shows a much more rapid epidemic, with Gompertz and Weibull distributions showing more gradual epidemic evolution. Finally, gamma, log-logistic, log-normal, and shifted log-logistic have the latest epidemic peaks, and are virtually indistinguishable from one another.

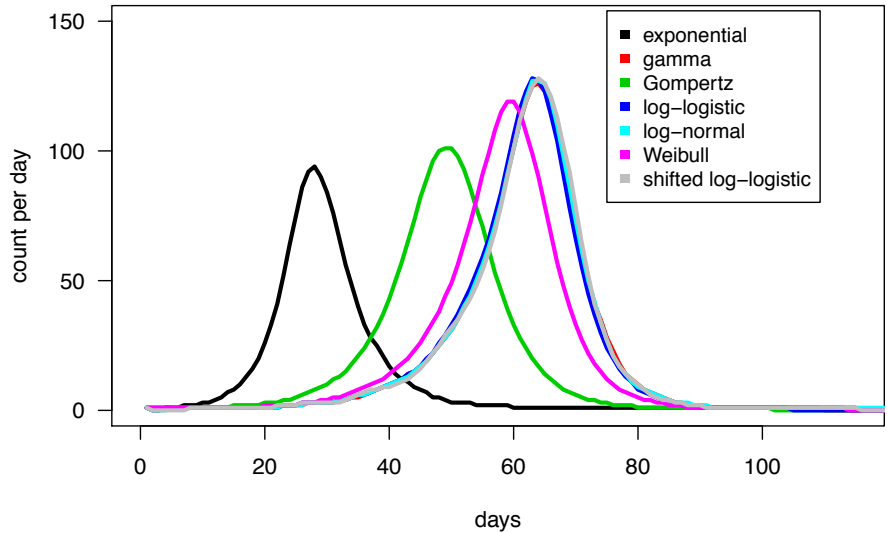

**Figure 12. Comparison of simulated *P. vivax* malaria epidemics.**
Note: $R_0 = 5$; mean values from 10,000 simulations for each standard distribution.

The mean total cases for each set of 10,000 simulations by underlying distributions are shown in Table 15. Comparison of these totals shows that within the 95% confidence intervals, the total number of cases within the epidemic is greater for the best-fitting log-logistic and shifted log-logistic distributions relative to exponential- and gamma-distributed incubation periods. Simulations with $R_0 = 50$ and 75 produced consistent results but with greater separation of the gamma, log-logistic, log-normal, and shifted-log-logistic epidemic curves (results not shown).

| Distribution | Total case count | (95% CI) |
|---|---|---|
| exponential | 1333 | (1330 to 1335) |
| gamma | 2329 | (2326 to 2332) |
| Gompertz | 1950 | (1947 to 1953) |
| log-logistic | 2306 | (2304 to 2310) |
| log-normal | 2328 | (2325 to 2331) |
| Weilbull | 2216 | (2214 to 2220) |
| **shifted log-logistic** | 2330 | (2328 to 2334) |

**Table 15. Total case counts from epidemic simulations, *P. vivax* malaria.**
Note: mean values and 95% CIs from 10,000 simulations for each distribution.

**4.5 Discussion and conclusions**

Although some of the earliest simulation models of malaria were directed towards *P. vivax* in epidemics, this parasite has received limited attention from modellers (MacDonald *et al*. 1968). The models that have appeared have used a range of distributions for the model parameters of incubation period and time-to-relapse. Some of the earliest comprehensive mathematical models for *P. vivax* did not consider distributional assumptions and relied on point estimates (de Zoysa *et al*. 1991); other mathematical models used a log-normal distribution for relapses and a single estimate of 15 days for incubation period (Ishikawa *et al*. 2003), implying an exponential distribution. A stochastic model of potential *P. vivax* transmission within Japan used a gamma distribution for the incubation period, an exponential distribution for short relapse periods, and a log-normal for longer relapses (Bitoh *et al*. 2011).

Two other comprehensive mathematical models implicitly assume exponential distributions for both incubation and times-to-relapse (Águas *et al*. 2012, Chamchod and Beier 2013). A recent comprehensive model including multiple relapse states used a 15-day incubation period in simulations to produce a mean relapse interval of 7.1 months for cases in India, with incubation as an exponential distribution, and relapses modelled using a gamma distribution (Roy *et al*. 2013).

Several studies have found that results from infectious disease models can be highly sensitive to accurate distributional assumptions (Lloyd 2001, Wearing *et al*. 2005); our study reinforces these conclusions in finding the 'default' exponential and gamma distributions, used for mathematical tractability, inadequately capture the complexity of experimental data (Sama *et al*. 2006, Reich *et al*. 2009). The results from our simulations concur with these statements and suggest that use of best-fitting distributions can lead to larger overall case-counts and slower epidemic evolution than would be predicted based upon exponential or gamma distributed incubation periods. As the underlying distributional assumptions have large and important impacts upon both the time-scale of epidemic evolution and total case counts in *P. vivax* epidemics, these parameters are therefore a critical component of accurate models.

A range of entomological, molecular, genetic, and epidemiological evidence suggests the existence of subspecies within *P. vivax* (Li *et al.* 2001, Neafsey *et al.* 2012, Lover and Coker 2013); however there has been limited consideration of this aspect of parasite biology in published models (Chamchod and Beier 2013). Few empirical data exists to support models that include explicit consideration of this aspect of the epidemiology; this study provides parameterization for subpopulations by climactic zone (temperate and tropical) as well as the postulated subspecies *P. vivax vivax* (E. hemisphere) and *P. vivax collinsi* (W. hemisphere) to inform region-based models towards global malaria elimination. Our results show that shifted log-logistic distributions adequately capture the incubation period for all regions except for the New World, temperate parasites, which show strong support for a shifted Weibull. However, as these parasite populations were eliminated in the early 20[th] century, they have limited relevance for modern modelling studies (Collins 2013a).

The results from the observational long-latent infections have several important implications. Although the biological underpinnings of relapse remain obscure (Markus 2012), there has been considerable debate that long-latencies may in fact be relapses after a sub-clinical primary infection (Horstmann 1973). The results from this study show that relapses exhibit quantitatively different behaviour at a population-level from these long-incubation periods, and this in turn suggests a closer biological link to 'normal' incubations than to relapses.

Secondly, the congruence of the distributions from experimental and observational studies suggests that results from observational studies, while inherently limited, may still adequately capture the natural history of infection with *P. vivax*. This finding may greatly expand the utility of available datasets to more completely explore the epidemiology of *P. vivax*.

In modelling the time-to-relapse, there is strong support for a mixture Gompertz for all event times except in the New World Tropical region, where a shifted Gompertz is supported. In addition to simplifying modelling, this concordance of distributions in different parasite populations suggests that hypnozoite activation may have a common underlying biological trigger, regardless of parasite genetics (Shanks and White 2013). While these results for temperate zone parasites are based primarily on now-eliminated Russian strains, the parasites currently circulating on the Korean peninsula have been reported to have similar relapse patterns (White 2011).

However, this study has several limitations. The times we have analysed are from adult, non-immune and mostly Caucasian subjects with uncertain inclusion or exclusion criteria, and may not represent the experience in high transmission settings due to the influence of immunological factors, as well as the poorly understood impact of mixed-species malaria infections (Zimmerman *et al*. 2004). While malariotherapy for neurosyphilis treatment has been shown to have minimal impacts on incubation periods, larger impacts were found for relapse periods in some sub-populations of *P. vivax* (Lover and Coker 2013).

A related study examined the length of *P. falciparum* infections found that the total duration of infections were best modelled using a Gompertz distribution (Sama *et al*. 2006). However, the existence of relapses makes defining a duration of infection with *P. vivax* difficult; multiple lines of evidence suggests that relapses within a single infection may be genetically distinct from the primary infection (Collins 2007).

Our results suggest that the 'default' distributions used in many modelling studies (exponential and gamma distributions), may be inadequate to fully capture the natural variability and complexity of event times in human infections with *Plasmodium vivax* malaria. Future modelling studies should consider the use of log-logistic and Gompertz distributions for incubation periods and relapse times respectively, and the region-specific distributions included in this work should be considered to accurately model regional variations in the epidemiology of this parasite. Future statistical and mathematical models of *P. vivax* transmission should incorporate the more complex distributions identified in this study to maximize the congruence with the true natural history and epidemiology of this important human pathogen.

# Chapter 5: Epidemiological impacts of mixed-strain infections in experimental human and murine malaria

## 5.1 Abstract

Infectious diseases within a single host may be composed of multiple strains or lineages of a pathogen with differing genetics or antigenicites. These pathogen populations may interact with one another, and range of effects have been documented that may impact clinical presentation, evolution of virulence, drug-resistance, and epidemiology of onward transmissions. In most endemic areas, many single-species human malaria infections are composed of multiple strains of *Plasmodium* sp., but reports on epidemiological impacts are sparse.

In this study we examine a limited set of human experimental infections including relapses, with mixed strains of *Plasmodium vivax*, using survival and binary regression to models to examine the population-level impacts of strain interactions relative to single strain infections. Using these same modelling frameworks, we then examine a wide range of experimental murine malaria infections with mixed-strains infections to assess the consistency of these animal models with human infections.

We find that while mixed-strain infections do show statistically significant differences from single strain infections, we also observe several types of divergent behaviour. In parasites that generally target younger red blood cells (reticulocytes), mixed-strain infections are significantly different from both single infections, while in parasites that infect all red blood cell age-classes, the mixed infection does not differ from single-strain infections with the more virulent parasite strains. These results are consistent with multiple mathematical models of intra-infection dynamics. We also find remarkable stability in observed differences in *P. chabaudi* murine infections in two laboratories across a decade.

These results suggest diverse strategies for virulence management and drug-resistance containment may be needed to address important differences in the biology of different parasite species, and we suggest how alternative analyses could potentially maximize results from these technically challenging experiments.

## 5.2 Introduction

Malaria is a major contributor to morbidity and mortality globally, with an estimated 207 million cases (95% CI: 135 to 287) and 627,000 deaths (95% CI: 473,000 to 789,000) in 2012 (World Health Organization 2013a). There are six species that generally infect humans, and the interactions between different species within a single infection can have important clinical implications (Zimmerman *et al*. 2004). Moreover, infections within an individual may be composed of multiple strains (clonal populations) of a single species, which may interact both with one another and with the host immune system.

The implications of these interactive populations have been explored through multiple lines of research, using two types of models: mathematical/statistical models, and experimental animal systems. A range of effects have been proposed or measured from these efforts; recent reviews have focused on both clinical (Balmer and Tanner 2011) and general ecological (Alizon *et al*. 2009) considerations. Briefly, the interaction between parasite populations may lead to parasite competition or to parasite mutualism; to the evolution of virulence and drug susceptibility; and may facilitate genetic exchange within infections (Balmer and Tanner 2011).

In a range of epidemiological settings, the majority of individual *P. vivax* and *P. falciparum* infections are composed of multiple strains. Recent pyrosequencing has demonstrated highly diverse suites of parasites within an infection (Juliano *et al*. 2010), but other studies have suggested that mixed-clone *P. falciparum* infections are generally very closely related, and only rarely consist of highly divergent parasites (Nkhoma *et al*. 2012).

Longitudinal studies from Papua New Guinea examining both *P. falciparum* and *P. vivax* populations suggest a cycling of different clonal suites, which are rapidly displaced by newer, less-related strains (Bruce *et al*. 2000, Mackinnon and Read 2004). Studies on population-level diversity suggests that parasite factors could be the single largest component of observed clinical differences in severity or virulence (Mackinnon *et al*. 2000); moreover, these infections may represent an underutilized tool to explore complex transmission dynamics (Bordes and Morand 2009, de Araujo *et al*. 2012).

However, the epidemiological (population-level) impacts of mixed-strain infections have received more limited attention due to difficulties in data capture and inherent ethical considerations in studies involving human subjects. In this study, we re-examine historical human experimental infections with *P. vivax* using modern analytical techniques, and then compare these results with re-analyses of data from murine model systems, to assess the consistency between these inherently limited human data and the much larger corpus of animal models and ecological theory.

*Strain theory and virulence*

The concept of microbial strains is pervasive throughout the medical and malaria literature, however, no consensus exists on what these clinical isolates truly represent (Balmer and Tanner 2011, McKenzie *et al*. 2008); in general they refer to clonal, or at least closely-related, populations. A wide range of observations have been made regarding the clinical impact of strains, and include reports of differences in virulence, clinical severity, transmissibility, and both the number and spacing of hypnozoite-derived relapses in *P. vivax* infections (McKenzie *et al*. 2008). Recent molecular and genetic methods have also added to this knowledge base by characterizing both inter- and intra-host infections; these studies suggest that while many strains/isolates represent a diversity of clonal populations, they generally produce stable clinical/immunological responses (McKenzie *et al*. 2008).

Previous analyses in murine systems have almost entirely focused on pooled haematological or parasitological outcomes as proxies for virulence (comparing means or geometric means from all surviving animals), but full multivariate analyses of outcomes or consideration of times-to-events have received very limited attention. While time itself is a critical component of virulence, analyses have generally focused on analysis at several (potentially arbitrary) time points. Finally, in many cases, censoring of subjects has not been considered- that is, animals are removed from analysis; beyond a potential for introducing biases, this greatly reduces the power of the analysis (Rothman *et al*. 2008). Finally, many studies have used simple p-values to evaluate significance, as is the norm in ecology; however, these do not allow consideration of the effect size (statistically significant effects may not be clinically/ biologically important). A fuller consideration of virulence, and of effect size measures may provide important additional information beyond these summary statistics.

Lastly, the concept of 'virulence' is complex and encompasses the outcomes of many different and interrelated interactions between the parasites and host. While a range of definitions have appeared (Poulin and Combes 1999, Casadevall and Pirofski 2001), herein we consider it as 'parasites populations that maximally exploit host resources' to capture the range of endpoints examined in this study, with a focus on direct parasite-induced mortality.

## 5.3 Methods

*Data sources*

The human study data included in this analysis was obtained during a series of rigorous human challenge experiments among prison volunteers in the 1940's - 50's in a program to develop novel antimalarial drugs; all case-patients were malaria-naïve white males. Infections were via mosquito challenge, there was complete follow-up, and patients were protected from super-infection due to the institutional nature of these populations. Analysis of incubation periods includes 44 patients inoculated with the St. Elizabeth strain; 131 with Chesson parasites; and nine patients with mixed infection. Analysis of relapses includes 38 patients with St. Elizabeth; 19 with Chesson, and nine patients with mixed infection (Table 16).

Full details for human challenge infections including literature search criteria and primary literature references can be found in (Lover and Coker 2013); data for the mixed strain challenge are from (Cooper *et al*. 1950). All of these infections were transmitted by infected vectors (generally *Anopheles quadrimaculatus*). Relapse times are presented as time from mosquito inoculation. All single-strain infections had symptomatic-only treatment with quinine; however, all mixed-strain inoculations had chloroquine exposure after the primary attack. The Chesson strain of *P. vivax* has been described as being extremely virulent, with rapid onset of symptoms and with a propensity for many closely spaced relapses (Ungureanu *et al*. 1976); the St.

Elizabeth is a milder 'temperate' strain first isolated in the southern US (Collins 2013a).

| End point | Treatment | n | % of total |
|---|---|---|---|
| *Incubation period* | Mixed | 9 | 4.9 |
| | Chesson | 131 | 71.2 |
| | St. Elizabeth | 44 | 23.9 |
| | **Total** | **184** | **100.0** |
| *Time-to-first relapse* | Mixed | 9 | 13.6 |
| | Chesson | 19 | 28.8 |
| | St. Elizabeth | 38 | 57.6 |
| | **Total** | **66** | **100.0** |

**Table 16. Study population, historical human challenge experiments with *P. vivax*.**

Data for the *P. yoelii* infections (Table 17) were extracted from (Hargreaves *et al*. 1975); although the authors refer to these parasites as *P. berghei yoelii*, a later revision elevated these parasites to the species-level as *P. yoelii* (Killick-Kendrick 1974, Perkins *et al*. 2007). CF1 mice were inoculated with $10^6$ parasitized RBCs, and $2 \times 10^6$ in the mixed infections. The mild strain of '17X' was obtained from a wild thicket rat; the virulent form was obtained after serial passages in rodents and *An. stephensi* (Hargreaves *et al*. 1975).

| Treatment | n | % of total |
|---|---|---|
| Mixed | 6 | 33.3 |
| '17X mild' | 6 | 33.3 |
| '17X virulent' | 6 | 33.3 |
| **Total** | **18** | **100.0** |

**Table 17. Study population, murine challenge experiments with *P. yoelii*.**

Data for the first set of *P. chabaudi* murine challenge experiments (Table 18) were taken from (Snounou *et al*. 1992). Male CBA/ca mice were inoculated with $10^4$ parasitized erythrocytes; mixed infections were inoculated with $1 \times 10^4$ of each strain.

| Treatment | n | % of total | Mortality % (95% CI) |
|-----------|-----|-----------|---------------------|
| AS | 10 | 10 | 20 (2.5 – 55.6) |
| CB | 10 | 10 | 80 (44.4 – 97.5) |
| DS | 10 | 10 | 60 (26.2 – 87.8) |
| AS + CB | 10 | 10 | 100 (69.2 – 100) |
| CB + DS | 10 | 10 | 100 (69.2 – 100) |
| AS + CB | 10 | 10 | 40 (12.2 – 73.8) |
| **Total** | **60** | **100.0** | - |

**Table 18. Study population, murine challenge experiments with *P. chabaudi* (set I).**
Source: (Snounou *et al*. 1992)

Data for the later murine challenges with *P. chabaudi* (Table 19) are from the work of Bell *et al*. and are available from the Drydad data repository (Bell *et al*. 2006, 2014). In these infections, C57B1/6J inbred female mice were inoculated with $10^6$ parasites; mixed infections received a total dose of 2 x$10^6$, however the authors cite prior studies suggesting this 2-fold difference has minimal impact on outcomes (Bell *et al*. 2006). The strains utilized are from diverse sources and were chosen specifically to mimic the diversity within wild parasite populations. Prior work suggested that the AJ and AT strains were more virulent relative to AS and CB; however, in the first experiment, CB was unexpectedly virulent and was replaced by CW for the second replicate. This analysis was focused on the initial acute phase of infections before specific immunological responses occur; therefore surviving mice were censored one day past the last recorded fatality (on day 15).

*Methods*

Where data allowed, preliminary analyses of the relationship between parasite strains and times-to-event used non-parametric analyses (Kaplan-Meier curves and related methods) to determine if statistically significant differences exist; differences were compared using the Peto-Peto test due to differences in censoring and non-proportional hazards (Klein and Kleinbaum 2005, Lachin 2011). These analyses were then followed by multivariate analysis (survival and/or log-binomial models for time-to-event or mortality) to estimate the magnitude of observed effects.

| Treatment | n | % of total | Mortality % (95% CI) |
|---|---|---|---|
| 1. AS | 10 | 10 | 0 (0 – 30.8) |
| 2. AJ | 10 | 10 | 70 (34.8 – 93.3) |
| 3. AT | 10 | 10 | 30 (6.7 – 65.2) |
| 4. CB | 5 | 5 | 60 (14.7 – 94.7) |
| 5. AS+AJ | 10 | 10 | 30 (6.7 – 65.2) |
| 6. AS+AT | 10 | 10 | 50 (18.7 – 81.3) |
| 7. AS+CB | 5 | 5 | 40 (5.3 – 85.3) |
| 8. AJ+AT | 10 | 10 | 50 (18.7 – 81.3) |
| 9. AJ+CB | 5 | 5 | 100 (47.8 – 100) |
| 10. AT+CB | 5 | 5 | 80 (28.4 – 99.5) |
| 11. CW | 5 | 5 | 0 (0 – 52.2) |
| 12. AS+CW | 5 | 5 | 20 (0.51 – 71.6) |
| 13. AJ+CW | 5 | 5 | 20 (0.51 – 71.6) |
| 14. AT+CW | 5 | 5 | 40 (5.3 – 85.3) |
| **Total** | **100** | **100.0** | **-** |

**Table 19. Study population, murine challenge experiments with *P. chabaudi* (set II).**
Note: Source (Bell *et al*. 2006).

A diverse range of parametric and semi-parametric survival models was explored in the analysis of these human and murine experimental data. However, extensive issues were observed with lack-of-fit, non-convergence due to collinear strains, crossing survival curves, or complete separation. Moreover, standard survival models are not applicable in the presence of crossing survival curves, as a single hazard ratio is an inadequate measure of a time-dependent risk of event.

Due to these analytical issues, a set of complementary analyses was performed so that all studies were analysed within a consistent analytical framework. First, pseudovalues were computed and used to compare differences in restricted mean survival times (Parner and Andersen 2010, Andersen and Perme 2010). This approach has been specifically suggested to assess differences in survival when hazard ratios may be an inaccurate measure of differences; the restricted mean survival times (RMST) were calculated at just before the end of follow-up time (symbolized as t*) (Royston and Parmar 2011). Secondly, the incidence rate ratios (that is, the ratio of total number of events per follow-up time, per strain) were compared using Poisson models.

Finally, where possible, full multivariate methods were used to assess the magnitude of strain differences, using the optimal models for each dataset. For the study of *P. yoelii* mortality, flexible parametric survival models were used to comprehensively model the underlying hazard function and to provide estimates of effect size with adjustment for covariates. These models are comprehensive alternatives to standard Cox survival models which allow greater flexibility in model-fitting and prediction (Royston and Parmar 2002, Royston and Lambert 2011).

For the first study of *P. chabaudi* mortality (Snounou *et al*. 1992), regressions for a binomial outcome (mortality) were used to quantify the differences in mortality between strains. As the outcome of mortality was common within this study, the odds ratios provided by standard logistic regression models would be biased estimates of risk; we therefore used alternative models for binary outcomes. Two models have been widely used: log-binomial and robust Poisson models; importantly, these provide risk ratios as opposed to the odds ratios produced in logistic regression. Both options provided consistent effect size estimates for these data; however, we report the Poisson models as they have been reported to be more robust to outliers and displayed fewer issues with model convergence (Zou 2004, Chen *et al*. 2014).

For the second study of *P. chabaudi* mortality (Bell *et al*. 2006), regressions for a mortality outcome were also used to quantify the differences in mortality between strains. Both model parameterizations in this case had convergence issues, so the 'COPY' method of Petersen and Deddens was used to facilitate model convergence and estimation of risk ratios (Petersen and Deddens 2006). This method consists of doubling the dataset, reversing the outcomes in all added cases, and then weighting the reversed outcomes at 0.001:0.999 relative to the original data (Cummings 2009).

For all models in this analysis model parsimony was assessed using Akaike and Bayesian information criteria (AIC/BIC); residuals and standard goodness-of-fit tests were used to assess adequacy of the robust Poisson and restricted mean survival models; the fit of the flexible parametric survival model was judged using overlaid comparisons with the non-parametric Kaplan-Meier curves. 'Robust' (Huber-White) error structure was used in all models to address the non-independence of subjects within study arms (Williams 2000). Models were assessed for proportional hazard violations using Schoenfeld residuals; however the models reported all have no proportional-hazard assumption.

All analyses were performed using Stata 13.1 (College Station, TX, USA) and all tests were two-tailed, with $\alpha = 0.05$.

## 5.4 Results

*Plasmodium vivax infections- incubations*

The results from Kaplan-Meier analysis of incubation periods (time from mosquito exposure to febrile illness) are shown in Figure 13 and Table 20. Infections with the Chesson strain had the shortest incubation periods, with intermediate (and crossing) curves from mixed infections, followed by the St. Elizabeth strain infections. The arithmetic median incubation periods were: Chesson 12 days (95% CI: 12 to 12), mixed infection 14 (12 to 14), and St. Elizabeth 15 (95% CI: 15 to 16), and the differences in restricted mean survival time (RMST) are all significant (mixed infection vs. St. Elizabeth; and Chesson vs. St. Elizabeth, both p < 0.001; Chesson vs.

mixed, p = 0.001).

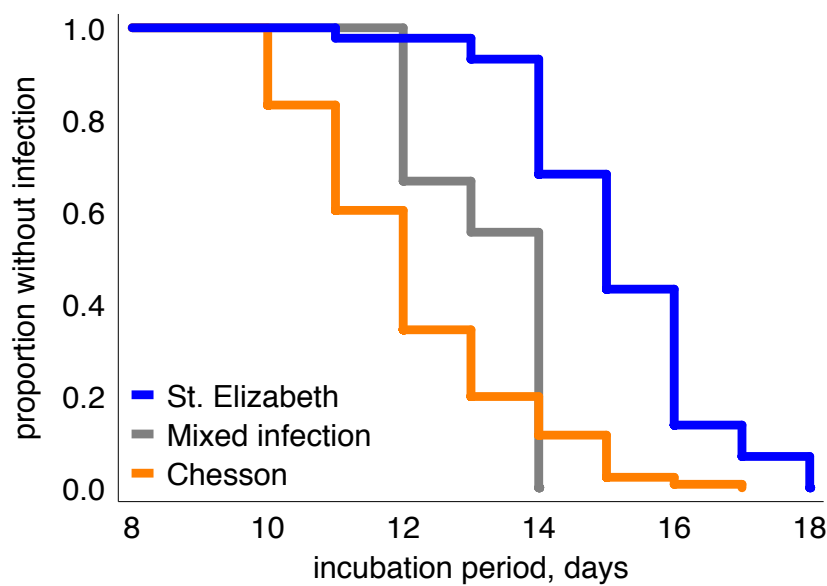

**Figure 13. Kaplan-Meier curves comparing incubation periods in single strain and mixed-strain infections in human challenge experiments with *P. vivax*.**
Note: N= 184

| Parasite strain | Incidence rate ratio, (95% CI) | p-value, IRR | RMST difference, (SE) | p-value, RMST |
|---|---|---|---|---|
| St. Elizabeth vs. Chesson (ref) | 0.80 (0.77 – 0.83) | **< 0.001** | 3.1 (0.25) | **0.001** |
| Chesson vs. mixed (ref) | 1.1 (1.03 – 1.15) | **< 0.001** | -1.1 (0.34) | **0.001** |
| St. Elizabeth vs. mixed (ref) | 0.87 (0.82– 0.91) | **< 0.001** | 2.0 (0.37) | **< 0.001** |

**Table 20. Comparison of incubation periods in human challenge experiments with mixed-strain infections of *P. vivax*.**
Notes: (N= 184) (RMST at t* = 18 days); entries in bold are significant with p < 0.05.

Comparisons between incidence rate ratios are consistent with the Kaplan-Meier

analysis: infections with the St. Elizabeth strain had an IRR of 0.80 (95% CI: 0.77 –

0.83; p < 0.001) relative to the more virulent Chesson strain. Relative to mixed inoculations, Chesson infections had an IRR of 1.1 (1.03 – 1.15; p < 0.001), with St.

Elizabeth having an IRR of = 0.87 (0.82 – 0.91; p < 0.001). That is, over the follow up period, case-patients infected with Chesson parasites had a ~10% faster rate, and those infected with St. Elizabeth parasites a ~13% slower rate, of presenting with febrile illness respectively, than case-patients infected with mixed-strain parasites.

*Plasmodium vivax infections- relapses*

Kaplan-Meier plots for time-to-first relapse (measured as the time from primary infection) are shown in Figure 14. Rapid times to relapse are apparent in infections with the Chesson strain, followed by an intermediate curve from mixed inoculations, with St. Elizabeth relapses having the slowest times-to-relapse, spaced out over ~10 months. The median uncensored relapse times are: Chesson 6.3 weeks (95% CI: 5.6 to 7.1); mixed 11.4 weeks (9.2 to 38.5); and St. Elizabeth 41.7 weeks (39.5 to 43.0).

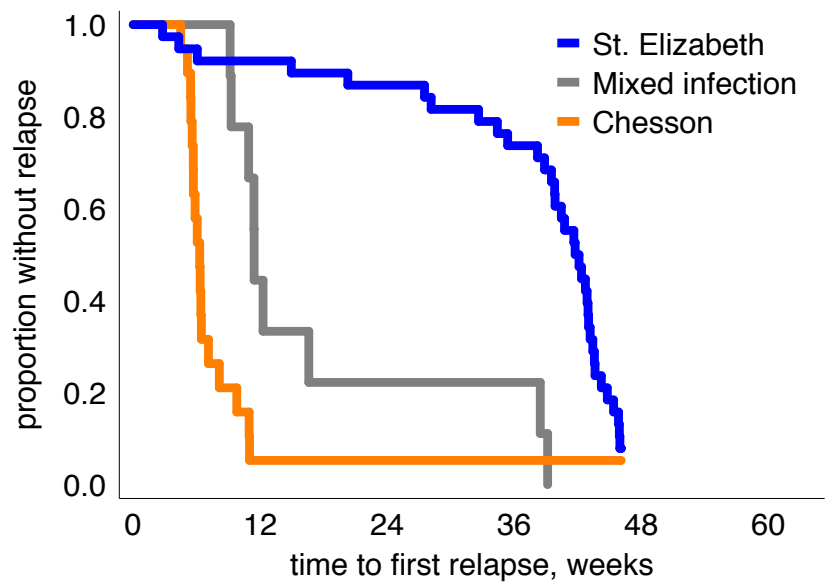

**Figure 14. Comparison of Kaplan-Meier curves, time-to-first relapse, human challenge infections with mixed-strain *P. vivax*.**
Note: N= 66.

Comparison of the incidence rates for relapses (Table 21) suggests that relative to the Chesson strain, infections with the St. Elizabeth strain have an incidence rate ratio of 0.23 (95% CI: 0.13 – 0.41, p < 0.001). In comparison to mixed strain infections, infections with the St. Elizabeth strain had an IRR of 0.44 (0.27 – 0.70, p = 0.001), while the Chesson infections do not achieve significance, with an IRR of 1.9 (0.92 –

3.9; p = 0.081). That is, in single strain infections with St. Elizabeth parasites, there is

~ 2.3 fold, and ~4.3 fold, slower rate of relapse relative to mixed infections and to

Chesson strain infections respectively, while mixed infections do not show a significant difference from infections with Chesson parasites.

| Parasite strain | Incidence rate ratio, (95% CI) | p-value, IRR | RMST difference (SE) | p-value, RMST |
|---|---|---|---|---|
| St. Elizabeth vs. Chesson (ref) | **0.23 (0.13 – 0.41)** | **< 0.001** | **28.1 (2.8)** | **< 0.001** |
| Chesson vs. mixed (ref) | 1.9 (0.92 – 3.9) | 0.081 | - 8.8 (4.4) | 0.042 |
| St. Elizabeth vs. mixed (ref) | **0.44 (0.27 – 0.70)** | **0.001** | **19.2 (4.3)** | **< 0.001** |

**Table 21. Comparison of time-to-first relapse (from parasite inoculation) in human challenge experiments with *P. vivax*.**
Notes: (N= 66) (RMST at t* = 45 weeks); entries in bold are significant with p < 0.05.

Comparison of the RMSTs at t*=45 weeks are significant for both St. Elizabeth

vs. Chesson (p < 0.001) and vs. mixed infection (p < 0.001); and the Chesson

infections show only a marginally significant difference from mixed strain infections

(p = 0.042).

*Plasmodium yoelii infections*

The results from survival analysis of mortality outcomes in experimental murine

*P. yoelii* infections are shown in Figure 15 and Table 22. There are clearly disenable

and statistically significant differences in the time-to-mortality between both of the

single strains and mixed infections, and between the two single-strain infections. In

Kaplan-Meier analysis, the Peto-Peto log-rank test indicates significant differences between the strains ('17X-mild' vs. '17X-virulent', p= 0.0010; '17X-virulent' vs.

mixed, p = 0.0103; '17X-mild' vs. mixed, p = 0.0034); comparison of the RMST

differences (Table 22) at t*= 22 days shows comparable p-values ('17X-mild' vs. '17X-virulent', p < 0.001; '17X-virulent' vs. mixed, p = 0.002; '17X-mild' vs. mixed, p = < 0.001). In flexible parametric survival models, with mixed infection as the reference, the '17X- virulent' strain showed an hazard ratio (HR) of 7.5 (95% CI: 1.5 to 38.0; p = 0.015); the '17X - mild' strain had an HR of 0.049 (95% CI: 0.011 to 0.21; p < 0.001). These imply that relative to mixed strain infections, mild strain infections have ~7.5 times lower risk, and virulent strain infections have ~20 times greater risk, of mortality in *P. yoelii* infections. The incidence rate ratios are also consistent with these other analyses, and suggest approximately 2-fold faster (vs. virulent) and 2-fold slower (vs. mild strain) mortality rates in mixed-strain infections.

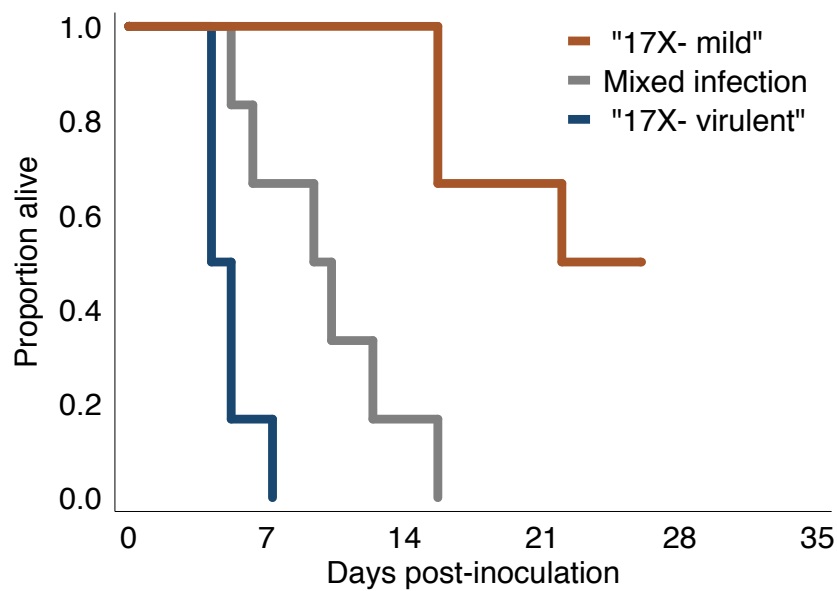

**Figure 15. Kaplan-Meier and flexible-parametric survival model curves comparing time-to-mortality in mixed-strain infections in murine challenge experiments with *P. yoelii.***
Note: (N= 18).

| Parasite strain | p value, Peto-Peto test | RMST difference (SE) | p-value, RMST | IRR (95% CI) | P values, IRR | Hazard ratio, 95% CI | p value, hazard ratio |
|---|---|---|---|---|---|---|---|
| '17X- mild' vs. '17X- virulent' (ref) | 0.0010 | 14.9 (1.3) | < 0.001 | 0.51 (0.36 – 0.72) | < 0.001 | 0.0065 (0.00091- 0.046) | < 0.001 |
| '17X- virulent' vs. mixed (ref) | 0.0103 | - 4.8 (1.6) | 0.002 | 1.96 (1.39 – 2.76) | < 0.001 | 7.51 (1.49 – 37.87) | 0.015 |
| '17X- mild' vs. mixed (ref) | 0.0034 | 10.0 (1.9) | < 0.001 | 0.23 (0.075 – 0.14) | 0.004 | 0.049 (0.011 - 0.21) | < 0.001 |

**Table 22. Comparison of time-to-mortality in murine challenge experiments with *P. yoelii*.**
**Notes:** (N=18) (RMST at t* = 22 days); IRR= incidence rate ratio; entries in bold are significant with p < 0.05.

*Plasmodium chabaudi infections*

The analysis of *P. chabaudi* data from Snounou *et al*. is shown in Table 23, with mortality as an endpoint. Overall, mixed-strain infections showed a significant difference in risk of mortality relative to single infections: any mixed infections vs. all single, (RR = 1.6; 95% CI: 1.1 to 2.3, p = 0.009). While mortality was analysed within a unified model for all infections, results are presented with differing reference categories within each triad (two single infections and the corresponding mixed infection). Shown first within each triad is the 'control' scenario- that is, a direct comparison between the two single strain infections, followed by comparisons of each single-strain infection to the mixed-strain scenario. In these data, only a single triad shows significant differences between the two 'pure' infections; a significant difference exists between single infections with AS (avirulent) and CB (virulent) strains; and the mixed infection remains significantly different from single-strain AS infections (RR = 0.20; 95% CI 0.58 to 0.70, p = 0.012). However, no evidence for difference was found in comparison with single CB infections (RR = 0.8; 0.59 to 1.1, p = 0.162).

| Parasite strain triad (more virulent strains in bold) | Comparison | Risk ratio (95% CI) | p-value, risk ratio |
|---|---|---|---|
| i. Mixed AS/**CB (ref)** | AS vs. CB (ref) | **0.25 (0.069 – 0.91)** | **0.035** |
| | AS vs. AS+CB (ref) | **0.20 (0.58 – 0.70)** | **0.012** |
| | CB vs. AS+CB (ref) | 0.80 (0.59 – 1.1) | 0.162 |
| ii. Mixed DS/**CB (ref)** | DS vs. CB (ref) | 0.75 (0.41 – 1.4) | 0.346 |
| | DS vs. DS+CB (ref) | **0.60 (0.36 – 0.99)** | **0.050** |
| | CB vs. DS+CB (ref) | 0.80 (0.59 – 1.1) | 0.162 |
| iii. Mixed AS/**DS (ref)** | AS vs. DS (ref) | 0.33 (0.09 – 1.3) | 0.111 |
| | AS vs. AS+DS (ref) | 0.33 (0.09 – 1.3) | 0.111 |
| | DS vs. AS+DS (ref) | 1.0 (0.49 – 2.1) | 1.0 |

**Table 23. Log-binomial models for mortality in mixed infections, murine challenge experiments (set I) with *P. chabaudi.***

Notes: (N= 60); entries in bold are significant with p < 0.05.

Within the second set of *P. chabaudi* infections, visual comparison of the Kaplan-Meier curves suggests highly complex behaviour between strains (Figure 16), as shown by the crossing of the survival curves in many of the infection triads. As standard log-rank tests are not applicable when survival curves cross, we have utilized the comparison of restricted mean survival times.

Overall, mixed infections showed no significant difference in time-to-mortality from single infections: any mixed infection vs. all single-strain infections (Peto-Peto test, p = 0.2141; RMST difference = -0.5(SE 0.5), p = 0.356); or in risk of mortality relative to single infections (RR = 1.4; 95% CI: 0.86 to 2.4, p = 0.168).

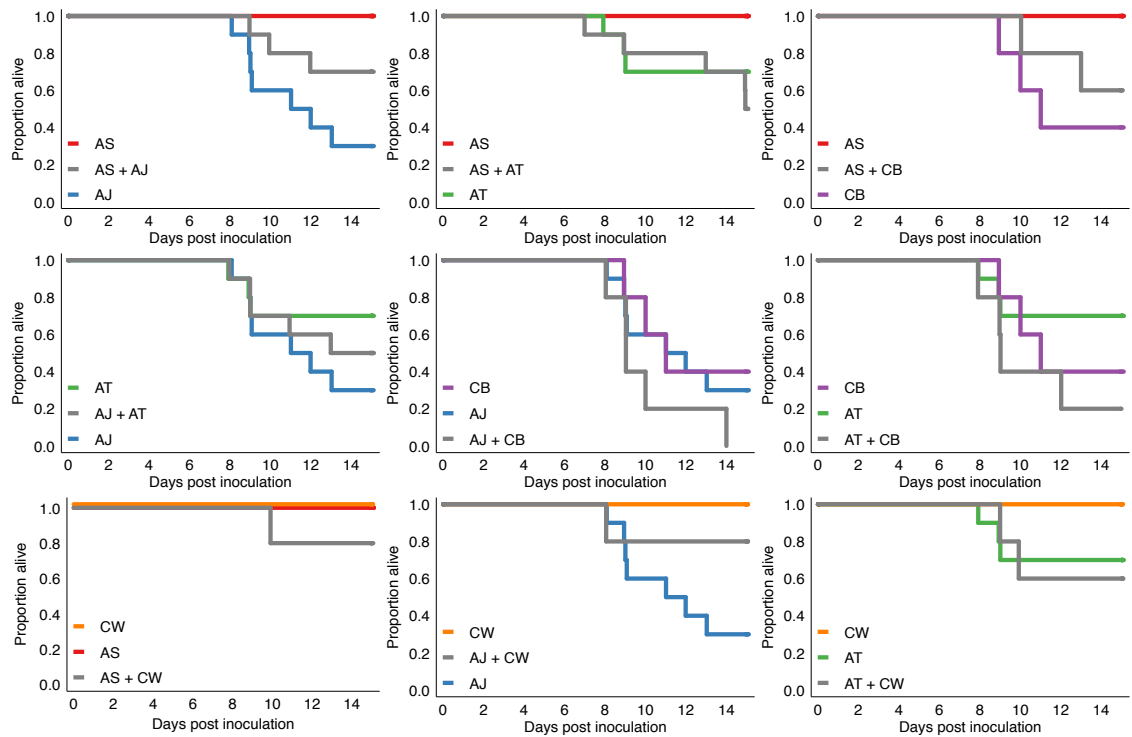

**Figure 16. Kaplan-Meier curves comparing time-to-mortality in single strain and mixed-strain infections in murine challenge experiments with *P. chabaudi.***
Note: N= 100.

Differences in mortality between strains of *P. chabaudi* are shown in Tables 24 and 25. As above, we present each infections triad with different reference groups. Several broad trends are apparent in these results. Specifically, in multivariate binary outcome models, mixed infections are indistinguishable from single infections with a more virulent strain, and are significantly different from single infections with less virulent strains. For example, in triad i. (Table 24), a significant difference exists between single infections with AS (avirulent) and AJ (virulent) strains; and the mixed infection remains significantly different from AS infections (RR = 0.0033; 95% CI 0.000092 to 0.012, p = < 0.001). However, no evidence for a significant difference was found when compared with single-strain AJ infections (RR = 2.3; 95% CI: 0.084 to 6.5, p = 0.105).

While the RMST differences are consistent with these risk ratios, the Peto-Peto log-rank test shows similar trends but does not reach significance (p = 0.068). In summary, we find that mixed-strain infections are indistinguishable from the single-strain infections with that more virulent strain, but are significantly different from single-strain infections with the less virulent strain, in cases where there are significant differences between the two single-strain infections. It should be noted that the extremely small reported RRs are due to zero events in some strain categories.

| Parasite strain triad (more virulent strains in bold) | Comparison | p-value, Peto-Peto test | RMST difference (SE) | p-value RMST difference | Risk ratio (95% CI) | p-value, risk ratio |
|---|---|---|---|---|---|---|
| i. Mixed AS/**AJ** | AS vs. AJ (ref) | **0.0015** | 3.4 (0.84) | **< 0.001** | **0.0015 (0.00054 – 0.0038)** | **< 0.001** |
| | AS vs. AS+AJ (ref) | 0.0679 | 1.4 (0.71) | **0.049** | **0.0033 (0.00092 – 0.012)** | **< 0.001** |
| | AJ vs. AS+AJ (ref) | 0.0792 | -2.0 (1.1) | 0.070 | 2.32 (0.84 – 6.5) | 0.105 |
| ii. Mixed AS/**AT** | AS vs. AT (ref) | 0.0685 | 1.9 (0.93) | **0.041** | **0.0033 (0.00091 – 0.012)** | **< 0.001** |
| | AS vs. AS+AT (ref) | **0.0130** | 1.6 (0.89) | 0.072 | **0.0020 (0.00068 – 0.0059)** | **< 0.001** |
| | AT vs. AS+AT (ref) | 0.5584 | - 0.30 (1.3) | 0.816 | 0.61 (0.19 – 1.88) | 0.381 |
| iii. Mixed AS/**CB** | AS vs. CB (ref) | **0.0062** | **3.0 (1.2)** | **0.012** | **0.0018 (0.00054 – 0.0056)** | **< 0.001** |
| | AS vs. AS+CB (ref) | **0.0350** | 1.4 (0.94) | 0.145 | **0.0026 (0.00063 – 0.011)** | **< 0.001** |
| | CB vs. AS+CB (ref) | 0.3547 | - 1.6 (1.4) | 0.266 | 1.50 (0.41 – 5.5) | 0.539 |
| iv. Mixed **AJ/AT** | AJ vs. AT (ref) | 0.1933 | - 1.5 (1.3) | 0.232 | 2.32 (0.83 – 6.55) | 0.109 |
| | AJ vs. AJ+AT (ref) | 0.4769 | - 0.90 (1.2) | 0.463 | 1.40 (0.67 – 2.9) | 0.371 |
| | AT vs. AJ+AT (ref) | 0.5080 | 0.60 (1.3) | 0.641 | 0.61 (0.19 – 1.9) | 0.379 |
| v. Mixed **AJ/CB** | AJ vs. CB (ref) | n/a | - 0.43 (1.4) | 0.762 | 1.2 (0.52 – 2.86) | 0.652 |
| | AJ vs. AJ+CB (ref) | 0.2534 | 1.6 (1.3) | 0.222 | 0.74 (0.45 – 1.2) | 0.184 |
| | CB vs. AJ+CB (ref) | 0.1293 | 2.0 (1.5) | 0.176 | 0.60 (0.30 – 1.2) | 0.163 |

**Table 24. Comparison of Kaplan-Meier estimator, restricted mean survival times, and risk ratios from binomial models for mortality in mixed infection in murine challenge experiments (set II) with *P. chabaudi*.**
Notes: (RMST at t* = 15 days); entries in bold are significant with p < 0.05); 'na' signifies test is not applicable due to crossing survival curves.

The sole outlier to these trends occurs in triad vii. (Table 25), which compares two avirulent strains: AS and CW. While there is no evidence for a difference between the two single-strain infections, the comparisons of the mixed infections are significant in the binomial response models. Both of the single strain infections had no mortality, while 20% of the mice infected with mixed-strain parasites died.

| Parasite strain triad (more virulent strains in bold) | Comparison | p value, Peto-Peto test | RMST Difference (SE) | p-value RMST difference | Risk ratio (95% CI) | p-value, risk ratio |
|---|---|---|---|---|---|---|
| vi. Mixed **AT/CB** | AT vs. CB (ref) | n/a | 1.1 (1.5) | 0.470 | 0.52 (0.16 – 1.7) | 0.290 |
| | AT vs. AT+CB (ref) | 0.1245 | 2.5 (1.5) | 0.099 | 0.39 (0.14 – 1.1) | 0.084 |
| | CB vs. AT+CB (ref) | 0.3602 | 1.4 (1.6) | 0.388 | 0.75 (0.32 – 1.7) | 0.503 |
| vii. Mixed AS/CW | AS vs. CW (ref) | 0.1573 | 0.33 (0.32) | 0.917 | 0.06 (0.20 – 4.5) | 0.955 |
| | AS vs. AS+CW (ref) | 0.1573 | 1.0 (0.95) | 0.279 | **0.0048 (0.00066 – 0.035)** | **< 0.001** |
| | CW vs. AS+CW (ref) | 0.3173 | 1.0 (0.90) | 0.266 | **0.0050 (0.00058 – 0.043)** | **< 0.001** |
| viii. Mixed **CW/AJ** | CW vs. AJ (ref) | **0.0230** | **3.4 (0.93)** | **< 0.001** | **0.0015 (0.00039 – 0.0057)** | **< 0.001** |
| | CW vs. CW+AJ (ref) | 0.3173 | 1.4 (1.3) | 0.266 | **0.0050 (0.00058- 0.042)** | **< 0.001** |
| | AJ vs. CW+AJ (ref) | na | -2.0 (1.6) | 0.209 | 3.3 (0.54 – 20.7) | 0.196 |
| ix. Mixed **CW/AT** | CW vs. AT (ref) | 0.1952 | 1.9 (1.1) | 0.067 | **0.0035 (0.00070 – 0.017)** | **< 0.001** |
| | CW vs. CW+AT (ref) | 0.1360 | 2.2 (1.2) | 0.071 | **0.0050 (0.00058 – 0.043)** | **< 0.001** |
| | AT vs. CW+AT (ref) | na | 0.33 (1.6) | 0.084 | 1.4 (0.19 – 10.9) | 0.729 |

**Table 25. Comparison of Kaplan-Meier estimator, restricted mean survival times, and risk ratios from binomial models for mortality in mixed infection in murine challenge experiments (set II) with *P. chabaudi* (continued).**
Notes: (RMST at t* = 15 days); entries in bold are significant with p < 0.05; 'na' signifies test is not applicable due to crossing survival curves.

Finally, to complement these two separate analyses of *P. chabaudi* infections, two further comparisons were made. All *P. chabaudi* infections from both studies were compared (N= 160) to assess the potential range in parasite virulence, and the overall impact of mixed-strain parasitemia (Table 26). A limited increase in the risk of mortality from any mixed-strain infections relative to any having any single-strain infection was observed (RR = 1.44; 95% CI 1.04 to 2.01, p = 0.026); as such, any infection with mixed parasite populations had a 44% greater risk of mortality relative to any single-strain infection, across a range of parasite clones with varying virulence.

| Parasite strain | Risk ratio | 95% CI, risk ratio | p-value, risk ratio |
|---|---|---|---|
| CW | 4.5 e-6 | 9.9 e-7 – 0.000020 | **< 0.001** |
| AS | ref. | ref. | ref. |
| AS + CW | 3.51 | 0.41 – 30.32 | 0.253 |
| AJ + CW | 3.51 | 0.41 – 30.32 | 0.253 |
| DS | 4.40 | 1.04 – 18.64 | **0.044** |
| AT | 4.70 | 0.99 – 22.26 | 0.051 |
| AS + DS | 4.94 | 1.21 – 20.13 | **0.026** |
| AJ + AS | 5.27 | 1.10 – 25.25 | **0.038** |
| CB | 6.54 | 1.70 – 25.10 | **0.006** |
| AT + CW | 7.03 | 1.35 – 36.48 | **0.020** |
| AS + CB | 8.00 | 2.22 – 28.81 | **0.001** |
| CB + DS | 8.22 | 2.22 – 30.51 | **0.002** |
| AS + AT | 8.78 | 2.18 – 35.35 | **0.002** |
| AJ + AT | 8.78 | 2.18 - 35.35 | **0.002** |
| AJ | 10.97 | 3.00 – 40.12 | **< 0.001** |
| AT + CB | 14.06 | 3.75 – 52.67 | **< 0.001** |
| AJ + CB | 17.57 | 5.06 – 61.06 | **< 0.001** |

**Table 26. Comparison of risk ratios for mortality by strains in murine challenge experiments with *P. chabaudi*, using robust Poisson regression.**
Notes: (N= 160), models adjusted for parasite dose and experiment number; entries in bold are significant with p < 0.05.

Secondly, a comparison of pooled data of the AS/CB triad from the two different sets of studies is presented in Table 27. This parasite pairing appears in both sets of experiments; as expected, parasite dose was not significant in these models. Within this triad, the risk of mortality in mixed infections was significantly greater relative to having either single-strain infection: RR = 2.5; 95% CI 1.6 to 3.6, p < 0.001. The difference between the single strain AS and CB infections remained highly significant AS vs. CB (ref) (RR = 0.13; 95% CI: 0.034 to 0.51, p= 0.003), as did the AS vs. mixed (ref) (RR= 0.10; 95% CI: 0.027 to 0.38, p= 0.001); while infection with CB parasites vs. mixed infection showed only a marginally significant difference (RR = 0.76; 95% CI: 0.59 to 0.998, p =0.048).

| Parasite strain triad (virulent strains in bold) | Comparison | Risk ratio (95% CI) | p-value, risk ratio |
|---|---|---|---|
| Mixed AS/**CB** | AS vs. CB (ref) | **0.13 (0.034 – 0.51)** | **0.003** |
| | AS vs. AS+CB (ref) | **0.10 (0.027 – 0.38)** | **0.001** |
| | CB vs. AS+CB (ref) | **0.76 (0.59 – 0.998)** | **0.048** |

**Table 27. Comparison of risk ratios from binomial models for mortality in mixed infection with AS and CB strains in murine challenge experiments (sets I and II) with *P. chabaudi*.**
Notes: (N= 57); entries in bold are significant with p < 0.05.

## 5.5 Discussion

*Human experimental studies*

Early published studies include a small set of co-inoculations done in Florida, which were then propagated via mosquitoes (Boyd *et al*. 1938, 1941). These studies appeared to show that co-inoculation led to a delay in parasite clearance, and impaired acquisition of immunity to either strain as assessed by subsequent re-infections; however, no measures of virulence were reported. These experiments did suggest that in mosquito-transmitted propagation of these mixed infections, either each single strain or the mixture could be transmitted; whether this was due to differences in circulating gametocyte populations or vector impacts was unresolved.

The authors of the original study with mixed *P. vivax* infections in human populations analysed in this work concluded that within wide individual variation, the overall pattern of all observed relapses in mixed-strain infections represented a combination of the two strain-specific responses; however, these conclusions were based solely on a qualitative assessment of the overall respective clinical courses (Cooper *et al*. 1950). Our analysis of the incubation period concurs with these observations, and we also find that the incubation period shows significant differences between the mixed and single-strain infections. Consideration of the time-to-first relapse, however, suggests that the mixed strain infections are only marginally different (RMST analysis) or indistinguishable (IRR analysis) from single-strain infections with the Chesson strain.

Although poorly understood, hypnozoite re-activation is likely a different biological process from incubation (Markus 2012), so the results from the relapse analysis may not be directly comparable to other measures of virulence. Qualitative comparison of the survival curves in Figure 13 suggests that the earlier part of the mixed-infection curve more closely follows, but is not equivalent to, that of Chesson strain infections. However, our analysis is unable to differentiate between two potential mechanisms: coordinated relapse between genetically diverse hypnozoites, or simply the first result of having two superimposed hypnozoite relapse curves.

Consensus about clonal diversity within relapses has been elusive (Imwong *et al*. 2007, 2012), and conclusions may be highly dependent on the specific probes used to compare parasite-relatedness (Restrepo *et al*. 2011). Some of these observed differences are also potentially due to primaquine treatment (de Araujo *et al*. 2012), as different strains may have differing drug susceptibility (Goller *et al*. 2007). However, it appears clear that some parasite sub-populations that are more prone to relapse, and that there can be extensive diversity of parasites within relapses (Lin *et al*. 2012). Critically, relapse epidemiology has been highlighted as a key gap in control of this parasite (Ferreira *et al*. 2007, Havryliuk and Ferreira 2009), and could present a critical barrier for the global elimination of all malaria species (Arnott *et al*. 2012).

The concordance of these epidemiological results with molecular studies provides evidence for a policy to prioritize surveillance and primaquine treatment specifically directed towards foci of long-relapse genotypes, thereby minimizing their impact on global *P. vivax* population structures towards global malaria elimination.

The late WE Collins (US-CDC) indicated that he was not aware of any unpublished studies with simultaneous mixed inoculations in primate models at the CDC malaria research labs; (Collins 2013b) the data in this current analysis therefore likely represent the sum of all that have been collected in both humans and non-human primates.

*Murine models*

In discussion of the mixed *P. yoelii* infections, the original study authors note that while the mild strain preferentially infected only reticulocytes even in fatal outcomes, and the virulent form 'overwhelmingly' infected mature erythrocytes, mixed infections initially showed parasitemias in both RBC stages, which then progressed in the four surviving animals to be predominately composed of infected reticulocytes (Hargreaves *et al*. 1975).

A range of modern studies with rodent malaria models has produced ambiguous results. Some studies have suggested intense competition between distinct clones so that potentially the more virulent clone in mixed-infections can fully dominates the clinical course; others have shown attenuation of more-virulent strains in competitive scenarios. Extensive modelling studies have suggested that general mixed infections may show two differing regimes: one in acute infections, and another suite of interactions in chronic infections as the host's immune systems begins to generate strain-specific response factors (Alizon and van Baalen 2008).

Bell and co-authors concluded that mixed infections with *P. chabaudi* have greater virulence than single clone infections, as measured by weight loss and red blood cell count (anaemia level) (Taylor *et al*. 1998). Our analyses of these data does not show a significant difference in mortality; however, a pooled analysis of both *P. chabaudi* experiments confirms the original authors' results using mortality as a direct outcome, and suggest a ~40% increased risk of mortality in mixed infections across a range of parasite clones.

However, our results suggest that within these parasite lineages, RBC and weight loss metrics may only capture a portion of the complexity in the measurement of 'virulence'; within parasite triads, we find mortality in mixed-strain infections is not increased, but *equivalent* to mortality in infections with the more virulent strain in *P. chabaudi* infections. This suggests that use of survival models allows more comprehensive population-level effects to be assessed across the entire time course of infections, and avoids the need to choose potentially biased time points for comparisons (Hosmer *et al*. 2011).

Our results reconcile these *P. chabaudi* experiments with mathematical models which have suggested that host mortality is strongly driven by the virulence of the more virulent strain within an infection (Alizon and van Baalen 2008), but suggests a fundamental limit may exist in virulence. Comparison of the risk ratios between the CB/AS triad (RR = 2.5) and all mixed/pure infections (RR = 1.4) among all studied strains, supports the findings of Alizon which suggest that the antigenic distance between strains has critical impacts (Alizon and van Baalen 2008). These mathematical models suggest that there may be a 'sweet-spot' where maximum competition occurs- if strains are highly antigenically divergent, they can parasitize essentially independent niches within the host with subsequent limited need for competition; whereas highly related parasites are too similar to be differentiated by general immune responses. It should be noted that the parasites in the work of Bell *et al*. are described as '…genetically (and antigenically) district at multiple loci' (Bell *et al*. 2006). While the phylogenetics of *P. chabaudi* has been examined, these laboratory isolates were specifically not included (Ramiro *et al*. 2012), so the antigenic distances are unknown.

The remarkable consistency of results with the CB/AS strains between the two different experiment sets should be noted, especially considering limited source data, a 10-year lag between experiments with associated parasite passages, a 100-fold difference in inocula, and Bell *et al*.'s report of a mutation in the *msp-1* gene (a key mediator of erythrocyte invasion) within the CB strain (Bell *et al*. 2006). This replication, even within inherent limitations of data reporting, greatly strengthens these results (Palmer 2000). It is also intriguing that the AS strain was originally created by forced selection for pyrimethamine resistance; the causal impact of this drug-resistance selection on the observed virulence however is unknown.

*Survivorship curves for semi-quantitative categorization of strains*

In the ecological and demographic literature, plots of survival within populations by age, called survivorship curves, allow populations to be categorized into three idealized life courses (referred to as type I, II and III curves) (Pianka 2011); see Figure 17. Type III curves are categorized by a large number of mortality events early in the time course; type II by a relatively constant rate of events; and type I by generally low 'failure' rate until the oldest age classes.

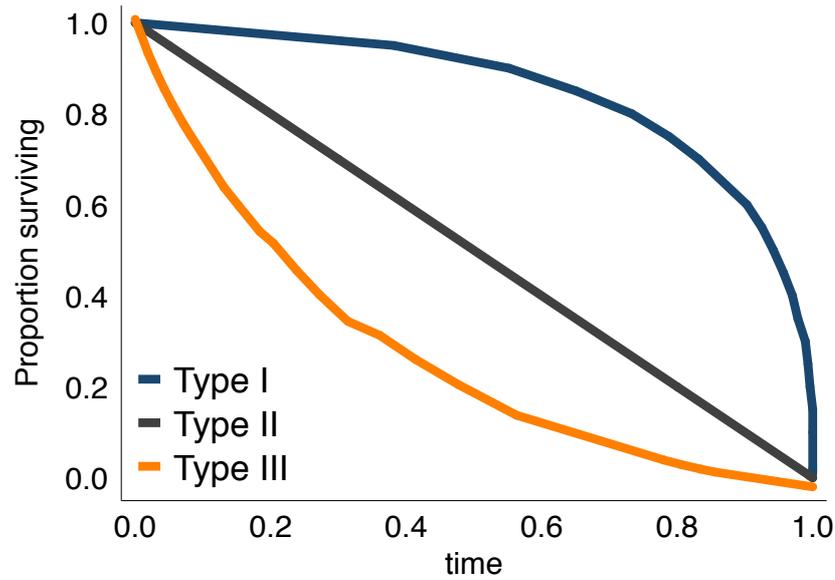

**Figure 17. Idealized ecological/demographic survivorship curves.**

Examination of the survival curves within this study (especially *P. yoelii*) and Figure 13 (with allowances for the crossing curves) closely mimic these classes of survivorship curves, although the underlying mechanisms are unrelated. We suggest that future studies should consider this general schema for classification of virulence for two reasons. First, it allows examination for population-level effects and 'smooths out' individual-level variation in responses. Secondly, it inherently captures the complex interplay between host and parasites to more comprehensively capture the composite endpoint of 'virulence'. Finally, several different metrics have been suggested to allow quantitative comparisons between the convexity of survivor curves, and have potentially utility in directly classifying the virulence of parasite populations (Anson 2002).

*Contrasts*

We find that amongst all the limited set of *P. vivax* (incubation) and *P. yoelii* infections, the survival curves at a population-level are intermediate between the clones, and are not dominated by the more virulent infection. However, the survival curves in mixed infections are more similar to the more virulent clones as assessed by log-rank values, and risk/hazard incidence ratios. This strongly suggests that highly virulent clones have their virulence attenuated by the presence of less-virulent clones. In line with ample evidence from models and evolutionary theory this is likely due to the competition for resources within the host bloodstream. However, these comparisons have been made with clones from the extremes of what is likely a wide continuum of virulence.

Our results from analysis of the much larger set of experiments using *P. chabaudi* suggests greater nuances in interactions between strains, and we find that mortality and time-to-event in mixed-strain infections are both indistinguishable from those of virulent-strain single infections. This is in sharp contrast to the results from *P. yoelii* experiments, with similar triad-wise sample sizes. A potential explanation for these differences comes from a range of models that have suggested different malaria species may exhibit differing specificity age-classes of red blood cells (McQueen and McKenzie 2004, 2006, Mideo *et al*. 2008). The blood stages parasites within the current study display differential blood cell preferences: *P. vivax* and *P. yoelii* generally infect the youngest RBC stages (reticulocytes) (McKenzie *et al*. 2002), whereas *P. chabaudi* (like *P. falciparum*) has limited specificity, and readily infects both reticulocytes and normocytes (Snounou *et al*. 1992).

These divergent dynamics suggest the possibility that virulence may have different underlying mechanisms in *P. vivax*-like and *P. falciparum*-like parasites.

This hypothesis is consistent with data suggesting that immunity and RBC dynamics, which differ greatly between the parasite 'types,' are both important contributors to competitive ability and parasite dynamics. This suggests that maximal competitive advantage may hit a fundamental limit, driven primarily by limited numbers of susceptible RBCs in *P. vivax*-type infections, so virulent strains can only out-compete to a certain and limited extent. In contrast, in *P. falciparum*-type infections, virulence in mixed infections is far less constrained and rapidly approaches that of 'pure' infections with virulent strains. The consistency of these effects across a range of parasite virulence in *P. chabaudi* isolates suggests these findings are robust.

However, other experiments and models of *P. chabaudi* reach differing conclusions and suggest the evolutionary trajectory of higher-virulence strains within a mixed strain infection are most impacted by both burst size (number of progeny per merozoite) in addition to differential interactions with host immunity (Mideo *et al*. 2011). The importance of burst size in human infections is circumstantially supported by an early study with the specific *P. vivax* strains included in this analysis: amongst 200 infected red blood cells examined from diverse infections for each strain, the St. Elizabeth strain had a median of 14 (95% CI: 14 to 14; range 9-20) merozoites; and the more virulent Chesson strain had a median of 16 merozoites (95% CI: 16 to 18; range 10 to 24); Mann-Whitney test for differences, p < 0.001; authors' calculations from data in (Wilcox *et al*. 1954). Similar values have been reported for other *P. vivax* strains with similar differences in virulence: a milder Dutch strain produced on average 12-13 merozoites, whereas the virulent Madagascar strain produced on average 17-18 merozoites per cycle (Swellengrebel and De Buck 1938).

These inconsistencies have been explicitly addressed in the elegant models of *P. vivax* blood-stage dynamics of McQueen, which suggest that RBC-stage specificities are of only limited importance, and that host immunity must be a critical component in keeping *P. vivax* from being as virulent/lethal as *P. falciparum* (McQueen 2010), although this paradigm is eroding (Baird 2013). In spite of the analytical limitations of these data, our results suggest that virulence proxies in murine malaria may not be fully capturing all types of virulence, and a broader selection of endpoints should be considered.

*Limitations*

There are several limitations of this analysis. Firstly the *P. vivax* and *P. yoelii* studies are based on analysis of secondary data with limited experimental detail in the original reports. Additionally, all murine experiments used blood-transfer via syringe and not transmission via infected vectors, which has the potential to impact parasite dynamics (Paul *et al*. 2004, Nkhoma *et al*. 2012). Moreover, different species of *Plasmodium* have also evolved independent pathways to counter the same stressor; for example, chloroquine resistance in *P. falciparum* and *P. vivax*, and parasite immune evasion strategies, both likely involve independent pathways in different species (Carlton *et al*. 2002, Lehane *et al*. 2012).

*Plasmodium vivax* and the murine studies were limited to reporting of events as integer days, while mortality events may have occurred at intermediate points; however randomization of reported times within 0.2 day did not impact the conclusions (data not shown). Additionally, the endpoint of the *P. chabaudi* studies was either mortality or euthanasia if extreme morbidity was observed; it is therefore possible that assessment of morbidity could have been variable between the experiments. Additionally, other authors have suggested that some *P. yoelii*-infected animals are able to tolerate very high parasitemia before succumbing several days later (Knowles and Walliker 1980).

Recent studies have also suggested that parasite passage through anopheline vectors can modulate virulence (Spence *et al*. 2013), and competitive advantage within a host also increases a parasite clone's ability to infect vectors and hence population dispersal in rodent models (de Roode *et al*. 2005). Therefore syringe-induced infections could represent extremes in virulence that may not reflect natural systems (Mideo *et al*. 2011). Lastly, many of the comparisons in the murine models were made using a small number of animals per treatment group, and not finding an effect in some groups is potentially due to sample size limitations.

## 5.6 Conclusions and public health impacts

These results tie together human and murine experimental studies, and also suggest that different dynamics could exist between malaria species dependent on blood-stage cell specificities. These 'slippages' suggest research areas that could provide important insights into intra-host malaria dynamics, and could be prioritized in planning new murine challenge studies. Our results suggest at least some of these conflicting results could be partially addressed by consideration of a more diverse set of endpoints beyond comparisons of mean anaemia or parasitemia between groups (Nakagawa and Cuthill 2007). While hampered by crossing survival curves and zero-event categories, which may be unavoidable in some systems, our analysis suggests possible fuller use of data could maximize the utility of these very complex and experimentally demanding experiments.

While the inherent limitations of some of these model systems have been recognized (White *et al*. 2010), they provide invaluable insight into parasite evolutionary dynamics within a host, and the main conclusion from a comprehensive review is that there should be closer linkages between human studies and animal models, to allow the development of truly synergistic research programs (Langhorne *et al*. 2011). While virulence management has been highlighted as a potential key contribution from evolutionary biology especially in malaria, it has been suggested this paradigm has had limited impact on clinical practice (Restif 2009). Our results highlight the critical interplay between these fields, and reinforce recent calls for greater evolutionary thinking in malaria control, drug policy, and 'containment' of resistance (Read *et al*. 2011). Additionally, it has been proposed that reductions in population-level prevalence of multiple-strain infections could decrease inherent pressures on the evolution of virulence and drug resistance, and this suggests that greater emphasis should be placed on surveillance of intra-infection diversity (Havryliuk and Ferreira 2009).

Finally, these results hint that evolutionary drivers of strain competition and virulence could differ between *P*. *vivax* and *P*. *falciparum*. This suggests several broad avenues for future studies in animal models. A head-to-head comparison in the same laboratory between reticulocyte-specific and red blood cell generalist parasites encompassing similar differences in virulence could provide greater evidence. These results could be critical importance for the planning of species-specific policies for drug-resistance mitigation, and towards global malaria elimination for all malaria species.

# Chapter 6: Note on the origin of the Madagascar strain of

# *Plasmodium vivax*

## 6.1 Introduction

Early malariotherapy studies suggested that many persons of African descent had inherent resistance to infections with *Plasmodium vivax*; the biological basis of this apparent immunity was later correlated with the 'Duffy antigen' (Zimmerman *et al*. 2013). For several decades, it was assumed that the widespread lack of this erythrocyte surface protein was an important barrier to *P. vivax* infections throughout sub-Saharan Africa. However, recent studies have established the potential for strains of *Plasmodium vivax* in Madagascar to infect Duffy-negative populations (Ménard *et al*. 2010), and sensitive PCR-based screening suggests a previously unrecognized burden of this parasite in multiple locations in sub-Saharan Africa (Fru-Cho *et al*. 2014, Ngassa Mbenda and Das 2014).

## 6.2 Letter

The identification of these *P. vivax* infections in Africa has highlighted a further major gap in vivax epidemiology: there is a dearth of data concerning any *P. vivax* parasites from Africa. However, one of the most widely studied strains used during the malariotherapy-era was the Madagascar strain, used at the site at Horton Hospital, UK from 1923- ca. 1970. This isolate has seen extensive study and re-analysis (James *et al*. 1936, Glynn and Bradley 1995), but there has been considerable debate about the origin of this strain, and two recent papers specifically suggest the origin may well be elsewhere (Baird 2013, Battle *et al*. 2014).

However, further detail is provided in the work of Percy G. Shute, the malariologist who personally interviewed and treated the index patient for this strain. In his comprehensive report (which is not readily available) related to this clinical isolate, he states:

> …the technique [of malariotherapy] was employed in England, with a variety of strains obtained by one of us (PGS) from sailors and others suffering from malaria and entering the Port of London. Eventually a ship arrived from Madagascar with several fever patients one board; of them - a Lascar (Indian) seaman - had a high fever and exhibited many *P. vivax* parasites in his blood. His blood was inoculated into a patient in a mental hospital in London; fever and parasites appeared 10 days later and gametocytes were detected some days afterwards. Mosquitoes *(Anopheles atroparvus)* were fed on him and they became infected. The infection was then passed to other patients and thus the « Madagascar strain » was established (…).

> The precise locality in Madagascar where the Indian seaman contracted the infection is impossible to determine. Ten days after his ship had left Madagascar, on the passage to England, he developed fever, but malaria was not diagnosed until he arrived at Tilbury (Port of London). He was then found to be suffering from a quotidian fever with temperatures rising to 40°- 41° C. He gave no previous history of malaria and this attack was unlikely to have been a relapse for the reason that his paroxysms occured [sic] daily, as in a primary attack; whereas the fever in relapses is tertian from the start. For these reasons, it is most probable that the designation « Madagascar strain » is geographically correct. (Shute *et al*. 1978).

These differences in fever periodicity between primary infections and relapses are strongly supported by other researchers and clinician in a range of settings e.g. (Noe *et al*. 1946, Covell and Nicol 1951, Adak *et al*. 2001). Moreover, meticulous study by James using the Madagascar strain itself reinforced Shute's observations (James *et al*. 1936). This clinical behaviour is likely due likely due to the slow establishment of well-synchronized parasite subpopulations in primary infections (Warrell and Gilles 2002), and the emergence of clonal hypnozoites in relapses (Chen *et al*. 2007) respectively.

## 6.3 Conclusions

These observations strongly suggest that the wealth of irreplaceable clinical and epidemiological data collected during malariotherapy treatments using the Madagascar strain have important and critical utility, and should be combined with modern phylogenetic and molecular studies (Li *et al*. 2001, Neafsey *et al*. 2012) to inform a research agenda into the epidemiology of African *P. vivax*. Finally, while the original geographic origin of *P. vivax* remains of considerable debate, recent studies suggest an African origin (Liu *et al*. 2014); and there have been multiple suggestions that there may be an unrecognized reservoir in Africa e.g. (Rosenberg 2007). A fuller and comprehensive understanding of these parasite populations, along their human and non-human primate reservoirs, will be a crucial next-step for global malaria elimination programs.

# Chapter 7: Conclusions

In 1985, the doyen of the 'golden era' of malariology noted:

> The old generation of 'blood, mud and sweat' malariologists has nearly disappeared and much of their experience has been largely forgotten. The author of the present book hopes that he has contributed to forging a link between the not too distant past and the new generation of specialists in tropical community health. (Bruce-Chwatt 1985)

## 7.1   Introduction

My initial reaction to reading about malariotherapy was one of amazement and incredulity- using one potentially fatal infectious disease to treat another? Did it really work? How did anyone know? The more I delved into the subject, the larger the breadth of the historical literature became, with hundreds of papers referencing hundreds of still-earlier works. It became readily apparent that many of these reports had had limited reconsideration in subsequent decades, with the conclusions simply cited as 'truth'. This led to another series of questions- how robust were these results and subsequent conclusions? Was it even possible to know at this remove, and was there analysable data that might add anything within a modern framework? Indeed, huge volumes of original working papers, colonial health board reports, plus WHO and League of Nations report series were readily available online and in the NUS archives.

While there has been some systems-level thinking in malaria control since Ronald Ross's earliest models, the contentious debates of the early 20th century regarding whether the proper focus should be on the parasite or upon the vector, highlight the serious divisions that have occurred within malaria control (Stapleton 2009). The burgeoning intersections between epidemiology, entomology, ecology, and parasitology have in recent years highlighted these false dichotomies, towards an understanding that the entire dynamic and continually shifting 'web of transmission' itself has to be unstitched. To this end, coordinated and synergistic combinations of all appropriate tools, deployed effectively, will be crucial for elimination. This thesis aimed to forge a small link in this chain of knowledge, and to highlight a few areas worthy of further exploration.

## 7.2    Summary of findings and implications for future work

Chapter 2 of this thesis has explored the measurable epidemiological impacts of geographic origin of *Plasmodium vivax* strains, and largely confirms results from earlier qualitative assessments of these effects, but finds subtleties in geographic epidemiology that been lost due to data aggregations. However, this analysis was limited by the historical and descriptive nature of these parasite strains; further work could try to connect these results to modern molecular and genetic studies, if samples from any of these parasites populations can be located and extracted, as recently suggested (Collins 2013a). Within the limitations of subsequent parasite evolution, this could allow a direct linkage between observed epidemiological measures and parasite data, to prioritize surveillance activities towards malaria elimination.

Chapter 3 of this work examines the evidence for sporozoite-dependent effects on incubation periods in *P. vivax* infections by re-analysis of historical challenge studies. In direct contrast to accumulated knowledge, we find extremely limited evidence for effects at any plausible naturally occurring exposures. In concurrence with other recent studies, these results suggest that parasite strain itself and parasite-vector interactions have larger quantitative effects than sporozoite dose. These findings highlight the complexity of the *Plasmodium-Anopheles* interface, and suggest that greater consideration of these interactions could be critical to ensure that human challenge studies for vaccine and malaria prophylaxis development programs have maximal congruence with natural transmission.

In Chapter 4, well-defined and readily implementable distributions that capture event times in experimental *P. vivax* infections are presented. These results suggest that the default parameterizations in many mathematical models of *P. vivax* transmission may incompletely capture the diversity within natural parasite populations, and that 'better' distributional choices can have large impacts on model outputs. Finally these results provide some evidence that long-latent infections have a closer tie to general incubation periods than they do to relapse. Future modelling studies should consider use of these optimal distributions, and the impact of these distributional choices should be explored within comprehensive mathematical and statistical models.

Chapter 5 of this thesis explores the impacts of mixed-strain infections in both human and murine experimental challenges. The results from *P.* vivax concur with some animal models and ecological theory, but also suggest that there are potentially large differences in virulence evolution between parasites that target normocytes and those that target reticulocytes. These differences could be explored in greater detail via head-to-head comparison of diverse murine malaria species, which could potentially provide an evidence base for parasite-specific virulence control measures.

Finally, Chapter 6 clarifies the most probable origin of the canonical Madagascar strain of *P. vivax*. This question has taken on renewed importance with the recent realization that this parasite may have far greater prevalence in Sub-Saharan Africa than previously believed, including a non-human primate reservoir (Liu *et al*. 2014). Linking the extensive historical data acquired using the Madagascar strain to modern

African *P. vivax* could have important implications for the future 'end-game' of malaria elimination in Africa.

Future research programs must confront the enigmatic biology of sporozoites in vectors and humans; and of gametocytes and hypnozoites in humans. Specifically, while extensive work is currently on-going to explore the fate of injected sporozoites in murine model systems, the inability to propagate *P. vivax* in laboratory settings will limit the study of vivax-specific sporozoite biology. The results in this thesis suggest important vector species-specific impacts on sporozoites that should be more fully explored, especially using non-human primate models (Joyner *et al*. 2015).

The timing and number of gametocytes produced in malaria infections are critically important factors in determining transmission success; while this life stage has had extensive study in *P. falciparum* infections, knowledge within *P. vivax* infections is far more limited. Moreover, production of gametocytes before or rapidly after the onset of clinical symptoms provides ample opportunity for continued transmission. While malariotherapy studies did not generally report gametocytemia data, studies with the Madagascar strain did find increases in gametocyte production in this strain after long-term mosquito-to-human propagation (Glynn and Bradley 1995); the evolutionary drivers and impact on transmission of these changes, and any geographic variation in gametocyte production amongst *P. vivax* strains remain important topics worthy of further study.

Finally, hypnozoite biology must progress beyond the current 'black-box.' The triggers for hypnozoite activation need to be identified; and population-level timespans necessary for surveillance in specific elimination settings, and for follow-up within clinical efficacy studies need to be carefully delineated, potentially as recently implemented in Mexico (Gonzalez-Ceron *et al*. 2013). Finally, recent modelling studies have found that either 3 or 6 classes of hypnozoites most closely explained observed epidemiology in India; however, the potential biological underpinnings of these results remain obscure, and suggest important areas for linking regional epidemiology with parasite biology (Roy *et al*. 2013).

## 7.3    Final statement

The chapters of this thesis provide a useful set of epidemiological studies to expand what is currently known about *Plasmodium vivax* with the potential to assist in developing an evidence-based research agenda for control of this parasite, and towards global elimination of all malaria species.

# Works cited

Adak, T., Valecha, N., and Sharma, V.P., 2001. *Plasmodium vivax* polymorphism in a clinical drug trial. *Clinical and Diagnostic Laboratory Immunology*, 8 (5), 891–894.

Águas, R., Ferreira, M.U., and Gomes, M.G.M., 2012. Modeling the effects of relapse in the transmission dynamics of malaria parasites. *Journal of Parasitology Research*, 2012, 1–8.

Alizon, S. and van Baalen, M., 2008. Multiple infections, immune dynamics, and the evolution of virulence. *The American Naturalist*, 172 (4), e150–e168.

Alizon, S., Hurford, A., Mideo, N., and van Baalen, M., 2009. Virulence evolution and the trade-off hypothesis: history, current state of affairs and the future. *Journal of Evolutionary Biology*, 22 (2), 245–259.

Andersen, P.K. and Perme, M.P., 2010. Pseudo-observations in survival analysis. *Statistical Methods in Medical Research*, 19 (1), 71–99.

Animut, A., Balkew, M., Gebre-Michael, T., and Lindtjørn, B., 2013. Blood meal sources and entomological inoculation rates of anophelines along a highland altitudinal transect in south-central Ethiopia. *Malaria Journal*, 12 (1), 76.

Anson, J., 2002. Of entropies and inequalities: Summary measures of the age distribution of mortality. *In*: J. Duchene, G. Wunsch, and M. Mouchart, eds. *The Life Table: Modelling Survival and Death*. Dordrecht: Springer.

Anstey, N.M., Douglas, N.M., Poespoprodjo, J.R., and Price, R.N., 2012. *Plasmodium vivax*: clinical spectrum, risk factors and pathogenesis. *In*: S.I. Hay, R.N. Price, and J.K. Baird, eds. *Advances in Parasitology*. Academic Press, 151–201.

De Araujo, F.C.F., de Rezende, A.M., Fontes, C.J.F., Carvalho, L.H., and Alves de Brito, C.F., 2012. Multiple-clone activation of hypnozoites is the leading cause of relapse in *Plasmodium vivax* infection. *PLoS ONE*, 7 (11), e49871.

Armengaud, A., Legros, F., Quatresous, I., Barre, H., Valayer, P., Fanton, Y., D'Ortenzio, E., and Schaffner, F., 2006. A case of autochthonous *Plasmodium vivax* malaria, Corsica, August 2006. *Euro Surveillance- European Communicable Disease Bulletin*, 11 (11), E061116.3.

Arnott, A., Barry, A.E., and Reeder, J.C., 2012. Understanding the population genetics of *Plasmodium vivax* is essential for malaria control and elimination. *Malaria Journal*, 10 (11).

Baird, J.K., 2007. Neglect of *Plasmodium vivax* malaria. *Trends in Parasitology*, 23 (11), 533–539.

Baird, J.K., 2013. Evidence and implications of mortality associated with acute *Plasmodium vivax* malaria. *Clinical Microbiology Reviews*, 26 (1), 36–57.

Baird, J.K., Schwartz, E., and Hoffman, S.L., 2007. Prevention and treatment of vivax malaria. *Current Infectious Disease Reports*, 9 (1), 39–46.

Balmer, O. and Tanner, M., 2011. Prevalence and implications of multiple-strain infections. *The Lancet Infectious Diseases*, 11 (11), 868–878.

Battle, K.E., Gething, P.W., Elyazar, I.R., Moyes, C.L., Sinka, M.E., Howes, R.E., Guerra, C.A., Price, R.N., Baird, J.K., and Hay, S.I., 2012. The global public health significance of *Plasmodium vivax*. *In*: S.I. Hay, R.N. Price, and J.K. Baird, eds. *Advances in Parasitology*. 1–111.

Battle, K.E., Karhunen, M.S., Bhatt, S., Gething, P.W., Howes, R.E., Golding, N., Boeckel, T.P.V., Messina, J.P., Shanks, G.D., Smith, D.L., Baird, J.K., and Hay, S.I., 2014. Geographical variation in *Plasmodium vivax* relapse. *Malaria Journal*, 13 (1), 144.

Bell, A.S., de Roode, J.C., Sim, D.G., and Read, A.F., 2014. Data from: Within-host competition in genetically diverse malaria infections: parasite virulence and competitive success. (Dryad Digital Repository). http://dx.doi.org/10.5061/dryad.bv188.

Bell, A.S., De. Roode, J.C., Sim, D., and Read, A.F., 2006. Within-host competition in genetically diverse malaria infections: parasite virulence and competitive success. *Evolution*, 60 (7), 1358–1371.

Bitoh, T., Fueda, K., Ohmae, H., Watanabe, M., and Ishikawa, H., 2011. Risk analysis of the re-emergence of *Plasmodium vivax* malaria in Japan using a stochastic transmission model. *Environmental Health and Preventive Medicine*, 16 (3), 171–177.

Bordes, F. and Morand, S., 2009. Parasite diversity: an overlooked metric of parasite pressures? *Oikos*, 118 (6), 801–806.

Bousema, T. and Drakeley, C., 2011. Epidemiology and infectivity of *Plasmodium falciparum* and *Plasmodium vivax* gametocytes in relation to malaria control and elimination. *Clinical Microbiology Reviews*, 24 (2), 377–410.

Boyd, M.F., 1940. The influence of sporozoite dosage in vivax malaria. *American Journal of Tropical Medicine and Hygiene*, s1-20 (2), 279–286.

Boyd, M.F. and Kitchen, S.F., 1937. A consideration of the duration of the intrinsic incubation period in vivax malaria in relation to certain factors affecting the parasites. *American Journal of Tropical Medicine and Hygiene*, 1 (3), 437–444.

Boyd, M.F., Kitchen, S.F., and Matthews, C.B., 1941. On the natural transmission of infection from patients concurrently infected with two strains of *Plasmodium vivax*. *American Journal of Tropical Medicine*, s1-21 (5), 645–652.

Boyd, M.F., Kupper, W.H., and Matthews, C.B., 1938. A deficient homologous immunity following simultaneous inoculation with two strains of *Plasmodium vivax*. *American Journal of Tropical Medicine*, s1-18 (5), 521–524.

Brachman, P., 1998. Reemergence of *Plasmodium vivax* malaria in the Republic of Korea. *Emerging Infectious Diseases*, 4 (4), 707–707.

Brasil, P., Costa, A. de P., Pedro, R., Bressan, C. da S., Silva, S. da, Tauil, P., and Daniel-Ribeiro, C., 2011. Unexpectedly long incubation period of *Plasmodium vivax* malaria, in the absence of chemoprophylaxis, in patients diagnosed outside the transmission area in Brazil. *Malaria Journal*, 10 (1), 122.

Bruce-Chwatt, L.J., 1965. Malaria research for malaria eradication. *Transactions of the Royal Society of Tropical Medicine and Hygiene*, 59, 105–144.

Bruce-Chwatt, L.J., 1984. Terminology of relapsing malaria: Enigma variations. *Transactions of the Royal Society of Tropical Medicine and Hygiene*, 78 (6), 844–845.

Bruce-Chwatt, L.J., 1985. *Essential Malariology*. 2nd edition. New York, NY: John Wiley & Sons.

Bruce, M.C., Galinski, M.R., Barnwell, J.W., Donnelly, C.A., Walmsley, M., Alpers, M.P., Walliker, D., and Day, K.P., 2000. Genetic diversity and dynamics of *Plasmodium falciparum* and *P. vivax* populations in multiply infected children with asymptomatic malaria infections in Papua New Guinea. *Parasitology*, 121 (03), 257–272.

Brunetti, R., Fritz, R.F., and Hollister Jr, A.C., 1954. An outbreak of malaria in California, 1952–1953. *American Journal of Tropical Medicine and Hygiene*, 3 (5), 779–788.

Carlton, J.M., Angiuoli, S.V., Suh, B.B., Kooij, T.W., Pertea, M., Silva, J.C., Ermolaeva, M.D., Allen, J.E., Selengut, J.D., Koo, H.L., Peterson, J.D., Pop, M., Kosack, D.S., Shumway, M.F., Bidwell, S.L., Shallom, S.J., van Aken, S.E., Riedmuller, S.B., Feldblyum, T.V., Cho, J.K., Quackenbush, J., Sedegah, M., Shoaibi, A., Cummings, L.M., Florens, L., Yates, J.R., Raine, J.D., Sinden, R.E., Harris, M.A., Cunningham, D.A., Preiser, P.R., Bergman, L.W., Vaidya, A.B., van Lin, L.H., Janse, C.J., Waters, A.P., Smith, H.O., White, O.R., Salzberg, S.L., Venter, J.C., Fraser, C.M., Hoffman, S.L., Gardner, M.J., and Carucci, D.J., 2002. Genome sequence and comparative analysis of the model rodent malaria parasite *Plasmodium yoelii yoelii*. *Nature*, 419 (6906), 512–519.

Casadevall, A. and Pirofski, L., 2001. Host-pathogen interactions: the attributes of virulence. *Journal of Infectious Diseases*, 184 (3), 337–344.

Chamchod, F. and Beier, J.C., 2013. Modeling *Plasmodium vivax*: relapses, treatment, seasonality, and G6PD deficiency. *Journal of Theoretical Biology*, 316, 25–34.

Chen, N., Auliff, A., Rieckmann, K., Gatton, M., and Cheng, Q., 2007. Relapses of *Plasmodium vivax* infection result from clonal hypnozoites activated at predetermined intervals. *Journal of Infectious Diseases*, 195 (7), 934–941.

Chen, W., Shi, J., Qian, L., and Azen, S.P., 2014. Comparison of robustness to outliers between robust Poisson models and log-binomial models when estimating relative risks for common binary outcomes: a simulation study. *BMC Medical Research Methodology*, 14 (1), 82.

Chernin, E., 1984. The malariatherapy of neurosyphilis. *Journal of Parasitology*, 70 (5), 611–617.

Coatney, G.R., Collins, W.E., and Contacos, P.G., 1971. *The primate malarias*. Washington, DC: US Government Printing Office.

Coatney, G.R., Cooper, W.C., and Ruhe, D.S., 1948. Studies in human malaria VI. The organization of a program for testing potential antimalarial drugs in prisoner volunteers. *American Journal of Hygiene*, 47 (1), 113–119.

Coatney, G.R., Cooper, W.C., Ruhe, D.S., Young, M.D., and Burgess, R.W., 1950. Studies in human malaria XVIII. The life pattern of sporozoite-induced St. Elizabeth strain vivax malaria. *American Journal of Hygiene*, 51 (2), 200–215.

Coatney, G.R., Cooper, W.C., and Young, M.D., 1950. Studies in human malaria XXX. A summary of 204 sporozoite-induced infections with the Chesson strain of *Plasmodium vivax*. *Journal of the National Malaria Society*, 9 (4), 381–96.

Collins, W., 2013a. Origin of the St. Elizabeth strain of *Plasmodium vivax*. *American Journal of Tropical Medicine and Hygiene*, 88 (4), 726–726.

Collins, W.E., 2002. Nonhuman primate models: II. Infection of *Saimiri* and *Aotus* monkeys with *Plasmodium vivax*. *In*: *Malaria Methods and Protocols*. New Jersey: Humana Press, 85–92.

Collins, W.E., 2007. Further understanding the nature of relapse of *Plasmodium vivax* infection. *Journal of Infectious Diseases*, 195 (7), 919–920.

Collins, W.E., 2013b. Personal communication: *P. vivax* and/or *P. cynomolgi* primate studies?. 4 July 2013.

Collins, W.E., Morris, C.L., Richardson, B.B., Sullivan, J.S., and Galland, G.G., 1994. Further studies on the sporozoite transmission of the Salvador I strain of *Plasmodium vivax*. *Journal of Parasitology*, 80 (4), 512–517.

Collins, W.E., Skinner, J.C., Pappaioanou, M., Broderson, J.R., Filipski, V.K., McClure, H.M., Strobert, E., Sutton, B.B., Stanfill, P.S., and Huong, A.Y., 1988. Sporozoite-induced infections of the Salvador I strain of *Plasmodium vivax* in *Saimiri sciureus boliviensis* monkeys. *Journal of Parasitology*, 74 (4), 582–585.

Collins, W.E., Sullivan, J.A.S., Morris, C.L., Galland, G.G., and Richardson, B.B., 1996. Observations on the biological nature of *Plasmodium vivax* sporozoites. *Journal of Parasitology*, 216–219.

Contacos, P.G., Collins, W.E., Jeffery, G.M., Krotoski, W.A., and Howard, W.A., 1972. Studies on the characterization of *Plasmodium vivax* strains from Central America. *American Journal of Tropical Medicine and Hygiene*, 21 (5), 707–712.

Cooper, W.C., Coatney, G.R., Culwell, W.B., Eyles, D.E., and Young, M.D., 1950. Studies in human malaria XXVI. Simultaneous infection with the Chesson and the St. Elizabeth strains of *Plasmodium vivax*. *Journal of the National Malaria Society*, 9 (2), 187–90.

Cotter, C., Sturrock, H.J., Hsiang, M.S., Liu, J., Phillips, A.A., Hwang, J., Gueye, C.S., Fullman, N., Gosling, R.D., and Feachem, R.G., 2013. The changing epidemiology of malaria elimination: new strategies for new challenges. *The Lancet*, 382 (9895), 900–911.

Covell, G. and Nicol, W.D., 1951. Clinical, chemotherapeutic and immunological studies on induced malaria. *British Medical Bulletin*, 8 (1), 51–55.

Cummings, P., 2009. Methods for estimating adjusted risk ratios. *Stata Journal*, 9 (2), 175–196.

Danis, K., Baka, A., Lenglet, A., Van Bortel, W., Terzaki, I., Tseroni, M., Detsis, M., Papanikolaou, E., Balaska, A., and Gewehr, S., 2011. Autochthonous *Plasmodium vivax* malaria in Greece, 2011. *Euro Surveillance- European Communicable Disease Bulletin*, 16 (42), 20.

Douglas, N.M., Anstey, N.M., Buffet, P.A., Poespoprodjo, J.R., Yeo, T.W., White, N.J., and Price, R.N., 2012. The anaemia of *Plasmodium vivax* malaria. *Malaria Journal*, 11 (1), 135.

Ferreira, M.U., Karunaweera, N.D., da Silva-Nunes, M., da Silva, N.S., Wirth, D.F., and Hartl, D.L., 2007. Population structure and transmission dynamics of *Plasmodium vivax* in rural Amazonia. *Journal of Infectious Diseases*, 195 (8), 1218–1226.

Frischknecht, F., Baldacci, P., Martin, B., Zimmer, C., Thiberge, S., Olivo-Marin, J.-C., Shorte, S.L., and Ménard, R., 2004. Imaging movement of malaria parasites during transmission by *Anopheles* mosquitoes. *Cellular Microbiology*, 6 (7), 687–694.

Fru-Cho, J., Bumah, V.V., Safeukui, I., Nkuo-Akenji, T., Titanji, V.P., and Haldar, K., 2014. Molecular typing reveals substantial *Plasmodium vivax* infection in asymptomatic adults in a rural area of Cameroon. *Malaria Journal*, 13 (1), 170.

Galinski, M.R. and Barnwell, J.W., 2008. *Plasmodium vivax*: who cares? *Malaria Journal*, 7 Suppl 1, S9.

Gelman, A., Carlin, J.B., Stern, H.S., and Rubin, D.B., 2003. *Bayesian data analysis*. Boca Raton, London, New York, Washington, DC: Chapman and Hall/CRC press.

Gething, P.W., Elyazar, I.R.F., Moyes, C.L., Smith, D.L., Battle, K.E., Guerra, C.A., Patil, A.P., Tatem, A.J., Howes, R.E., Myers, M.F., George, D.B., Horby, P., Wertheim, H.F.L., Price, R.N., Mueller, I., Baird, J.K., and Hay, S.I., 2012. A long neglected world malaria map: *Plasmodium vivax* endemicity in 2010. *PLoS Neglected Tropical Diseases*, 6 (9), e1814.

Geweke, J., 1992. Evaluating the accuracy of sampling-based approaches to the calculation of posterior moments. *In*: *Bayesian Statistics*. Oxford: Clarendon Press, 169–193.

Glynn, J. and Bradley, D., 1995. Inoculum size, incubation period and severity of malaria. Analysis of data from malaria therapy records. *Parasitology (Cambridge)*, 110, 7–19.

Glynn, J.R., 1994. Infecting dose and severity of malaria: a literature review of induced malaria. *Journal of Tropical Medicine and Hygiene*, 97 (5), 300–316.

Goller, J.L., Jolley, D., Ringwald, P., and Biggs, B.-A., 2007. Regional differences in the response of *Plasmodium vivax* malaria to primaquine as anti-relapse therapy. *American Journal of Tropical Medicine and Hygiene*, 76 (2), 203–207.

Gonzalez-Ceron, L., Mu, J., Santillán, F., Joy, D., Sandoval, M.A., Camas, G., Su, X., Choy, E.V., and Torreblanca, R., 2013. Molecular and epidemiological characterization of *Plasmodium vivax* recurrent infections in southern Mexico. *Parasites & Vectors*, 6, 109.

Granados, T.J., 2003. Economics, demography, and epidemiology: an interdisciplinary glossary. *Journal of Epidemiology and Community Health*, 57 (12), 929.

Guerra, C.A., Howes, R.E., Patil, A.P., Gething, P.W., Van Boeckel, T.P., Temperley, W.H., Kabaria, C.W., Tatem, A.J., Manh, B.H., Elyazar, I.R.F., Baird, J.K., Snow, R.W., and Hay, S.I., 2010. The international limits and population at risk of *Plasmodium vivax* transmission in 2009. *PLoS Neglected Tropical Diseases*, 4, e774.

Guilbride, D.L., Guilbride, P.D.L., and Gawlinski, P., 2012. Malaria's deadly secret: a skin stage. *Trends in Parasitology*, 28 (4), 142–150.

Hanna, J.N., Ritchie, S.A., Brookes, D.L., Montgomery, B.L., Eisen, D.P., and Cooper, R.D., 2004. An outbreak of *Plasmodium vivax* malaria in Far North Queensland, 2002. *Medical Journal of Australia*, 180 (1).

Harcourt, B.E., 2011. Making willing bodies: the University of Chicago human experiments at Stateville Penitentiary. *Social Research*, 78 (2), 443–478.

Hargreaves, J., Yoeli, M., and Nussenzweig, R.S., 1975. Immunological studies in rodent malaria. I: Protective immunity induced in mice by mild strains of *Plasmodium* *berghei yoelii* against a virulent and fatal line of this *Plasmodium*. *Annals of Tropical Medicine and Parasitology*, 69 (3), 289–299.

Harkness JM, 1996. Nuremberg and the issue of wartime experiments on US prisoners: The Green committee. *Journal of the American Medical Association*, 276 (20), 1672–1675.

Havryliuk, T. and Ferreira, M.U., 2009. A closer look at multiple-clone *Plasmodium vivax* infections: detection methods, prevalence and consequences. *Memorias do Instituto Oswaldo Cruz*, 104 (1), 67–73.

Herrera, S., Fernandez, O., Manzano, M.R., Murrain, B., Vergara, J., Blanco, P., Palacios, R., Velez, J.D., Epstein, J.E., Chen-Mok, M., Reed, Z.H., and Arevalo-Herrera, M., 2009. Successful sporozoite challenge model in human volunteers with *Plasmodium vivax* strain derived from human donors. *American Journal of Tropical Medicine and Hygiene*, 81 (5), 740–746.

Herrera, S., Solarte, Y., Jordan-Villegas, A., Echavarria, J.F., Rocha, L., Palacios, R., Ramirez, O., Velez, J.D., Epstein, J.E., Richie, T.L., and Arevalo-Herrera, M., 2011. Consistent safety and infectivity in sporozoite challenge model of *Plasmodium vivax* in malaria-naive human volunteers. *American Journal of Tropical Medicine and Hygiene*, 84 (Suppl 2), 4–11.

Horstmann, P., 1973. Delayed attacks of malaria in visitors to the tropics. *BMJ*, 3 (5877), 440–442.

Hosmer, D.W., Hosmer, T., Le Cessie, S., and Lemeshow, S., 1997. A comparison of goodness-of-fit tests for the logistic regression model. *Statistics in Medicine*, 16 (9), 965–980.

Hosmer, D.W., Lemeshow, S., and May, S., 2011. *Applied survival analysis: regression modeling of time to event data*. 2nd ed. Hoboken, N.J.: John Wiley & Sons.

Howes, R.E., Battle, K.E., Satyagraha, A.W., Baird, J.K., and Hay, S.I., 2013. G6PD deficiency: global distribution, genetic variants and primaquine therapy. *In*: S.I. Hay, R. Price, and J. Baird, eds. *Advances in Parasitology*. Elsevier, 133–201.

Howes, R.E., Dewi, M., Piel, F.B., Monteiro, W.M., Battle, K.E., Messina, J.P., Sakuntabhai, A., Satyagraha, A.W., Williams, T.N., Baird, J.K., and Hay, S.I., 2013. Spatial distribution of G6PD deficiency variants across malaria-endemic regions. *Malaria Journal*, 12 (1), 418.

Huwaldt, J.A., 2012. *Plot Digitizer,* http://plotdigitizer.sourceforge.net/.

Imwong, M., Boel, M.E., Pagornrat, W., Pimanpanarak, M., McGready, R., Day, N.P.J., Nosten, F., and White, N.J., 2012. The first *Plasmodium vivax* relapses of life are usually genetically homologous. *Journal of Infectious Diseases*, 205 (4), 680–683.

Imwong, M., Snounou, G., Pukrittayakamee, S., Tanomsing, N., Kim, J.R., Nandy, A., Guthmann, J.-P., Nosten, F., Carlton, J., Looareesuwan, S., Nair, S., Sudimack, D., Day, N.P.J., Anderson, T.J.C., and White, N.J., 2007. Relapses of *Plasmodium vivax* infection usually result from activation of heterologous hypnozoites. *Journal of Infectious Diseases*, 195 (7), 927–933.

Ishikawa, H., Ishii, A., Nagai, N., Ohmae, H., Harada, M., Suguri, S., and Leafasia, J., 2003. A mathematical model for the transmission of *Plasmodium vivax* malaria. *Parasitology International*, 52 (1), 81–93.

James, S., Nicol, W., and Shute, P., 1936. Clinical and parasitological observations on induced malaria. *Proceedings of the Royal Society of Medicine*, 29 (8), 879–894.

James, S.P., 1931. Some general results of a study of induced malaria in England. *Transactions of the Royal Society of Tropical Medicine and Hygiene*, 24 (5), 477–525.

Jin, Y., Kebaier, C., and Vanderberg, J., 2007. Direct microscopic quantification of dynamics of *Plasmodium berghei* sporozoite transmission from mosquitoes to mice. *Infection and Immunity*, 75 (11), 5532–5539.

Joyner, C., Barnwell, J.W., and Galinski, M.R., 2015. No more monkeying around: primate malaria model systems are key to understanding *Plasmodium vivax* liver-stage biology, hypnozoites, and relapses. *Frontiers in Microbiology*, 6.

Juliano, J.J., Porter, K., Mwapasa, V., Sem, R., Rogers, W.O., Ariey, F., Wongsrichanalai, C., Read, A., and Meshnick, S.R., 2010. Exposing malaria in-host diversity and estimating population diversity by capture-recapture using massively parallel pyrosequencing. *Proceedings of the National Academy of Sciences*, 107 (46), 20138–20143.

Killick-Kendrick, R., 1974. Parasitic protozoa of the blood of rodents: a revision of *Plasmodium berghei*. *Parasitology*, 69 (02), 225–237.

Kim, S.-J., Kim, S.-H., Jo, S.-N., Gwack, J., Youn, S.-K., and Jang, J.-Y., 2013. The long and short incubation periods of *Plasmodium vivax* malaria in Korea: the characteristics and relating factors. *Infection & Chemotherapy*, 45 (2), 184–193.

Klein, M. and Kleinbaum, D.G., 2005. *Survival Analysis*. 2nd ed. New York: Springer Science+Business Media, Inc.

Knowles, G. and Walliker, D., 1980. Variable expression of virulence in the rodent malaria parasite *Plasmodium yoelii yoelii*. *Parasitology*, 81 (01), 211–219.

Krotoski, W.A., 1989. The hypnozoite and malarial relapse. *Progress in Clinical Parasitology*, 1, 1–19.

Krotoski, W.A., Garnham, P.C.C., Cogswell, F.B., Collins, W.E., Bray, R.S., Gwadz, R.W., Killick-Kendrick, R., Wolf, R.H., Sinden, R., Hollingdale, M., Lowrie, R.C., Koontz, L.C., and Stanfill, P.S., 1986. Observations on early and late post-sporozoite tissue stages in primate malaria. IV. Pre-erythrocytic schizonts and/or hypnozoites of Chesson and North Korean strains of *Plasmodium vivax* in the chimpanzee. *American Journal of Tropical Medicine and Hygiene*, 35 (2), 263–274.

Lachin, J.M., 2011. *Biostatistical methods the assessment of relative risks*. Hoboken, N.J.: Wiley.

Langhorne, J., Buffet, P., Galinski, M., Good, M., Harty, J., Leroy, D., Mota, M.M., Pasini, E., Renia, L., Riley, E., Stins, M., and Duffy, P., 2011. The relevance of non-human primate and rodent malaria models for humans. *Malaria Journal*, 10 (1), 23.

Lehane, A.M., McDevitt, C.A., Kirk, K., and Fidock, D.A., 2012. Degrees of chloroquine resistance in *Plasmodium* – Is the redox system involved? *International Journal for Parasitology: Drugs and Drug Resistance*, 2, 47–57.

Li, J., Collins, W.E., Wirtz, R.A., Rathore, D., Lal, A., and McCutchan, T.F., 2001. Geographic subdivision of the range of the malaria parasite *Plasmodium vivax*. *Emerging Infectious Diseases*, 7 (1), 35–42.

Lindsey, J.C. and Ryan, L.M., 1998. Methods for interval-censored data. *Statistics in Medicine*, 17 (2), 219–238.

Lin, J.T., Juliano, J.J., Kharabora, O., Sem, R., Lin, F.-C., Muth, S., Ménard, D., Wongsrichanalai, C., Rogers, W.O., and Meshnick, S.R., 2012. Individual *Plasmodium vivax msp1* variants within polyclonal *P. vivax* infections display different propensities for relapse. *Journal of Clinical Microbiology*, 50 (4), 1449–1451.

Liu, W., Li, Y., Shaw, K.S., Learn, G.H., Plenderleith, L.J., Malenke, J.A., Sundararaman, S.A., Ramirez, M.A., Crystal, P.A., Smith, A.G., Bibollet-Ruche, F., Ayouba, A., Locatelli, S., Esteban, A., Mouacha, F., Guichet, E., Butel, C., Ahuka-Mundeke, S., Inogwabini, B.-I., Ndjango, J.-B.N., Speede, S., Sanz, C.M., Morgan, D.B., Gonder, M.K., Kranzusch, P.J., Walsh, P.D., Georgiev, A.V., Muller, M.N., Piel, A.K., Stewart, F.A., Wilson, M.L., Pusey, A.E., Cui, L., Wang, Z., Färnert, A., Sutherland, C.J., Nolder, D., Hart, J.A., Hart, T.B., Bertolani, P., Gillis, A., LeBreton, M., Tafon, B., Kiyang, J., Djoko, C.F., Schneider, B.S., Wolfe, N.D., Mpoudi-Ngole, E., Delaporte, E., Carter, R., Culleton, R.L., Shaw, G.M., Rayner, J.C., Peeters, M., Hahn, B.H., and Sharp, P.M., 2014. African origin of the malaria parasite *Plasmodium vivax*. *Nature Communications*, 5.

Lloyd, A.L., 2001. Destabilization of epidemic models with the inclusion of realistic distributions of infectious periods. *Proceedings of the Royal Society B: Biological Sciences*, 268 (1470), 985–993.

Lover, A.A. and Coker, R.J., 2013. Quantifying effect of geographic location on epidemiology of *Plasmodium vivax* malaria. *Emerging Infectious Diseases*, 19 (7), 1058–1065.

Lu, F., Gao, Q., Chotivanich, K., Xia, H., Cao, J., Udomsangpetch, R., Cui, L., and Sattabongkot, J., 2011. *In vitro* anti-malarial drug susceptibility of temperate *Plasmodium vivax* from Central China. *American Journal of Tropical Medicine and Hygiene*, 85 (2), 197–201.

MacDonald, G., Cuellar, C.B., and Foll, C.V., 1968. The dynamics of malaria. *Bulletin of the World Health Organization*, 38 (5), 743–755.

Mackinnon, M.J., Gunawardena, D.M., Rajakaruna, J., Weerasingha, S., Mendis, K.N., and Carter, R., 2000. Quantifying genetic and nongenetic contributions to malarial infection in a Sri Lankan population. *Proceedings of the National Academy of Sciences*, 97 (23), 12661–12666.

Mackinnon, M.J. and Read, A.F., 2004. Virulence in malaria: an evolutionary viewpoint. *Philosophical Transactions of the Royal Society of London. Series B: Biological Sciences*, 359 (1446), 965–986.

Mandal, S., Sarkar, R.R., and Sinha, S., 2011. Mathematical models of malaria - a review. *Malaria Journal*, 10 (1), 202.

Marcsisin, S.R., Sousa, J.C., Reichard, G.A., Caridha, D., Zeng, Q., Roncal, N., McNulty, R., Careagabarja, J., Sciotti, R.J., Bennett, J.W., and others, 2014. Tafenoquine and NPC-1161B require CYP 2D metabolism for anti-malarial activity: implications for the 8-aminoquinoline class of anti-malarial compounds. *Malaria journal*, 13 (1), 2.

Markus, M.B., 2012. Dormancy in mammalian malaria. *Trends in Parasitology*, 28 (2), 39–45.

Mason, J., 1975. Patterns of *Plasmodium vivax* recurrence in a high-incidence coastal area of El Salvador, C.A. *American Journal of Tropical Medicine and Hygiene*, 24 (4), 581–585.

Mayne, B., 1933. The injection of mosquito sporozoites in malaria therapy. *Public Health Reports*, 48 (31), 909–916.

McKenzie, F.E., 2000. Why model malaria? *Parasitology Today*, 16 (12), 511–516.

McKenzie, F.E., Jeffery, G.M., and Collins, W.E., 2002. *Plasmodium vivax* blood-stage dynamics. *Journal of Parasitology*, 88 (3), 521–535.

McKenzie, F.E., Smith, D.L., O'Meara, W.P., and Riley, E.M., 2008. Strain theory of malaria: the first 50 years. *Advances in Parasitology*, 66, 1–46.

McQueen, P.G., 2010. Population dynamics of a pathogen: the conundrum of vivax malaria. *Biophysical Reviews*, 2, 111–120.

McQueen, P.G. and McKenzie, F.E., 2004. Age-structured red blood cell susceptibility and the dynamics of malaria infections. *Proceedings of the National Academy of Sciences*, 101 (24), 9161–9166.

McQueen, P.G. and McKenzie, F.E., 2006. Competition for red blood cells can enhance *Plasmodium vivax* parasitemia in mixed-species malaria infections. *American Journal of Tropical Medicine and Hygiene*, 75 (1), 112–125.

Medica, D.L. and Sinnis, P., 2005. Quantitative dynamics of *Plasmodium yoelii* sporozoite transmission by infected anopheline mosquitoes. *Infection and Immunity*, 73 (7), 4363–4369.

Ménard, D., Barnadas, C., Bouchier, C., Henry-Halldin, C., Gray, L.R., Ratsimbasoa, A., Thonier, V., Carod, J.-F., Domarle, O., Colin, Y., Bertrand, O., Picot, J., King, C.L., Grimberg, B.T., Mercereau-Puijalon, O., and Zimmerman, P.A., 2010. *Plasmodium vivax* clinical malaria is commonly observed in Duffy-negative Malagasy people. *Proceedings of the National Academy of Sciences*, 107 (13), 5967–5971.

Ménard, R., Tavares, J., Cockburn, I., Markus, M., Zavala, F., and Amino, R., 2013. Looking under the skin: the first steps in malarial infection and immunity. *Nature Reviews Microbiology*, 11 (10), 701–712.

Mideo, N., Barclay, V.C., Chan, B.H.K., Savill, N.J., Read, A.F., and Day, T., 2008. Understanding and predicting strain-specific patterns of pathogenesis in the rodent malaria *Plasmodium chabaudi*. *The American Naturalist*, 172 (5), E214–E238.

Mideo, N., Savill, N.J., Chadwick, W., Schneider, P., Read, A.F., Day, T., and Reece, S.E., 2011. Causes of variation in malaria infection dynamics: insights from theory and data. *The American Naturalist*, 178 (6), E174–E188.

Moon, K.T., Kim, Y.K., Ko, D.H., Park, I., Shin, D.C., and Kim, C., 2009. Recurrence rate of vivax malaria in the Republic of Korea. *Transactions of the Royal Society of Tropical Medicine and Hygiene*, 103 (12), 1245–1249.

Mouchet, J., Carnevale, P., and Manguin, S., 2008. *Biodiversity of malaria in the world*. Paris: John Libbey Eurotext.

Mueller, I., Galinski, M.R., Baird, J.K., Carlton, J.M., Kochar, D.K., Alonso, P.L., and del Portillo, H.A., 2009. Key gaps in the knowledge of *Plasmodium vivax*, a neglected human malaria parasite. *The Lancet Infectious Diseases*, 9 (9), 555–566.

Murray, C.J.L. and Global Burden of Disease Group, 2014. Global, regional, and national incidence and mortality for HIV, tuberculosis, and malaria during 1990–2013: a systematic analysis for the Global Burden of Disease Study 2013. *The Lancet*, 384 (9947), 1005–1070.

Myatt, A.V. and Coatney, G.R., 1954. Present concepts and treatment of *Plasmodium vivax* malaria. *Archives of Internal Medicine*, 93 (2), 191.

Nájera, J.A., 1989. Malaria and the work of WHO. *Bulletin of the World Health Organization*, 67 (3), 229–243.

Nájera, J.A., 1999. *Malaria control: achievements, problems and strategies*. Geneva: World Health Organization, WHO/CDS/RBM/99.10; WHO/MAL/99.1087 No. WHO/CDS/RBM/99.10; WHO/MAL/99.1087.

Nakagawa, S. and Cuthill, I.C., 2007. Effect size, confidence interval and statistical significance: a practical guide for biologists. *Biological Reviews*, 82 (4), 591–605.

Neafsey, D.E., Galinsky, K., Jiang, R.H.Y., Young, L., Sykes, S.M., Saif, S., Gujja, S., Goldberg, J.M., Young, S., and Zeng, Q., 2012. The malaria parasite *Plasmodium vivax* exhibits greater genetic diversity than *Plasmodium falciparum*. *Nature Genetics*, 44 (9), 1046–1050.

Ngassa Mbenda, H.G. and Das, A., 2014. Molecular evidence of *Plasmodium vivax* mono and mixed malaria parasite infections in Duffy-negative native Cameroonians. *PLoS ONE*, 9 (8), e103262.

Nishiura, H., Lee, H.W., Cho, S.H., Lee, W.G., In, T.S., Moon, S.U., Chung, G.T., and Kim, T.S., 2007. Estimates of short-and long-term incubation periods of *Plasmodium vivax* malaria in the Republic of Korea. *Transactions of the Royal Society of Tropical Medicine and Hygiene*, 101 (4), 338–343.

Nkhoma, S.C., Nair, S., Cheeseman, I.H., Rohr-Allegrini, C., Singlam, S., Nosten, F., and Anderson, T.J.C., 2012. Close kinship within multiple-genotype malaria parasite infections. *Proceedings of the Royal Society B: Biological Sciences*, 279 (1738), 2589–2598.

Noe, W.L., Greene, C.C., and Cheney, G., 1946. The natural course of chronic southwest Pacific malaria. *American Journal of the Medical Sciences*, 211, 215–219.

Obadia, T., Haneef, R., and Boëlle, P.-Y., 2012. The R0 package: a toolbox to estimate reproduction numbers for epidemic outbreaks. *BMC Medical Informatics and Decision Making*, 12 (1), 147.

Palmer, A.R., 2000. Quasireplication and the contract of error: lessons from sex ratios, heritabilities and fluctuating asymmetry. *Annual Review of Ecology and Systematics*, 31, 441–480.

Pampana, E., 1969. *A textbook of malaria eradication*. London: Oxford University Press.

Parner, E.T. and Andersen, P.K., 2010. Regression analysis of censored data using pseudo-observations. *Stata Journal*, 10 (3), 408–422.

Paul, R.E., Diallo, M., and Brey, P.T., 2004. Mosquitoes and transmission of malaria parasites – not just vectors. *Malaria Journal*, 3 (1), 39.

Perkins, S.L., Sarkar, I.N., and Carter, R., 2007. The phylogeny of rodent malaria parasites: Simultaneous analysis across three genomes. *Infection, Genetics and Evolution*, 7 (1), 74–83.

Petersen, M.R. and Deddens, J.A., 2006. Letter: Easy SAS calculations for risk or prevalence ratios and differences. *American Journal of Epidemiology*, 163 (12), 1158–1159.

Pianka, E.R., 2011. *Evolutionary Ecology*. Eric R. Pianka.

Pongsumpun, P. and Tang, I.M., 2007. Mathematical model for the transmission of *Plasmodium vivax* malaria. *International Journal of Mathematical Models and Methods in Applied Sciences*, 3, 117–121.

Poulin, R. and Combes, C., 1999. The concept of virulence: interpretations and implications. *Parasitology Today*, 15 (12), 474–475.

Prajapati, S.K., Joshi, H., Shalini, S., Patarroyo, M.A., Suwanarusk, R., Kumar, A., Sharma, S.K., Eapen, A., Dev, V., and Bhatt, R.M., 2011. *Plasmodium vivax* lineages: geographical distribution, tandem repeat polymorphism, and phylogenetic relationship. *Malaria Journal*, 10 (1), 1–9.

Price, R.N., Douglas, N.M., and Anstey, N.M., 2009. New developments in *Plasmodium vivax* malaria: severe disease and the rise of chloroquine resistance. *Current Opinion in Infectious Diseases*, 22 (5), 430–435.

Price, R.N., Tjitra, E., Guerra, C.A., Yeung, S., White, N.J., and Anstey, N.M., 2007. Vivax malaria: Neglected and not benign. *American Journal of Tropical Medicine and Hygiene*, 77 (6 Suppl), 79–87.

Ramasamy, R., Ramasamy, M.S., Wijesundera, D.A., Wijesundera, A.P., Dewit, I., Ranasinghe, C., Srikrishnaraj, K.A., and Wickremaratne, C., 1992. High seasonal malaria transmission rates in the intermediate rainfall zone of Sri Lanka. *Annals of Tropical Medicine and Parasitology*, 86 (6), 591–600.

Ramiro, R.S., Reece, S.E., and Obbard, D.J., 2012. Molecular evolution and phylogenetics of rodent malaria parasites. *BMC Evolutionary Biology*, 12 (1), 219.

Rattanarithikul, R., Konishi, E., and Linthicum, K.J., 1996. Detection of *Plasmodium vivax* and *Plasmodium falciparum* circumsporozoite antigen in anopheline mosquitoes collected in southern Thailand. *American Journal of Tropical Medicine and Hygiene*, 54 (2), 114–121.

R Core Team, 2013. *R: A language and environment for statistical computing*. Vienna, Austria: Foundation for Statistical Computing.

Read, A.F., Day, T., and Huijben, S., 2011. The evolution of drug resistance and the curious orthodoxy of aggressive chemotherapy. *Proceedings of the National Academy of Sciences*, 108 (Suppl 2), 10871–10877.

Reich, N.G., Lessler, J., Cummings, D.A.T., and Brookmeyer, R., 2009. Estimating incubation period distributions with coarse data. *Statistics in Medicine*, 28 (22), 2769–2784.

Restif, O., 2009. Evolutionary epidemiology 20 years on: challenges and prospects. *Infection, Genetics and Evolution*, 9 (1), 108–123.

Restrepo, E., Imwong, M., Rojas, W., Carmona-Fonseca, J., and Maestre, A., 2011. High genetic polymorphism of relapsing *P. vivax* isolates in northwest Colombia. *Acta Tropica*, 119 (1), 23–29.

De Roode, J.C., Pansini, R., Cheesman, S.J., Helinski, M.E.H., Huijben, S., Wargo, A.R., Bell, A.S., Chan, B.H.K., Walliker, D., and Read, A.F., 2005. Virulence and competitive ability in genetically diverse malaria infections. *Proceedings of the National Academy of Sciences*, 102 (21), 7624–7628.

Rosenberg, R., 2007. *Plasmodium vivax* in Africa: hidden in plain sight? *Trends in Parasitology*, 23 (5), 193–196.

Rosenberg, R., 2008. Malaria: some considerations regarding parasite productivity. *Trends in Parasitology*, 24 (11), 487–491.

Rothman, K.J., Greenland, S., and Lash, T.L., 2008. *Modern epidemiology*. Third edition. Philadelphia: Lippincott Williams & Wilkins.

Roy, M., Bouma, M.J., Ionides, E.L., Dhiman, R.C., and Pascual, M., 2013. The potential elimination of *Plasmodium vivax* malaria by relapse treatment: insights from a transmission model and surveillance data from NW India. *PLoS Negl Trop Dis*, 7 (1), e1979.

Royston, P. and Lambert, P.C., 2011. *Flexible parametric survival analysis using Stata: beyond the Cox model*. 1st ed. College Station, Texas: Stata Press.

Royston, P. and Parmar, M.K.B., 2002. Flexible parametric proportional-hazards and proportional-odds models for censored survival data, with application to prognostic modelling and estimation of treatment effects. *Statistics in Medicine*, 21 (15), 2175–2197.

Royston, P. and Parmar, M.K.B., 2011. The use of restricted mean survival time to estimate the treatment effect in randomized clinical trials when the proportional hazards assumption is in doubt. *Statistics in Medicine*, 30 (19), 2409–2421.

Russell, P.F., 1963. *Practical malariology*. 2nd ed. London, New York: Oxford University Press.

Russell, P.F., West, L.S., and Manwell, R.D., 1946. *Practical malariology*. 1st ed. Philadelphia: W. B. Saunders.

Sama, W., Dietz, K., and Smith, T., 2006. Distribution of survival times of deliberate *Plasmodium falciparum* infections in tertiary syphilis patients. *Transactions of the Royal Society of Tropical Medicine and Hygiene*, 100 (9), 811–816.

Sartwell, P.E., 1950. The distribution of incubation periods of infectious disease. *American Journal of Epidemiology*, 51 (3), 310–318.

Schwartz, E., Parise, M., Kozarsky, P., and Cetron, M., 2003. Delayed onset of malaria—implications for chemoprophylaxis in travelers. *New England Journal of Medicine*, 349 (16), 1510–1516.

Shanks, D.G., 2012. Control and elimination of *Plasmodium vivax*. *In*: S.I. Hay, R.N. Price, and J.K. Baird, eds. *Advances in Parasitology*. Oxford; London: Academic Press, 301–341.

Shanks, G.D. and White, N.J., 2013. The activation of vivax malaria hypnozoites by infectious diseases. *The Lancet Infectious Diseases*, 13 (10), 900–906.

Sheehy, S.H., Spencer, A.J., Douglas, A.D., Sim, B.K.L., Longley, R.J., Edwards, N.J., Poulton, I.D., Kimani, D., Williams, A.R., Anagnostou, N.A., Roberts, R., Kerridge, S., Voysey, M., James, E.R., Billingsley, P.F., Gunasekera, A., Lawrie, A.M., Hoffman, S.L., and Hill, A.V.S., 2013. Optimising controlled human malaria infection studies using cryopreserved *P. falciparum* parasites administered by needle and syringe. *PLoS ONE*, 8 (6), e65960.

Shute, P.G., Garnham, P.C., and Maryon, M., 1978. The Madagascar strain of *Plasmodium vivax*. *Archives de l'Institut Pasteur de Madagascar*, 47 (1), 173–183.

Shute, P.G., Lupascu, G.H., Branzei, P., Maryon, M., Constantinescu, P., Bruce-Chwatt, L.J., Draper, C.C., Killick-Kendrick, R., and Garnham, P.C.C., 1976. A strain of *Plasmodium vivax* characterized by prolonged incubation: the effect of numbers of sporozoites on the length of the prepatent period. *Transactions of the Royal Society of Tropical Medicine and Hygiene*, 70 (5), 474–481.

Sivagnanasundram, C., 1973. Reproduction rates of infection during the 1967-1968 *P. vivax* malaria epidemic in Sri Lanka (Ceylon). *The Journal of Tropical Medicine and Hygiene*, 76 (4), 83–86.

Snounou, G., Bourne, T., Jarra, W., Viriyakosol, S., Wood, J.C., and Brown, K.N., 1992. Assessment of parasite population dynamics in mixed infections of rodent plasmodia. *Parasitology*, 105 (03), 363–374.

Snounou, G. and Pérignon, J.-L., 2013. Malariotherapy – insanity at the service of malariology. *In*: S.I. Hay, J.K. Baird, and R.N. Price, eds. *Advances in Parasitology*. Elsevier, 223–255.

Song, J.Y., Park, C.W., Jo, Y.M., Kim, J.Y., Kim, J.H., Yoon, H.J., Kim, C.H., Lim, C.S., Cheong, H.J., and Kim, W.J., 2007. Two cases of *Plasmodium vivax* malaria with the clinical picture resembling toxic shock. *American Journal of Tropical Medicine and Hygiene*, 77 (4), 609–611.

Spence, P.J., Jarra, W., Lévy, P., Reid, A.J., Chappell, L., Brugat, T., Sanders, M., Berriman, M., and Langhorne, J., 2013. Vector transmission regulates immune control of *Plasmodium* virulence. *Nature*, 498 (7453), 228–231.

Spiegelhalter, D.J., Best, N.G., Carlin, B.P., and Van Der Linde, A., 2002. Bayesian measures of model complexity and fit. *Journal of the Royal Statistical Society: Series B (Statistical Methodology)*, 64 (4), 583–639.

Stapleton, D.H., 2009. Historical perspectives on malaria: the Rockefeller antimalaria strategy in the 20th century. *Mount Sinai Journal of Medicine: A Journal of Translational and Personalized Medicine*, 76 (5), 468–473.

Swellengrebel, N.H. and De Buck, A., 1938. *Malaria in the Netherlands*. 1st ed. Amsterdam: Scheltema & Holkema Ltd.

Tanner, M.A. and Wong, W.H., 1987. The calculation of posterior distributions by data augmentation. *Journal of the American Statistical Association*, 82 (398), 528–540.

Targett, G.A.T., Moorthy, V.S., and Brown, G.V., 2013. Malaria vaccine research and development: the role of the WHO MALVAC committee. *Malaria Journal*, 12 (1), 362.

Ta, T.H., Hisam, S., Lanza, M., Jiram, A.I., Ismail, N., and Rubio, J.M., 2014. First case of a naturally acquired human infection with *Plasmodium cynomolgi*. *Malaria Journal*, 13 (1), 68.

Taylor, L.H., Mackinnon, M.J., and Read, A.F., 1998. Virulence of mixed-clone and single-clone infections of the rodent malaria *Plasmodium chabaudi*. *Evolution*, 583–591.

Tiburskaja, N.A. and Vrublevskaja, O.S., 1977. *The course of infection caused by the North Korean strain of* Plasmodium vivax. Geneva: World Health Organization, No. WHO/MAL/77.985.

Ungureanu, E., Killick-Kendrick, R., Garnham, P.C.C., Branzei, P., Romanescu, C., and Shute, P.G., 1976. Prepatent periods of a tropical strain of *Plasmodium vivax* after inoculations of tenfold dilutions of sporozoites. *Transactions of the Royal Society of Tropical Medicine and Hygiene*, 70 (5), 482–483.

Vanderberg, J.P., 2014. Imaging mosquito transmission of *Plasmodium* sporozoites into the mammalian host: Immunological implications. *Parasitology International*, 63 (1), 150–164.

Verhave, J.P., 2013. Experimental, therapeutic and natural transmission of *Plasmodium vivax* tertian malaria: scientific and anecdotal data on the history of Dutch malaria studies. *Parasites & Vectors*, 6 (1), 19.

Warrell, D.A. and Gilles, H.M., eds., 2002. *Essential Malariology*. 4th ed. London: Hodder Arnold Publishers.

Wearing, H.J., Rohani, P., and Keeling, M.J., 2005. Appropriate models for the management of infectious diseases. *PLoS Med*, 2 (7), e174.

Weijer, C., 1999. Another Tuskegee? *American Journal of Tropical Medicine and Hygiene*, 61 (Suppl 1), 1–3.

Wernsdorfer, W.H. and Gregor, I.M., eds., 1988. *Malaria: Principles and Practice of Malariology*. Edinburgh, New York: Churchill Livingstone.

White, N.J., 2011. Determinants of relapse periodicity in *Plasmodium vivax* malaria. *Malaria Journal*, 10 (1), 297.

White, N.J. and Imwong, M., 2012. Relapse. *In*: S.I. Hay, R. Price, and J. Baird, eds. *Advances in Parasitology*. Oxford; London: Academic Press, 113–150.

White, N.J., Turner, G.D.H., Medana, I.M., Dondorp, A.M., and Day, N.P.J., 2010. The murine cerebral malaria phenomenon. *Trends in Parasitology*, 26 (1), 11–15.

WHO, 2014. *Global strategic plan for* P. vivax *control and elimination*. *http://www.who.int/malaria/mpac/mpac_mar2014_global_strategic_plan_p_vivax.pdf [accessed Sept 2014]*.

WHO Malaria Policy Advisory Committee and Secretariat, 2013. Malaria Policy Advisory Committee to the WHO: conclusions and recommendations of March 2013 meeting. *Malaria Journal*, 12 (1), 213.

Whorton, C.M., Kirschbaum, W., Pullman, T.N., Jones, R., Craige, B., and Alving, A.S., 1947. The Chesson strain of *Plasmodium vivax* malaria I. Factors influencing the incubation period. *Journal of Infectious Diseases*, 80 (3), 223–227.

Wilcox, A., Jeffery, G.M., and Young, M.D., 1954. The Donaldson strain of malaria II. morphology of the erythrocytic parasites. *American Journal of Tropical Medicine and Hygiene*, 3 (4), 638–649.

Williams, R.L., 2000. A note on robust variance estimation for cluster-correlated data. *Biometrics*, 56 (2), 645–646.

Winckel, C., 1955. Long latency in *Plasmodium vivax* infections in a temperate zone. *Documenta de Medicina Geographica et Tropica*, 7 (3), 292–298.

Winckel, C.W.F., 1941. Are the experimental data of therapeutic malaria applicable to conditions obtaining in nature? *American Journal of Tropical Medicine and Hygiene*, 1 (6), 789–794.

World Health Organization, 1969. *Parasitology of malaria: Report of a WHO scientific group*. Geneva: World Health Organization, No. 433, WHO Technical Report Series.

World Health Organization, 2013a. *World Malaria Report 2013*. Geneva.

World Health Organization, 2013b. Plasmodium vivax *control & elimination: development of global strategy and investment case*. Geneva: WHO.

Yang, B., 1996. Experimental and field studies on some biological characteristics of *Plasmodium vivax* isolated from tropical and temperate zones of China. *Chinese Medical Journal*, 109 (4), 266–271.

Young, M.D., Ellis, J.M., and Stubbs, T.H., 1947. Some characteristics of foreign vivax malaria induced in neurosyphilitic patients. *American Journal of Tropical Medicine and Hygiene*, s1-27 (5), 585–596.

Zimmerman, P.A., Ferreira, M.U., Howes, R.E., and Mercereau-Puijalon, O., 2013. Red blood cell polymorphism and susceptibility to *Plasmodium vivax*. *In*: S.I. Hay, J.K. Baird, and R.N. Price, eds. *Advances in Parasitology*. Elsevier, 27–76.

Zimmerman, P.A., Mehlotra, R.K., Kasehagen, L.J., and Kazura, J.W., 2004. Why do we need to know more about mixed *Plasmodium* species infections in humans? *Trends in Parasitology*, 20 (9), 440–447.

Zou, G., 2004. A modified Poisson regression approach to prospective studies with binary data. *American Journal of Epidemiology*, 159 (7), 702–706.

De Zoysa, A.P., Mendis, C., Gamage-Mendis, A.C., Weerasinghe, S., Herath, P.R., and Mendis, K.N., 1991. A mathematical model for *Plasmodium vivax* malaria transmission: estimation of the impact of transmission-blocking immunity in an endemic area. *Bulletin of the World Health Organization*, 69 (6), 725.

# Appendices

## Appendix A  - Chapter 2 additional files

Supplemental information for Chapter 2 can be found at:

http://wwwnc.cdc.gov/eid/article/19/7/12-1674-techapp1.pdf

## Appendix B  - Chapter 4 additional files

Supplemental information for Chapter 4 can be found at:

http://www.biomedcentral.com/imedia/1158833051144777/supp1.docx

# Appendix C  - Copyrighted material release

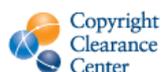

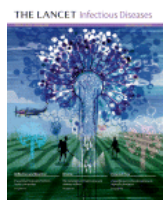

| | |
|---|---|
| **Title:** | Key gaps in the knowledge of Plasmodium vivax, a neglected human malaria parasite |
| **Author:** | Ivo Mueller,Mary R Galinski,J Kevin Baird,Jane M Carlton,Dhanpat K Kochar,Pedro L Alonso,Hernando A del Portillo |
| **Publication:** | The Lancet Infectious Diseases |
| **Publisher:** | Elsevier |
| **Date:** | September 2009 |

Copyright © 2009 Elsevier Ltd. All rights reserved.

Logged in as:
Andrew Lover

LOGOUT

## Order Completed

Thank you very much for your order.

This is a License Agreement between Andrew A Lover ("You") and Elsevier ("Elsevier"). The license consists of your order details, the terms and conditions provided by Elsevier, and the payment terms and conditions.

Get the printable license.

| | |
|---|---|
| License Number | 3481120404906 |
| License date | Oct 02, 2014 |
| Licensed content publisher | Elsevier |
| Licensed content publication | The Lancet Infectious Diseases |
| Licensed content title | Key gaps in the knowledge of Plasmodium vivax, a neglected human malaria parasite |
| Licensed content author | Ivo Mueller,Mary R Galinski,J Kevin Baird,Jane M Carlton,Dhanpat K Kochar,Pedro L Alonso,Hernando A del Portillo |
| Licensed content date | September 2009 |
| Licensed content volume number | 9 |
| Licensed content issue number | 9 |
| Number of pages | 12 |
| Type of Use | reuse in a thesis/dissertation |
| Portion | figures/tables/illustrations |
| Number of figures/tables /illustrations | 1 |
| Format | both print and electronic |
| Are you the author of this Elsevier article? | No |
| Will you be translating? | No |
| Title of your thesis/dissertation | Epidemiology of latency and relapse in Plasmodium vivax malaria |
| Expected completion date | Jan 2015 |
| Estimated size (number of pages) | 175 |
| Elsevier VAT number | GB 494 6272 12 |
| Permissions price | 0.00 USD |
| VAT/Local Sales Tax | 0.00 USD / 0.00 GBP |